\documentclass[12pt,a4paper]{article}
\pdfoutput=1
\usepackage{amssymb,amsmath}
\usepackage{graphicx}
\usepackage{latexsym}
\usepackage[colorlinks,allcolors=blue,linktocpage]{hyperref}
\usepackage{color}
\usepackage{rotating}
\usepackage{wrapfig}

\newcommand{\ba}{\begin{array}}
\newcommand{\ea}{\end{array}}
\newcommand{\be}{\begin{equation}}
\newcommand{\ee}{\end{equation}}
\newcommand{\bea}{\begin{eqnarray}}
\newcommand{\eea}{\end{eqnarray}}
\newcommand{\bg}{\begin{gather}}
\newcommand{\eg}{\end{gather}}
\newcommand{\bseq}{\begin{subequations}}
\newcommand{\eseq}{\end{subequations}}

\renewcommand{\Im}{\mathop{\rm Im}\nolimits}

\makeatletter

\def\compoundrel#1\over#2{\mathpalette\compoundreL{{#1}\over{#2}}}
\def\compoundreL#1#2{\compoundREL#1#2}
\def\compoundREL#1#2\over#3{\mathrel
         {\vcenter{\hbox{$\m@th\buildrel{#1#2}\over{#1#3}$}}}}
\makeatother

\numberwithin{equation}{section}

\begin{document}
$\mbox{ }$

\vspace{-30mm}
\begin{flushright}
INR--TH--2015--023
\end{flushright}

\vspace{5mm}
\begin{center}

    {\LARGE Semiclassical description of soliton--antisoliton pair
      production in particle collisions\\
\bigskip
{\large
    S.V.~Demidov\footnote{{\bf E-mail}: demidov@ms2.inr.ac.ru}, 
    D.G.~Levkov\footnote{{\bf E-mail}: levkov@ms2.inr.ac.ru}}
    \\
    \medskip
    {\small{\em 
	Institute for Nuclear Research of the Russian Academy of Sciences, }}\\
      {\small{\em
	  60th October Anniversary prospect 7a, Moscow 117312, Russia
      }
      }}
      \\
  \end{center}

\begin{abstract}
  We develop a consistent semiclassical method to calculate the
  probability of topological soliton--antisoliton pair production in
  collisions of elementary particles. In our method one adds an 
  auxiliary external field pulling the soliton and antisoliton in the
  opposite directions. This transforms the original scattering
  process into a Schwinger pair creation of the solitons induced
  by the particle collision. One describes the Schwinger process
  semiclassically and recovers the original scattering probability in
  the   limit of vanishing external field. We illustrate the method in
  $(1+1)$--dimensional scalar field model where the suppression
  exponents of soliton--antisoliton production in the
  multiparticle and two--particle collisions are computed numerically. 
\end{abstract}

PACS numbers: 11.15.Kc, 05.45.Yv

\newpage 
\tableofcontents

\section{Introduction}
\label{sec:1}
An intriguing possibility of observing nonperturbative phenomena in
particle collisions and in the early Universe triggered development of
powerful semiclassical methods for description of false
vacuum decay~\cite{Okun, Stone, Coleman}, instanton--like
transitions~\cite{Belavin:1975fg, Rubakov_Shaposhnikov} and 
multiparticle production~\cite{Son:1995wz,
  Landau_khlebnikov}. Although these processes occur with exponentially
small rates~\cite{banks, unitarity, voloshin_induced,
  kuznetsov_tinyakov, rst, rst2, Demidov:2015} in weakly interacting 
theories, they may be important for
baryogenesis~\cite{Kuzmin:1985mm, Rubakov_Shaposhnikov}, stability of
the Higgs vacuum~\cite{Krive:1976sg}, or violation of the baryon and
lepton numbers at future colliders~\cite{Ringwald:1989ee}.

\begin{figure}[t]
\centerline{\includegraphics[width=9cm]{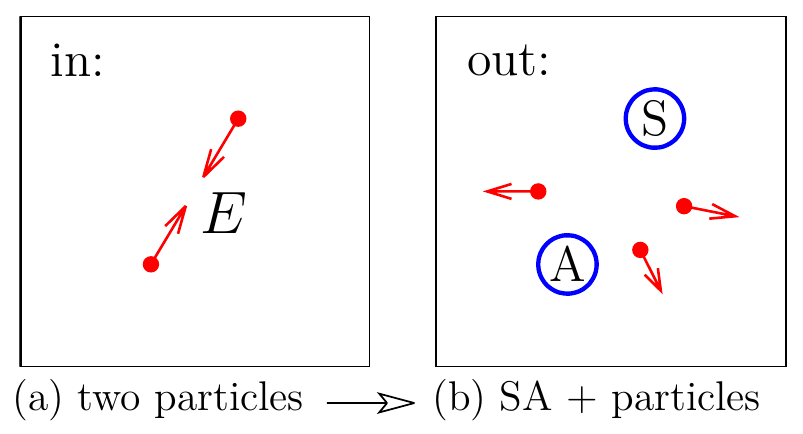}\hspace{10mm}
\includegraphics[width=5.5cm]{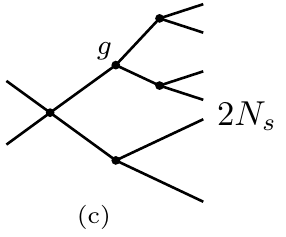}}

\caption{(a), (b) Production of the soliton--antisoliton (SA) pair in the
  two--particle collision. (c) A tree--level diagram with $2N_S$
  particles in the final state.\label{fig:collision}}
\end{figure}

We consider nonperturbative transitions of a new type: classically
forbidden creation of topological soliton--anti\-so\-li\-ton pairs in
$N$--particle collisions, see the sketch of the $N=2$ process in
Figs.~\ref{fig:collision}a,b. At weak coupling the probabilities of
these transitions are expected to be exponentially suppressed. Indeed,
the 
solitons can  be crudely regarded~\cite{Drukier:1981fq} as bound
states of $N_S \propto 1/g^{2}$ particles, where $g\ll 1$ is the
coupling constant. The cross section of their pair production in
the collision of 
few particles involves the   multiparticle factor~\cite{Son:1995wz}
${\cal  P}_{\mathrm{few}}\sim g^{4N_S}\, (2N_S)!\, \sim
\exp(-\mbox{const}/g^2)$ due to roughly $(2N_S)!$ tree--level diagrams 
exemplified in Fig.~\ref{fig:collision}c. This argument,
however hand--waving\footnote{In particular, loop corrections 
  are shown~\cite{Son:1995wz} to be large at  $g^2 N_S \gtrsim   1$.}, 
suggests a general expression for the probability
\begin{equation}
\label{eq:2}
{\cal P}_N(E)\sim {\rm e}^{-F_N(E)/g^2}\;
\end{equation}
of the inclusive process $N \mbox{ particles} \to \mbox{solitons}  +
\mbox{particles}$; here 
$F_N(E)$ is the multiparticle suppression exponent. The
suppression should disappear at sufficiently large number of colliding
particles $N \gtrsim 2N_S$ when creation of the solitons 
proceeds classically. Presently the form~\eqref{eq:2} is 
supported by unitarity arguments~\cite{unitarity} and
recent calculation~\cite{Papageorgakis:2014dma} in a quantum
mechanical model describing the soliton  moduli space. However, no
reliable field theoretical method for evaluating the exponent $F_N(E)$
is known. Below we propose a general semiclassical method of this kind
which is applicable, in particular, to the processes with $N=2$
initial particles.   

Recently several dynamical mechanisms for enhancing the rate of
soliton--antisoliton production in particle collisions were
proposed~\cite{Romanczukiewicz:2005rm}. They are supported by classical
arguments which cannot be directly extended  to quantum
level. Our method for
computing the exponent in Eq.~(\ref{eq:2}) will be valuable for tests
of these mechanisms and their phenomenological applications.

So far the processes in Figs.~\ref{fig:collision}a,b eluded
semiclassical treatment. The reason can be traced back to the
attraction between the topological soliton and
antisoliton~\cite{Manton:1977er}: 
taken at rest, they accelerate towards each  
other and annihilate\footnote{We do not consider integrable models
  where annihilation of the soliton and antisoliton may be 
  forbidden by conservation laws.} into many particles. Thus, the
potential barrier between the soliton--antisoliton pair and 
multiparticle states is absent. As a consequence, classically forbidden
transitions from particles to solitons cannot be directly described by
the powerful semiclassical methods developed for tunneling
processes.  

In this paper which extends Ref.~\cite{Demidov:2011dk} we solve the above
difficulty by relating production of the solitons from particles
to their Schwinger pair creation in external field. We illustrate
our method with explicit numerical calculations and present details of
the numerical techniques. 

The above relation is established in the following way. By
definition the topological soliton and antisoliton can be equipped
with the topological charges $\pm q$; we denote by $J_\mu$ the
respective topological current,
$\partial_\mu J^\mu = 0$. We introduce an external $U(1)$ field
$A_\mu^{\mathrm{ext}}$ coupled to this current\footnote{The
  interaction in Eq.~(\ref{eq:12}) does not need to be consistent at
  the quantum level.   It is introduced as an auxiliary semiclassical
  tool and will be switched off in the end of the calculations.}, 
\begin{equation}
\label{eq:12}
\Delta {S} = - \int J^\mu A_\mu^{\mathrm{ext}}\, d^d x\;,
\end{equation}
where $d$ is the number of spacetime dimensions.
We consider the field $A_\mu^{\mathrm{ext}} = (-{\cal  E}_ix^i,\; 0)$ with 
the constant ``electric'' component $\mathbf{\cal E}$ which 
pulls the soliton and antisoliton in the
opposite  
\begin{figure}
\centerline{\includegraphics[width=1.7cm,angle=90]{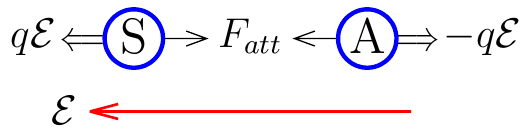}}
\vspace{-2mm}
\caption{Soliton--antisoliton pair in the external field $\mathbf{\cal
    E}$.\label{fig:SA_field} } 
\end{figure}
directions, see Fig.~\ref{fig:SA_field}. 
Then the solitons are Schwinger pair produced like the opposite
charges in the ordinary electric field. In our 
method one starts from the 
well--known 
semiclassical solution describing spontaneous creation of the soliton
pair at ${{\cal E} \ne 0}$. Then one adds $N$ particles with energy $E$
to the initial state and obtains solutions for the collision--induced
Schwinger processes, cf.~\cite{induced_schwinger}. Finally, one
raises the particle energy above the 
kinematic threshold $2M_S$ of soliton--antisoliton production, where
$M_S$ is the soliton mass, and  takes the limit
of vanishing external field  ${\cal E} \to 0$. In this way one arrives
at the semiclassical solutions describing creation of the solitons
from particles. The semiclassical suppression exponent $F_N(E)$ is
computed using these solutions.

To be concrete, imagine that our method is applied to pair production
of 't Hooft--Polyakov monopoles~\cite{monopole} in particle collisions. The
modification term~(\ref{eq:12}) in this case couples 
gauge--invariant magnetic current~\cite{Rajaraman:1982is} $J_\mu$
to a small external field ${\cal E}$. One can show~\cite{Affleck:1981}
that the latter is equivalent to external magnetic field. Then the
monopole--antimonopole pairs are created in the Schwinger 
process in accordance with the  
electric--magnetic duality. One therefore obtains the semiclassical
solutions relevant for the monopole--antimonopole production in
particle collisions from the  solutions~\cite{Affleck:1981,
  induced_schwinger1} describing their creation in the magnetic field.  

The main technical difficulty of our method is related to 
the semiclassical description of the collision--induced Schwinger
processes. Below we use the approach of~\cite{Rubakov:1992ec,
  kuznetsov_tinyakov, rst} which involves a family of
complex semiclassical solutions satisfying a certain boundary value
problem. The suppression exponent $F_N(E)$ is calculated as a
functional on these solutions.  In the end of the calculation we send
${\cal E}\to 0$.  

An obstacle to the semiclassical method appears in the
phenomenologically interesting case of $N=2$ colliding
particles: the two--particle initial state cannot be described
semiclassically. We overcome this obstacle by 
appealing to the Rubakov--Son--Tinyakov conjecture\footnote{An
  alternative strategy involving singular solutions is suggested in~\cite{Landau,  voloshin_induced, 
    Landau_khlebnikov}.}~\cite{Rubakov:1992ec} which states that the
multiparticle exponent $F_N(E)$ does not depend on the initial
particle number $N$ at
semiclassically small values of the latter i.e.\ at ${N \ll 1/g^2}$, 
see~\cite{rst_confirmations,Rebbi_QM,rst_confirmations2} for
confirmations. This means that the  
two--particle exponent $F_2(E)$ coincides with $F_N(E)$ in the 
limit $g^2 N \to 0$. Calculating semiclassically\footnote{Recall that
  the semiclassical description is valid at $N \gg 1$, which can hold 
  even in the limiting region $g^2 N  \ll 1$.} the
latter exponent and taking the limit, we obtain $F_2(E)$. 

\begin{figure}[t]
\begin{center}

\vspace{-.3\unitlength}
\begin{minipage}{5\unitlength}
\vspace{.45\unitlength}
\includegraphics[width=5\unitlength]{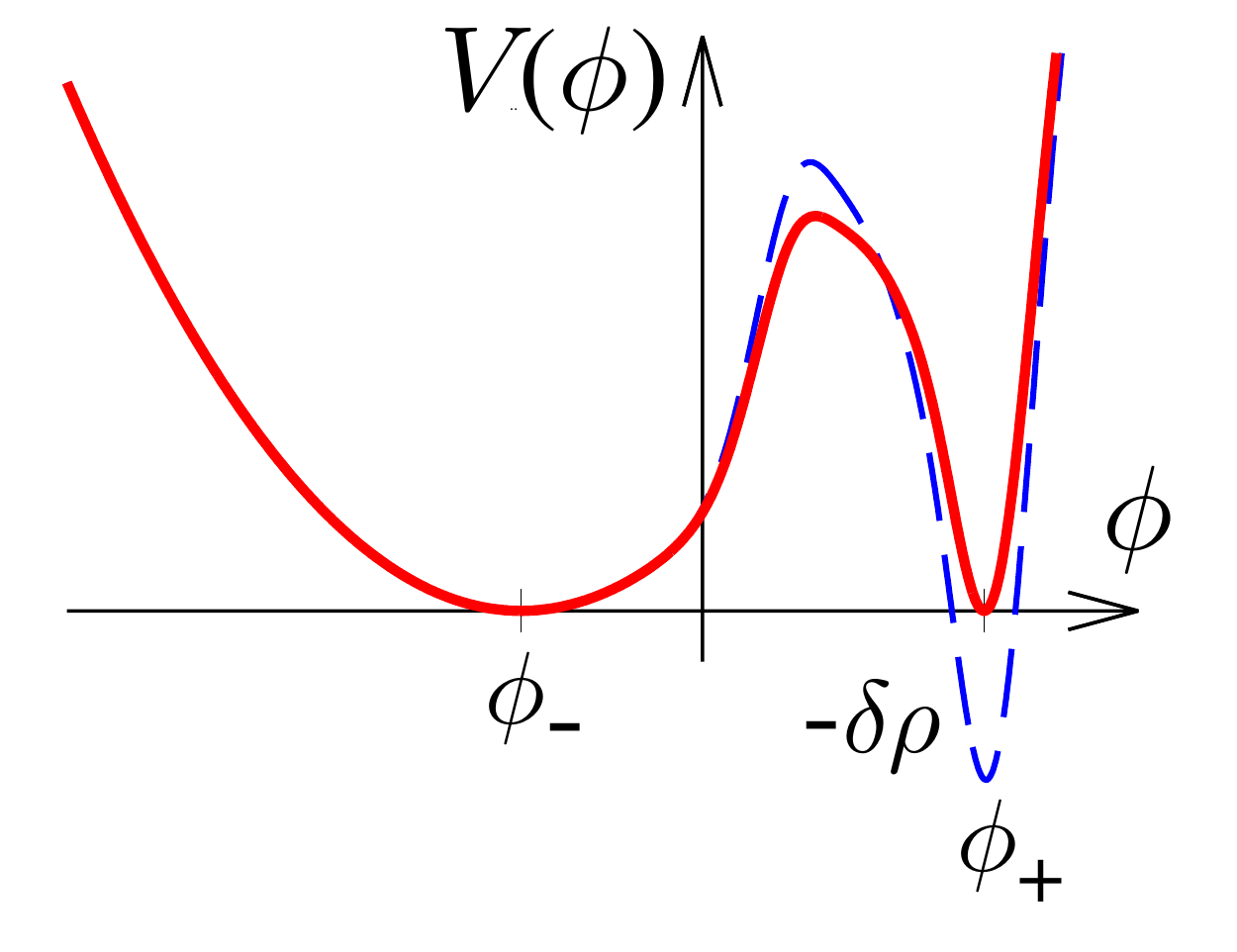}
\end{minipage}\hspace{0.3\unitlength}
\begin{minipage}{6.25\unitlength}

\vspace{.93\unitlength}
\includegraphics[width=6.25\unitlength]{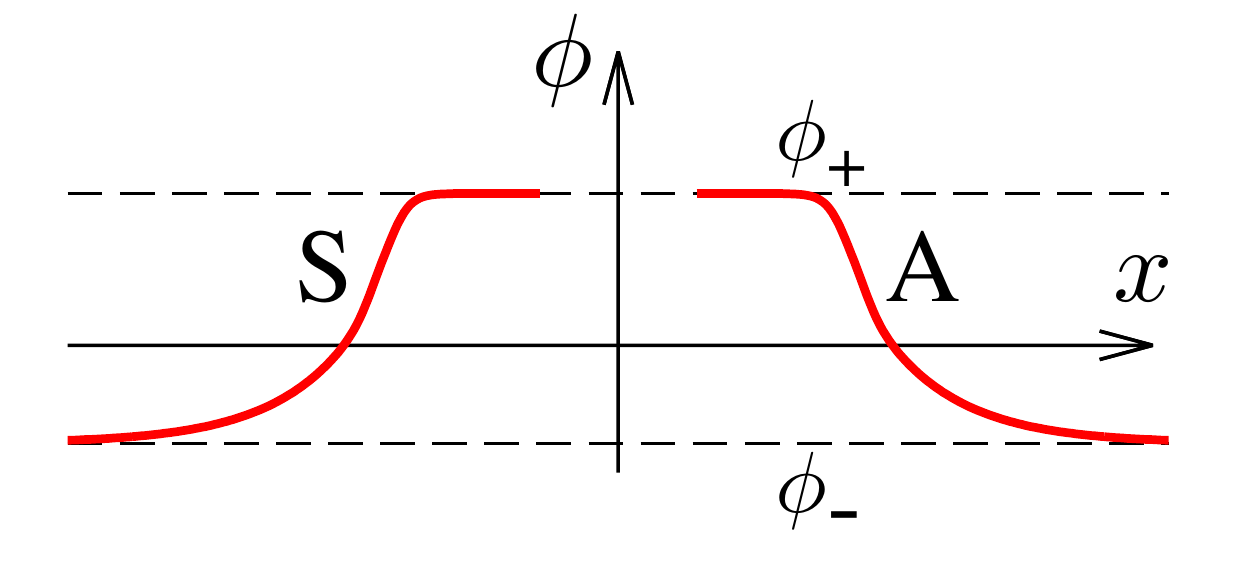}
\end{minipage}\hspace{.4\unitlength}
\begin{minipage}{3.7\unitlength}

\includegraphics[width=3.7\unitlength]{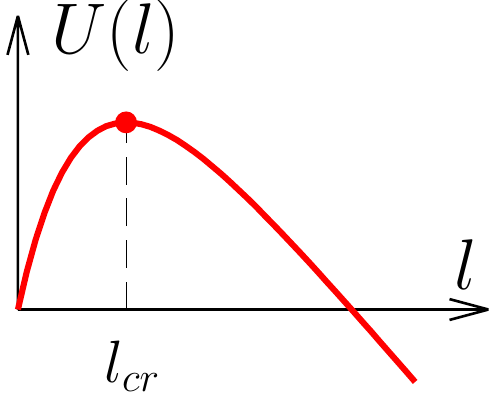}
\end{minipage}

\vspace{-4mm}
\hspace{0.9\unitlength}(a) \hspace{4.9\unitlength} (b) \hspace{4.7\unitlength} (c)
\end{center}

\vspace{-3mm}
\caption{\label{fig1}  (a) Scalar potential $V(\phi)$ (solid line) and
  its modification (dashed line). (b)~Soliton (S) and antisoliton (A)
  solutions. (c)~Potential energy of their interaction.}  
\end{figure}

We illustrate the above semiclassical method in the
$(1+1)$--dimensional scalar field model,
\begin{equation}
\label{eq:1}
S = \frac{1}{g^2}\int dt\, dx\left[\frac12 \left(\partial_{\mu}\phi\right)^2 -
  V(\phi)\right]\;,
\end{equation}
where the coupling constant\footnote{Field rescaling
  $\phi \to g\phi$ brings this constant in front of the interaction
  terms in the action.} $g$ is small, $x^\mu = (t,x)$, and the
scalar potential $V(\phi)$ has two degenerate vacua $\phi =
\phi_{\pm}$, see Fig.~\ref{fig1}a, solid line.
\unitlength=0.95cm
The model~\eqref{eq:1} possesses a pair of kink--like static solutions
interpolating between the vacua: topological soliton (S) and
antisoliton (A) shown in Fig.~\ref{fig1}b. Below we consider 
their pair production in the $N$--particle collisions. 

\begin{sloppy}

Following the general strategy outlined above, we introduce an
external field 
${A_\mu^{\mathrm{ext}} = (-{\cal E}x,\, 0)}$ coupled to the 
topological current\footnote{The related topological charge is
  $\int J_0\, dx \equiv \phi(x = +\infty) - \phi(x = -\infty)$.}
$J^\mu = \epsilon^{\mu\nu} \, \partial_\nu\phi$, 
where ${\cal E}$ is negative in accordance with
Fig.~\ref{fig:SA_field} and $\epsilon^{\mu\nu}$ is the 
antisymmetric symbol in two dimensions. Integrating by parts, we
rewrite the modification term~(\ref{eq:12}) as
\begin{equation}
\label{eq:32}
\Delta S =  -{\cal E}
  \int \phi \; dt \, dx   \;.
\end{equation} 
This interaction can be absorbed in the redefinition of the scalar
potential $V \to V +g^2 {\cal E} \phi$. After that $\phi_-$ and 
$\phi_+$ become false  and true vacua, $V(\phi_-) > V(\phi_+)$. One
concludes that the Schwinger pair production of the kink--like
solitons in the external field ${\cal E}$ is equivalent~\cite{Stone}
to the spontaneous false vacuum decay via the nucleation of true
vacuum bubbles. The latter is a celebrated tunneling process with
well--known semiclassical description~\cite{Coleman}. 

\end{sloppy}

In what follows we modify the scalar potential $V(\phi)$ giving 
negative energy density $V(\phi_+) = -\delta \rho$ to the true vacuum, 
$\delta \rho \propto |{\cal E}|$. We calculate the rate of false
vacuum decay accompanied by the $N$--particle
collision~\cite{Voloshin:1986zq, voloshin_induced, kuznetsov_tinyakov,
Demidov:2015}
and remove the modification $\delta \rho\to 0$ in the end of the
calculation.   

Let us visualize the potential barrier between the
soliton--antisoliton pair and multiparticle sector of the modified
system. Kink--like soliton and antisoliton at distance $l$
attract each other with Yukawa force $F_{att} \propto  \mathrm{e}^{-
  m_+ l}$, where $m_+^2 = V''(\phi_+)$,
cf.~\cite{Manton:1977er}. Besides, they are pushed apart by the
pressure $\delta \rho$. This leads to the potential energy of the
static soliton--antisoliton pair 
\begin{equation}
\label{eq:33} 
U(l) \sim - U_0 \, \mathrm{e}^{-m_+ l} - l\, \delta \rho + \mbox{const}
\end{equation} 
shown in Fig.~\ref{fig1}c. The barrier $U(l)$ separates the
multiparticle sector of the theory from the static
soliton--antisoliton pairs with ${l>l_{cr}}$, where $l_{cr} \sim
|{\log(\mathrm{const}\cdot 
  \delta\rho)|/m_+}$ is the critical distance corresponding to the
barrier top.  The pairs with $l<l_{cr}$ annihilate and therefore
should be attributed to the multiparticle sector. Unstable static
solution  $\phi_{cb}(x)$ describing the soliton and antisoliton at a
distance $l = 
l_{cr}$ is the famous critical bubble~\cite{Okun, Stone, Coleman}; its
energy $E_{cb}$ determines the height of the potential barrier
between the solitons and multiparticle states. As
$\delta \rho \to 0$, the critical bubble turns into infinitely
separated soliton and antisoliton at rest: $l_{cr}\to +\infty$ and
$E_{cb} \to 2M_{S}$. In this limit the potential barrier is 
hidden below the kinematic threshold~$E=2M_{S}$ of soliton pair
production. 

\begin{figure}[t]
\begin{center}
(a)\hspace{7.9cm}(b)

\begin{minipage}{0.46\textwidth}
\includegraphics[width=\textwidth]{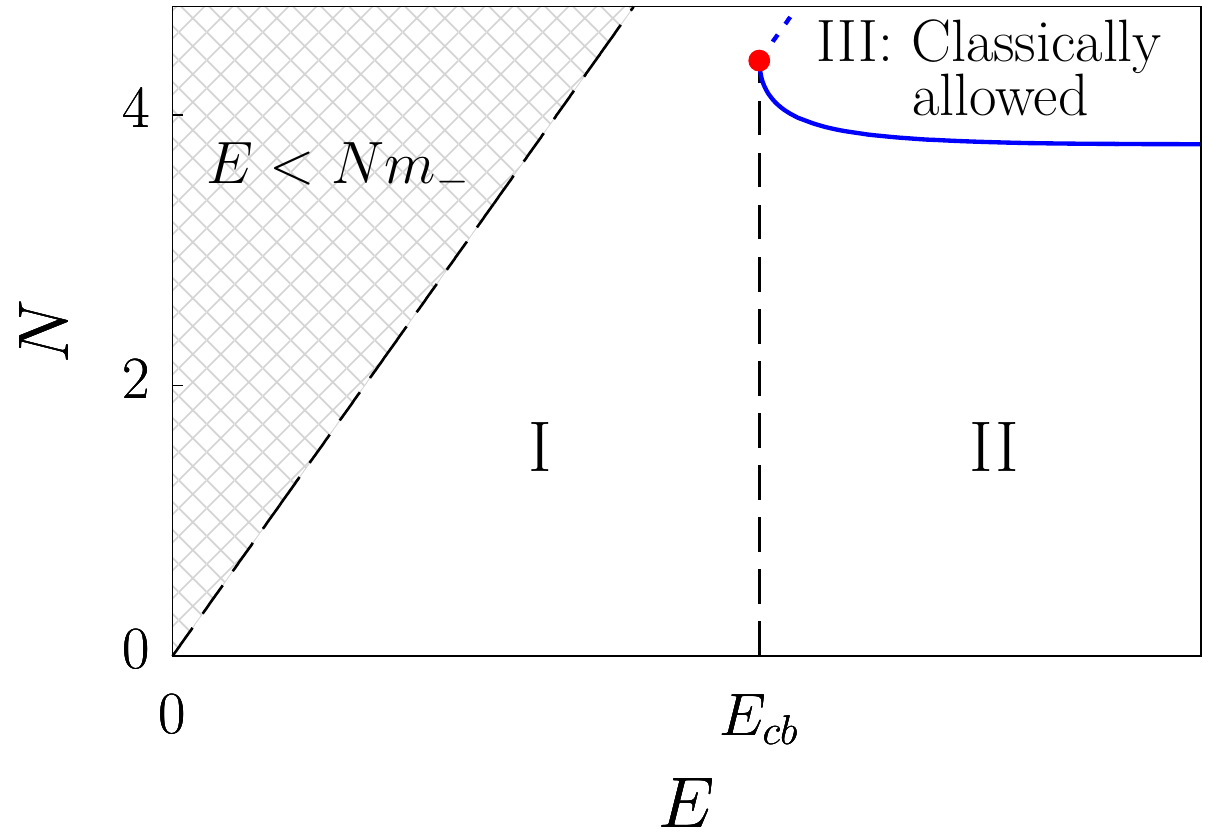}
\end{minipage}\hspace{5mm}
\begin{minipage}{0.497\textwidth}
\includegraphics[width=\textwidth]{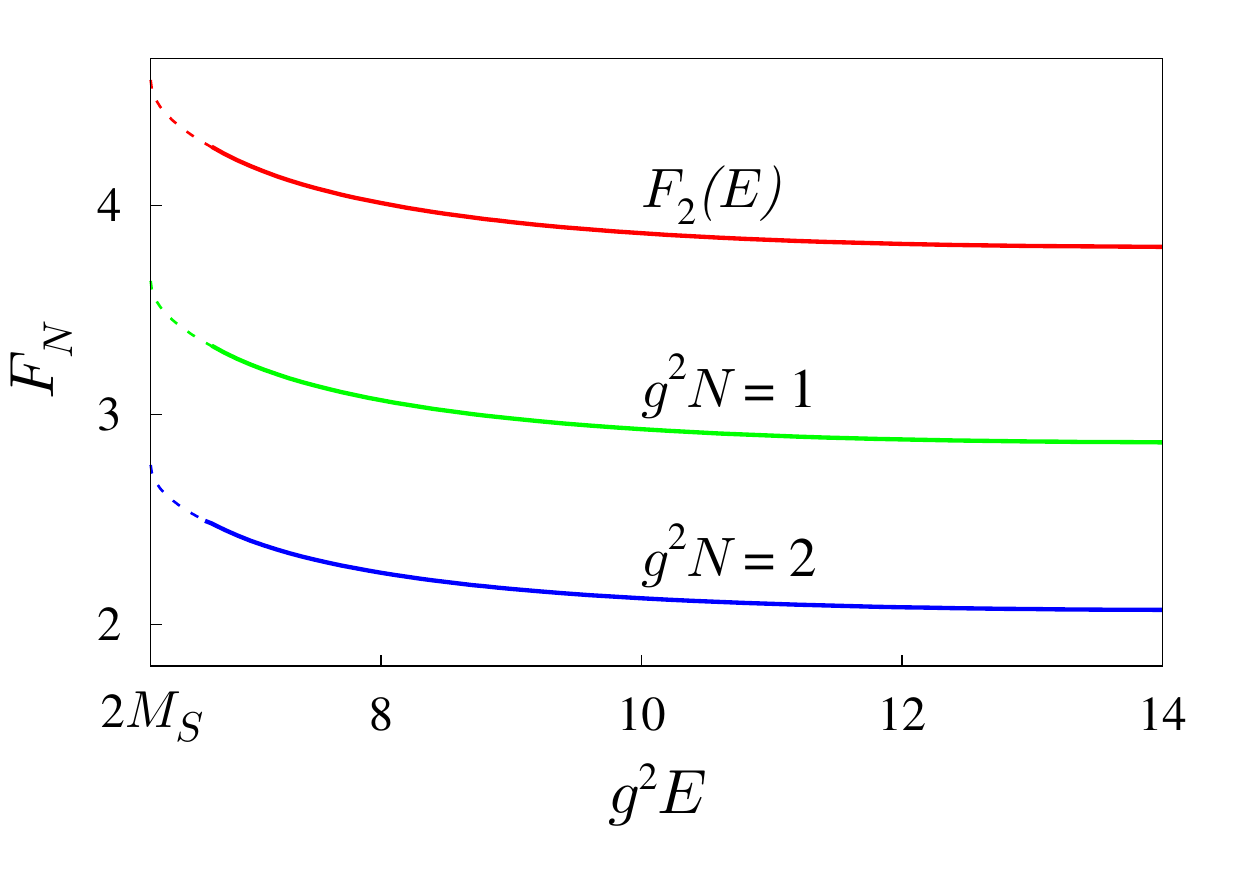}

\vspace{1.2mm}
\end{minipage}

\vspace{-7mm}
\end{center}
\caption{(a) Regimes of transitions in the $(E,\, N)$ plane of initial data. (b)
  Numerical results for the suppression exponent $F_N(E)$ of
  soliton--antisoliton production in the model~(\ref{eq:1}).\label{EN_schema} }
\end{figure}

False vacuum decay at $E=N=0$ (no initial particles) is described
by the bounce solution~\cite{Coleman}. In the main body of the
paper we numerically\footnote{We use the scalar potential shown in
  Fig.~\ref{fig1}a; expression for $V(\phi)$ will be given in the main
  text.} find similar semiclassical solutions $\phi_s(t,\, x)$ for the
decay accompanied by the collision of $N$ particles at energy
$E$. We observe several regimes of transitions summarized in
Fig.~\ref{EN_schema}a.  
The shaded region $E<Nm_{-}$, where $m_-$ is the particle mass in the
false vacuum, is kinematically forbidden. In the
opposite case of high energies and $N \gtrsim 4/g^2$ (region III
in Fig.~\ref{EN_schema}a) transitions proceed classically and
we obtain $F_N(E) = 0$. The latter classical regime is investigated
in~\cite{Demidov:2011eu}, see also~\cite{Rebbi:1996zx}. 
\begin{sloppy}
 
At smaller multiplicities the false vacuum decay is classically 
forbidden, ${F_N(E) > 0}$. We find that the properties of the
respective semiclassical solutions $\phi_s(t,\, x)$ are qualitatively
different at energies below and above the height $E_{cb}$
of the potential barrier (regions I and II in
Fig.~\ref{EN_schema}a). Solutions with $E<E_{cb}$ 
resemble the bounce: they describe formation of large true vacuum
bubbles and imply infinite suppression\footnote{This property can be
  deduced from the thin--wall approximation~\cite{Voloshin:1986zq,
    Rubakov:1992gi} which predicts large suppression at
  low energies but breaks down at~$E\gtrsim 2M_S$.} $F_N(E)\propto
\delta \rho^{-1}$ in the limit $\delta \rho \to 0$. High--energy solutions 
from the region II in Fig.~\ref{EN_schema}a have smaller spatial extent. 
They approach the critical bubble $\phi_{cb}(x)$
plus outgoing waves as $t\to +\infty$. Thus, at infinitesimally small
$\delta \rho$ they describe creation of widely separated
soliton--antisoliton pairs at rest. We explicitly demonstrate that the
respective value of the suppression exponent $F_N(E)$ approaches a
finite value as~$\delta \rho \to 0$. 

\end{sloppy}
Our numerical results for the suppression exponent of producing
the soliton pair from particles at $\delta \rho=0$ are shown in 
Fig.~\ref{EN_schema}b. The two--particle exponent $F_2(E)$ is obtained
by evaluating the additional limit $g^2 N \to 0$. 
One observes that at high energies $F_N(E)$ is finite and (almost)
energy--independent which is the expected
behavior~\cite{unitarity, voloshin_induced, Demidov:2015}. 

The paper is organized as follows. We introduce the semiclassical
method in Sec.~\ref{sec:2} and numerical technique in
Sec.~\ref{sec:numerical-method}. Numerical solutions and results
for the suppression exponent are described in Sec.~\ref{sec:3}. We
conclude in Sec.~\ref{sec:6}. Technical details  are presented in 
Appendices.

\section{The semiclassical boundary value problem}
\label{sec:2}
Consider false vacuum decay induced by a collision of $N$ particles
at energy $E$. The inclusive probability of this process equals
\begin{equation}
\label{eq:3}
{\cal P}_N(E) = \sum_{i,f}\left|\langle f |\hat{U}(t_f,\, t_i)|i;\, E,
N\rangle\right|^2 \sim {\rm e}^{-F_{N}(E)/g^2},
\end{equation}
where $\hat{U}$ is the quantum evolution operator and the
infinite time limits $t_{i,\, f} \to \mp \infty$ are assumed. The sum
in Eq.~(\ref{eq:3}) runs over all 
$N$--particle initial states with energy $E$ in the false vacuum and
final states $|f\rangle$ containing a bubble  of true vacuum. In the approximate
equality we introduced the suppression exponent $F_N(E)$,
c.f.\ Eq~(\ref{eq:2}).  

At small $g^2$ the action (\ref{eq:1}) is parametrically large and any
path integral with $\exp(iS)$ in the integrand can be evaluated in
the saddle--point approximation. In Appendix \ref{sec:derivation} we
review the semiclassical method for calculating the probability
(\ref{eq:3}): 
introduce a path integral for ${\cal P}_N(E)$  and derive equations
for its 
(generically complex) saddle--point configuration 
$\phi_s(t,\, x)$, see Refs.~\cite{Rubakov:1992ec,
  Rebbi:1996qt, Rubakov_Shaposhnikov} for details. The leading
semiclassical exponent $\exp(-F_N/g^2)$ is given by the value of the
integrand at $\phi = \phi_s$.

Let us describe the saddle--point equations for $\phi_s(t,\, x)$ which will
be solved in the next Sections. This complex configuration should
extremize the 
action~(\ref{eq:1}) i.e.\ satisfy the classical field equation   
\begin{subequations}
\label{eq:Ttheta}
\begin{equation}
\label{eq:eq}
\partial_{t}^{2}\phi_s - \partial_{x}^2\phi_s + V'(\phi_s) = 0\;,
\end{equation}
where $V' \equiv dV/d\phi$. We will consider Eq.~(\ref{eq:eq}) along the
complex--time contour in Fig.~\ref{fig3}a corresponding to evolution of
particles prior to collision (part AB of the contour),  
\begin{figure}[t]

\vspace{5mm}
\hspace{3.97cm}(a) \hspace{7.7cm} (b)

\vspace{-5mm}
\centerline{
\hspace{-5mm}
\begin{minipage}{0.6\textwidth}
\includegraphics[width=\textwidth]{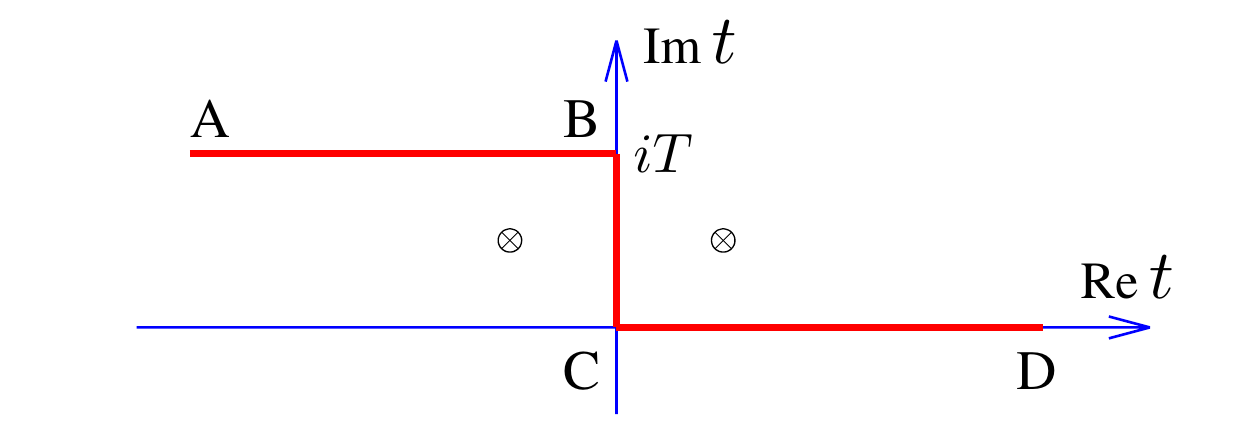}
\vspace{12mm}
\end{minipage}
\begin{minipage}{0.42\textwidth}
\includegraphics[width=\textwidth]{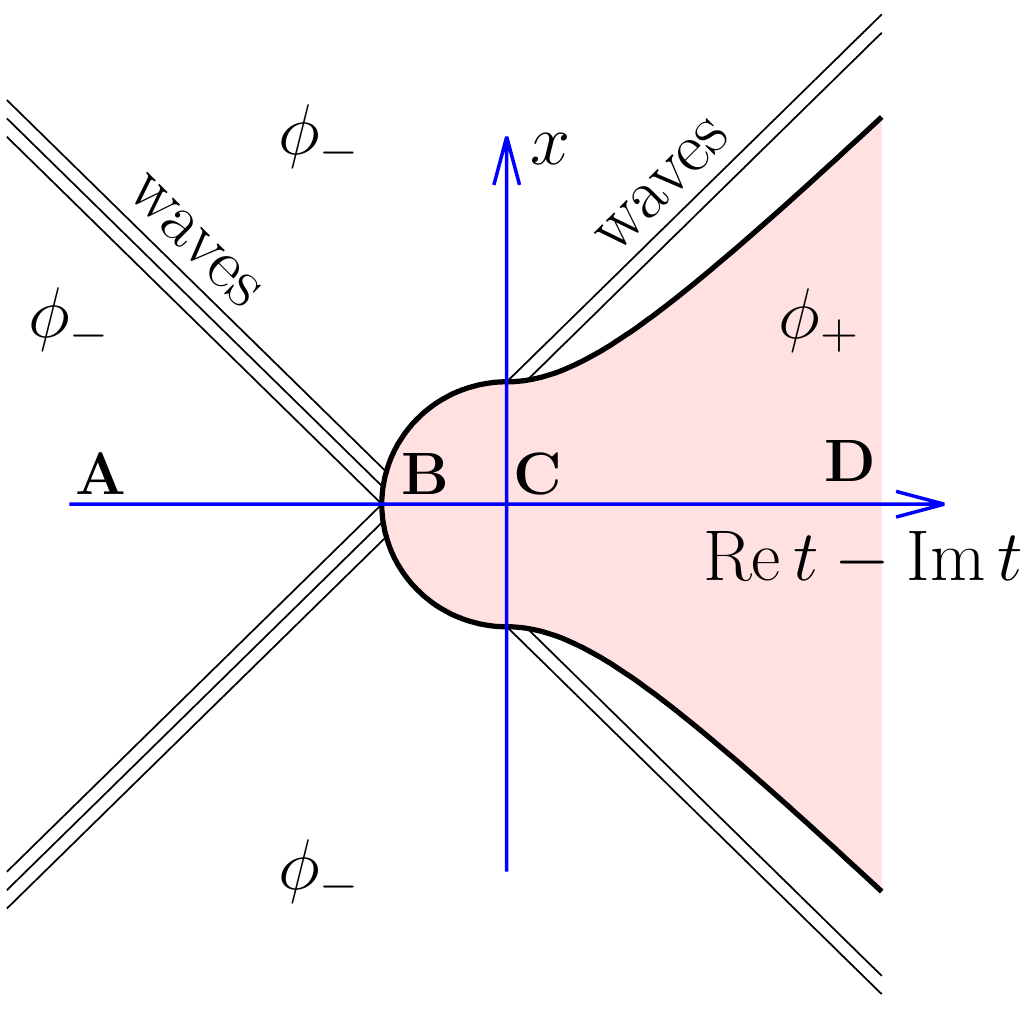}
\end{minipage}\hspace{5mm}
}
\caption{(a) Contour in complex time for the semiclassical boundary
  value problem. Crossed circles represent singularities of the
  solution. (b)~Schematic form of the semiclassical 
  solution $\phi_s(t,x)$.  Pink/gray region corresponds to $\phi\approx
  \phi_+$.\label{fig3}}
\end{figure}
tunneling via formation of the true vacuum bubble (part BC) and expansion
of the bubble to infinite size (part CD). The expected structure of
the semiclassical solution along the contour is visualized in
Fig.~\ref{fig3}b. The duration $T$ of Euclidean evolution is a
parameter of the solution.

The field equation is supplemented with the boundary conditions related to 
the initial and final states of the process. Namely, in the asymptotic future
the solution should be real,
\begin{equation}
\label{000}
\Im \phi_s,\; \mathrm{Im}\, \partial_t \phi_s \to 0 \qquad \qquad
    {\rm as}\qquad  t\to +\infty\;,
\end{equation}
due to the inclusive final state in
Eq.~(\ref{eq:3}). In the asymptotic past i.e.\ at $t = iT + t'$ and $t'
\to -\infty$,  the configuration $\phi_s$ should evolve linearly 
in the false vacuum,
\begin{equation}
\label{111}
\phi_s \to \phi_- + \int\limits_{0}^{\infty}
\frac{dk}{\sqrt{\pi\omega_k}} \;\cos{(kx)} 
\left( f_k\, {\rm e}^{-i\omega_{k}t^{\prime}} + g_{k}^{*}\, {\rm
    e}^{i\omega_{k}t^{\prime}} \right) \qquad\qquad \mbox{as} 
\qquad
t'\to -\infty\;,
\end{equation}
where $\omega^2_{k} = k^2 + m_{-}^{2}$ and we expect that the
relevant semiclassical solutions are $x\to -x$ symmetric
like the soliton pair in Fig.~\ref{fig1}b. The waves in Eq.~(\ref{111}) represent free on--shell 
particles in the initial state of the process. The initial
saddle--point condition relates negative-- and positive--frequency  
components of these waves,
\begin{equation}
\label{222}
f_{k} = {\rm e}^{-\theta}g_{k}\;,
\end{equation}
\end{subequations}
where $\theta$ is the second parameter of the solution. In the special 
case $\theta = 0$ Eq.~(\ref{222}) implies real--valued $\phi_s$ at
$t'\to -\infty$. The initial state in this case
is inclusive i.e.\ optimized with respect to the multiplicity $N$ at
fixed energy $E$, cf.\ Eq.~(\ref{000}). The semiclassical solutions at   
$\theta=0$ are called {\it periodic
  instantons}~\cite{Khlebnikov:1991th}, we will consider them in the 
Sec.~\ref{sec:3}. At $\theta>0$ the solutions are complex at
$t' \to -\infty$. Moreover, at $\theta \to +\infty$ Eq.~(\ref{222})
reduces to the Feynman boundary condition $f_k = 0$. The initial state
in this limit is 
indistinguishable from the vacuum and therefore contains
semiclassically small number of particles, $N \ll 1/g^2$.

Equations \eqref{eq:Ttheta} form the $T/\theta$ boundary value
problem~\cite{Rubakov:1992ec} which will be used to find the
semiclassical solutions $\phi_s$. The parameters $T$ and $\theta$ of
the solutions are Lagrange multipliers conjugate to the energy $E$ and 
initial multiplicity $N$, respectively. The latter quantities are
given by the standard expressions
\begin{equation}
\label{E_N}
g^2 E=2\int\limits_0^\infty dk \, \omega_{k}f_k g_k^{*}, \qquad\qquad
g^2 N =2 \int\limits_0^\infty dk \, f_k g_k^{*}.
\end{equation}
In what follows we find $\phi_s$ for all possible values of $T$ and
$\theta$  and compute $(E,\, N)$ by Eqs.~\eqref{E_N}. 

The suppression exponent in Eq.~(\ref{eq:3}) is evaluated as a
functional on the semiclassical solution $\phi_s(t,\,x)$, 
\begin{equation}
\label{s_exp}
F_N(E) = 2g^2{\rm Im}\,S[\phi_s] -g^2 (2ET+ N\theta) + 
2\mathrm{Im} \int\limits_0^\infty dx \, (\phi_s - \phi_-) \partial_t \phi_s \Big|_{t_i}\;,
\end{equation}
where the last two terms represent contributions from the nontrivial
initial state of the process. Note that the semiclassical
parameter $g$ does 
not enter the boundary value problem~(\ref{eq:Ttheta}). Thus, the
exponent~\eqref{s_exp} depends on the combinations $g^2 E$ and $g^2 N$ 
rather than individually on $g$, $E$ and $N$,
cf.~Eqs.~\eqref{E_N}. This feature is in agreement with the
Rubakov--Son--Tinyakov conjecture~\cite{Rubakov:1992ec} discussed in
the Introduction: the semiclassical exponent does not depend on $N$ at
$N \ll 1/g^2$ if the limit
\begin{equation}
\label{eq:4}
F_2(E) = \lim_{g^2N\to 0} F_{N}(E)
\end{equation}
exists. Then $F_{2}(E)$ is the suppression exponent of the
two--particle processes.  

We remark that the method of Lagrange multipliers implies the
following Legendre transforms for $T$ and $\theta$,
\begin{equation}
\label{eq:6}
2g^2 T = -\partial_E F_N(E)\;, \qquad \qquad 
g^2 \theta = -\partial_N F_N(E)\;,
\end{equation}
see Appendix \ref{sec:derivation} for derivation. Below we will keep
in mind that $T$ and $\theta$ are proportional to the partial
derivatives of $F_N(E)$.  

Let us comment on the delusively simple final state of the
process. First, recall that the configurations $\phi_s(t,\,
x)$ should describe false vacuum decay i.e.\ contain the bubble of true
vacuum~--- expanding or critical --- at $t\to +\infty$ (pink/gray region in
Fig.~\ref{fig3}b). This
property is nontrivial and will be used  as a selection rule for the
relevant semiclassical solutions. Second, one might think that the
semiclassical solutions are real at the real time axis: starting from
real $\phi_s$ and $\partial_t\phi_s$ in the asymptotic future and
solving the classical field equation backwards in time, one arrives at
real $\phi_{s}(t,\, x)$ at $t\in \mathbb{R}$. This simple logic is, 
however, not applicable if the solution approaches the critical bubble 
$\phi_{cb}(x)$ in the asymptotic
future~\cite{Bezrukov:2003yf}. Indeed, the latter is unstable and
contains exponentially growing and decaying modes $\delta  
\phi_{\pm}(t,x) \propto \mathrm{e}^{\pm |\omega_-| t}$ in the spectrum
of linear perturbations. At large positive times one obtains $\phi_s
\approx\phi_{cb} + A_{-} \delta \phi_- + \mbox{real waves}$, and
the coefficient $A_-$ is complex in general. Then the overall solution,
although complex--valued at the real time axis, satisfies
Eq.~(\ref{000}) asymptotically. We will demonstrate that the
semiclassical solutions with $E>E_{cb}$ approach the critical bubble in
the infinite future. In this case the asymptotic reality (\ref{000})
cannot be imposed at finite real $t$; we overcome this difficulty
using the methods of~\cite{Bezrukov:2003yf, epsilon,
  rst_confirmations2}. 

To summarize, we arrived at the practical method for evaluating the
suppression exponent of soliton pair production in particle
collisions. One starts by describing the induced false vacuum decay at
$\delta \rho > 0$: finds a family of the solutions $\phi_s(t,\, x)$ to
Eqs.~(\ref{eq:Ttheta}), relates their parameters $(T,\,\theta)$ to
$(E,\, N)$ by Eqs.~\eqref{E_N} and computes the suppression exponent
$F_N(E)$ using Eq.~\eqref{s_exp}.  The exponent of the
soliton--antisoliton 
production is recovered in the limit $\delta \rho \to 0$ above the
kinematic threshold $E>2M_S$. Besides, in the limit $g^2 N\to 
0$ one obtains the exponent $F_2(E)$ of the two--particle processes,
see Eq.~\eqref{eq:4}.

\section{Numerical methods}
\label{sec:numerical-method}

\subsection{Choosing the potential}
\label{sec:choosing-potential}
Nonlinear semiclassical equations (\ref{eq:Ttheta}) do not admit
analytic treatment and we solve them numerically. To this end we 
specify the scalar potential in dimensionless units,
\begin{equation}
\label{V_phi}
V(\phi) = \frac{1}{2}(\phi+1)^2\left[1 -
  vW\left(\frac{\phi-1}{a}\right)\right]\;, \qquad\;\; \mbox{where} \;\;\;\;\;
W(u) = {\rm e}^{-u^2}\left( u + u^3 + u^5\right)
\end{equation}
and $a=0.4$. This function has a double--well form with false and
true vacua at ${\phi = \phi_- \equiv -1}$ and $\phi = \phi_+ > 0$,
respectively. We 
have $V(\phi_-)=0$ and fix the energy density $V(\phi_+) \equiv
-\delta \rho$ of the true vacuum by tuning the parameter $v$. In
particular, at $v \approx 0.75$ the vacua are degenerate; the 
respective scalar masses are $m_- \approx 1$, $m_{+} \approx
7.6$. Below we start from the solutions at $\delta \rho>0$ (larger
$v$), then send $\delta \rho \to 0$. Function~\eqref{V_phi} is shown in
Fig.~\ref{fig1}a at $\delta\rho = 0$ and $\delta \rho = 1$ (solid and
dashed lines, respectively). The soliton and antisoliton
configurations in the case of degenerate vacua are demonstrated in
Fig.~\ref{fig1}b.  

Let us explain the choice \eqref{V_phi} of the scalar
potential. First, we do not consider the standard $\phi^4$ theory
because evolution of the kink--antikink pairs there is known to
exhibit chaotic behavior~\cite{Campbell:1983xu}. That chaos is related
to the fact that the spectrum of linear perturbations around the
$\phi^4$ kink 
contains two localized modes representing its spatial translations and
periodic vibrations of its form. The modes accumulate energy during  
the kink--antikink evolution which is therefore described by
two\footnote{The other two coordinates representing 
  center--of--mass motion and $P$--odd vibrations of the
  kink--antikink pair decouple due to momentum conservation and $x\to
  -x$ symmetry.} collective coordinates. This evolution is chaotic like 
the majority of two--dimensional mechanical motions. Irregular
dynamics is difficult~\cite{chaos, shudo_stokes, Onishi,
  Levkov:2007ce} for the semiclassical methods, we avoid it 
by changing the form of the scalar potential. We checked that the
model~(\ref{V_phi}) is not chaotic\footnote{Another type of irregular
  behavior~\cite{Dorey:2011yw} appearing at $m_{+}<m_{-}$ is also 
  not met in our model.},  
i.e.\ there is a single localized mode in the spectrum of linear
perturbations around the soliton.  

Second, the potential (\ref{V_phi}) is almost quadratic at $\phi < 0$
and essentially nonlinear at positive $\phi$. This property is
convenient for numerical methods. Indeed, the initial conditions
(\ref{111}), (\ref{222}) should be imposed in the asymptotic past where
$\phi_s(t,\, x) < 0$ describes wave packets linearly evolving in the
false vacuum (diagonal lines in the left part of Fig.~\ref{fig3}b). In
the model~(\ref{V_phi}) the wave packets remain free almost up to
their collision point, and the initial condition can be imposed at
relatively small negative times (realistically, at $\mathrm{Re}\, t_i
\sim {-(6\div 7) / m_-}$). Long nonlinear evolution in other models
can be costly for numerical methods. Note that this problem with slow
linearization is specific to (1+1)--dimensional systems and should be
absent in higher dimensions.

\subsection{Discretization}
\label{sec:lattice-system}
\begin{figure}[t]
\centerline{\includegraphics[width=.65\textwidth]{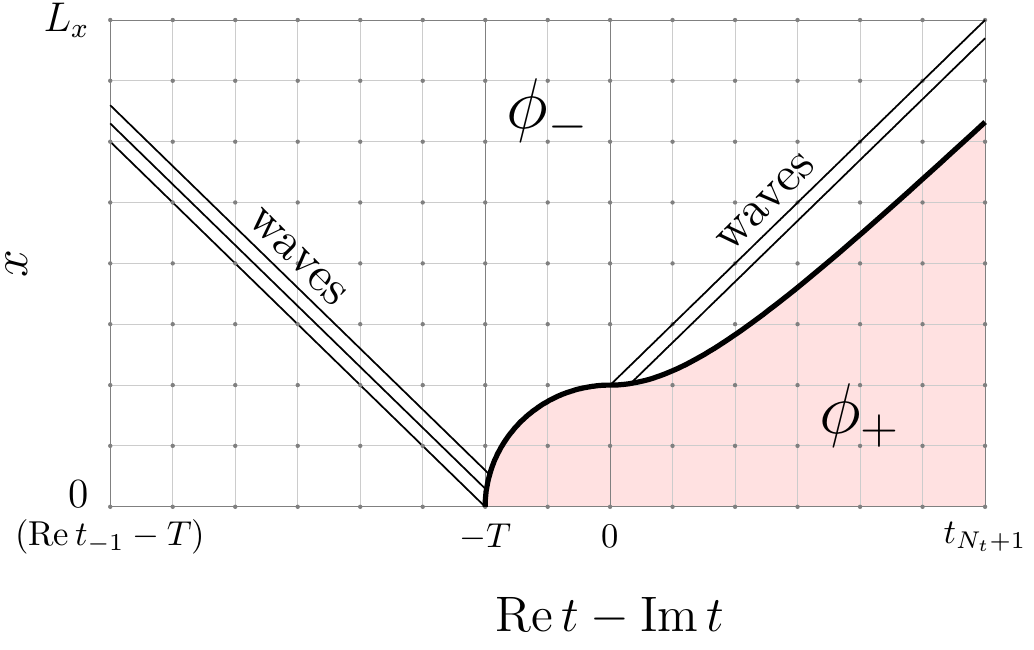}}

\vspace{-3mm}
\caption{Rectangular lattice superimposed with the schematic form of the
  semiclassical solution $\phi_{j,i}$. The time sites $t_j$ cover the
  contour ABCD in Fig.~\ref{fig3}a. Solution at $x<0$ is
  reconstructed using the $x\to -x$ symmetry.\label{fig:lattice}}   
\end{figure}
We discretize the semiclassical equations~(\ref{eq:Ttheta}) using the
rectangular $(N_{t}+3) \times N_x$ lattice in
Fig.~\ref{fig:lattice}. The sites $t_j$ and $x_i$ of the lattice cover
the contour ABCD in Fig.~\ref{fig3} and the interval ${0 \leq x\leq 
  L_x}$, respectively. The solution at $x<0$ can be
restored from the reflection symmetry $\phi_s(t,\, x) =
\phi_s(t,\, -x)$. Our lattice is as close to uniform as
possible. Namely, $x_i = i\cdot \Delta x$ with integer 
${0 \leq i   \leq N_x-1}$ and spacing $\Delta x = L_x/(N_x-1)$. The time
spacings $\Delta t_j \equiv  t_{j+1} - t_j$, where $-1 \leq j  \leq
N_t+1$, are constant at the real and imaginary parts of the
contour. Typically, we choose $|\Delta t_j| \ll \Delta x$. This
property 
will be useful for formulating the boundary conditions. Our
unknowns $\phi_{j,\, i} \equiv \phi_s(t_j,\, x_i)$ are the (complex)
values of the semiclassical solution at the lattice sites.

We replace the derivatives in the action (\ref{eq:1}) with the 
second--order finite--difference expressions
\begin{equation}
\label{eq:16}
\partial_t\phi\big|_{j+1/2,\, i} \to  \frac{\phi_{j+1,\, i} - \phi_{j,\,
    i}}{\Delta t_{j}} \;, \qquad\qquad 
\partial_x\phi\big|_{j,\, i+1/2} \to \frac{\phi_{j,\, i+1} - \phi_{j,\,
    i}}{\Delta x} \;,
\end{equation}
taking values at the links; below we mark all link quantities with
half--integer indices. We also trade the time and space integrals for
the second--order trapezoidal sums, e.g. 
\begin{equation}
\label{eq:17}
\int dt \, f(t) \to \sum_{j=-1}^{N_t} \Delta t_j \, f_{j+1/2} =
\sum_{j=-1}^{N_t+1} \Delta \bar{t}_j \, f_{j}\;,
\end{equation}
where the first and second expressions are used for the link and site
quantities, respectively. The last sum involves the ``averaged'' 
spacings $\Delta \bar{t}_j = (\Delta t_j + \Delta 
t_{j-1})/2$ and $\Delta \bar{t}_{-1} = \Delta t_{-1}/2$, $\Delta
\bar{t}_{N_t+1} = \Delta t_{N_t}/2$. 

Substitutions~\eqref{eq:16},
\eqref{eq:17} turn the action into a function $S(\phi)$ of
$(N_t+3)\times N_x$ complex variables $\phi_{j,\, i}$. For
completeness we present this function in Appendix~\ref{App:B}.
The lattice field equations are then obtained by extremizing $S(\phi)$
with respect to the field values at the interior sites of the contour
ABCD:
\begin{equation}
\label{eq:35}
{\partial S / \partial \phi_{j,\, i} =   0}\; , \qquad \qquad  0
\leq j \leq N_t 
\qquad \mbox{and all $i$}\;.
\end{equation}
Note that the equations at the last spatial site $i=N_x-1$
correspond to the Neumann boundary condition $\partial_x \phi = 0$
imposed at $x=L_x$. Note also that Eqs.~\eqref{eq:35} can be derived
by writing the lattice path integral for the probability and
evaluating it in the saddle--point approximation, like in the
continuous approach of Appendix~\ref{sec:derivation}.  

The final boundary condition~(\ref{000}) implies that the field is real
at the last two time sites,
\begin{equation}
\label{eq:24}
\mathrm{Im}\, \phi_{N_t,\, i} = \mathrm{Im}\, \phi_{N_t+1,\, i} = 0
\qquad\qquad \mbox{for all $i$} \;,
\end{equation}
cf.\ Eq.~(\ref{eq:16}). 
In Sec.~\ref{sec:2} we remarked that this condition cannot be
used for solutions approaching the critical bubble at~$t\to
+\infty$. That case will be considered separately. 

Discretization of the initial conditions~(\ref{111}), (\ref{222}) is
far less trivial.  Let us first simplify the discussion and consider
the system in  discrete 
space $\{x_i\}$ and continuous time $t$. Recall that the
semiclassical evolution is linear in the beginning of the process
$t\sim t_{-1}$. We therefore introduce a deviation of the field from
the false vacuum $\psi_i(t) = [\phi(t,\,   x_i) - \phi_-] \sqrt{\Delta
  \bar{x}_i}$ and leave only quadratic terms in the
action. We arrive at the system of $N_x$ coupled  oscillators,
\begin{equation}
\label{eq:20}
g^2 S^{(2)} = \int dt \left\{ \sum_{i=0}^{N_x-1}(\partial_t
\psi_i)^2 - \sum_{i,\, i' =0}^{N_x-1} \psi_i H_{i,\, i'}
\psi_{i'}\right\}\;,
\end{equation}
where the real symmetric matrix $H_{i,\, i'}$ is a discretized version
of the operator $(-\partial_x^2 + m_-^2)$, see  
Appendix~\ref{App:B} for the explicit form. Linear evolution in
Eq.~(\ref{eq:20}) is solved by decomposing $\psi_i(t)$ in
real--valued eigenvectors $\xi_i^{(n)}$ of $H$ with 
eigenvalues $\omega_n^2$, 
\begin{equation}
\label{eq:21}
\psi_i(t) = \sum_{n=1}^{N_x} \frac{\xi^{(n)}_i}{\sqrt{2\omega_n}} \left(f_n\,
\mathrm{e}^{-i\omega_n (t-iT)} + g_n^*\, \mathrm{e}^{i\omega_n
  (t-iT)}\right)\;. 
\end{equation}
Here we introduced the integration constants $f_n$, $g_n$ and 
shifted the time by $iT$ to keep contact with  Eq.~(\ref{111}). By
direct inspection one observes that the the index $n$, vectors
$\xi_i^{(n)}$ and constants $f_n$, $g_n^*$ are the lattice analogs of
$k$, $\cos(kx)$ and $f_k$,  $g_k^*$ in Eq.~(\ref{111}),
respectively. Thus, the discrete 
version of the initial condition (\ref{222}) is $f_n = \mathrm{e}^{-\theta}
g_n$. The quantities entering this condition are extracted from $\psi_i(t)$ as
\begin{equation}
\label{eq:22}
f_n = \frac{1}{\sqrt{2\omega_n}} \sum_{i} \xi_i^{(n)} \left( \omega_n \psi_i + i
\partial_t\psi_i \right)\Big|_{t_{-1}}\;, \qquad 
g_n = \frac{1}{\sqrt{2\omega_n}} \sum_{i} \xi_i^{(n)} \left( \omega_n \psi_i^* + i
\partial_t\psi_i^* \right)\Big|_{t_{-1}}\;,
\end{equation}
where we used Eq.~(\ref{eq:21}) assuming normalization $\sum_{i=0}^{N_x-1}\,
\xi_i^{(n)}  \xi_i^{(n')} = \delta_{n,\, n'}$ and omitting irrelevant
phase factor. To obtain the lattice form of the initial condition, we
discretize the time derivatives in Eqs.~(\ref{eq:22}). The simplest
way is to perform the substitutions (\ref{eq:16}), (\ref{eq:17}) in the
quadratic action (\ref{eq:20}), and then define $\partial_t \psi_i$ at the
very first time site as
\begin{equation}
\label{eq:19}
\partial_t \psi_i\big|_{t_{-1}} \equiv -\frac{g^2}2\;  \frac{\partial S^{(2)}}{\partial
  \psi_{i}(t_{-1})}  = \frac{\psi_i(t_0) - \psi_i(t_{-1})}{\Delta t_{-1}}
+ \Delta\bar{t}_{-1}\sum_{i'} H_{i,\, i'} \psi_{i'}(t_{-1})\;,
\end{equation}
cf.\ the derivation of the initial conditions in
Appendix~\ref{sec:derivation}.

We see that $f_n$ and $g_n$ in Eqs.~(\ref{eq:22}), \eqref{eq:19}
are the linear combinations of $\psi_i(t_{-1})$,
$\psi_i(t_0)$ and their complex conjugates. This means that the initial condition $f_n
= \mathrm{e}^{-\theta}\, g_n$ can be explicitly rewritten as a set of $2N_x$  
real equations on ${\Psi = \{ \psi_i(t_{-1}), \, \psi_{i'}(t_{0})\}}$,  
\begin{equation}
\label{eq:23}
M_R\cdot \mathrm{Re}\, \Psi + M_I \cdot \mathrm{Im}\, \Psi = 0\;.
\end{equation}
Constant
$2N_x \times 2N_x$ matrices $M_R$, $M_I$ can be deduced from
Eqs.~(\ref{eq:22}), see Appendix \ref{App:B}. In the numerical code we
explicitly 
compute\footnote{This is done once for each lattice.} the
eigenvectors and eigenvalues of $H_{i,\, i'}$, construct matrices
$M_R$, $M_I$, and add Eqs.~(\ref{eq:23}) to the system of discrete
equations.

Discretization leaves us with $(N_t+1)\times N_x$ complex field
equations~(\ref{eq:35})  and $4N_x$ real boundary conditions~(\ref{eq:24}),
(\ref{eq:23}). This is precisely the required number of equations to
determine $(N_t+3)\times N_x$ complex unknowns $\phi_{j,\,
  i}$. The semiclassical equations will be solved in the subsequent Sections.

Finally, let us introduce lattice expressions for the energy, 
initial particle number and suppression exponent. The full nonlinear
energy 
\begin{equation}
\label{eq:25}
g^2 E = \int_0^L dx \left[ (\partial_t \phi)^2 + (\partial_x \phi)^2 +
2V(\phi)\right] 
\end{equation}
is directly discretized using the
substitutions~(\ref{eq:16}),~(\ref{eq:17}). We monitor its
conservation to control the discretization errors. The initial 
energy and particle number~(\ref{E_N})  in the discrete case have the 
form, 
\begin{equation}
\label{eq:26}
g^2 E = 2\sum_{n=1}^{N_x} \omega_n \, f_n g_n^*\;, \qquad \qquad g^2 N
= 2 \sum_{n=1}^{N_x} f_n g_n^*\;,
\end{equation}
see Appendix~\ref{App:B}. 
The suppression exponent $F_N(E)$ is computed using 
the discrete action $S(\phi_{j,\, i})$ and quantum
numbers~\eqref{eq:26} in Eq.~\eqref{s_exp}; the last integral in this
expression is 
discretized in the standard way.

\subsection{Fixing the time translations}
\label{sec:discretization}
The continuous semiclassical equations
(\ref{eq:Ttheta}) preserve time\footnote{The spatial translations
  are fixed by $x\to -x$ symmetry imposed on the semiclassical
  solutions.} shifts. Namely, ${\phi_s(t - t_0,\, x)}$ with real $t_0$
satisfies the equations if $\phi_s(t,\, x)$ does. This feature, if
left unnoticed, leads to a numerical artifact. Indeed, since the time
translations are explicitly broken in the lattice system, the position
$t_0$ of the numerical solution is out of direct control:
it is  fixed by the discretization 
and finite--volume effects. Depending on the lattice
parameters, it may become arbitrary large and cause divergence of
the numerical procedure. 

To cure the above artifact, we notice~\cite{rst} that due to the
continuous time--shift symmetry one of the lattice equations starts to
depend on the others at $\Delta t\to 0$, $t_{-1} \to -\infty$. Indeed, in
this limit the total energy of the solution  is conserved. It is real
due to the final boundary  
conditions~(\ref{eq:24}) and, on the other hand, proportional to the
sum $\sum_n \, \omega_n f_n g_n^*$ in Eq.~(\ref{eq:26}). Now, consider
the initial conditions $f_n = \mathrm{e}^{-\theta} g_n$ which imply,
in particular, $N_x$ relations $\mathrm{arg}\, f_n = \mathrm{arg}\,
g_n$. One of the the latter is redundant because the sum $\sum_n \,
\omega_n f_n g_n^*$ is automatically real. We can exclude the
redundant equation without affecting the solution.

Following this line of reasoning, we drop the phase of the initial
condition 
$f_{n_0} = \mathrm{e}^{-\theta} g_{n_0}$ at a given $n=n_0$ relating
only the absolute values
\begin{equation}
\label{eq:31}
|f_{n_0}| = \mathrm{e}^{-\theta} |g_{n_0}|\;.
\end{equation}
To keep the number of lattice equations equal to the number
of unknowns, we add one condition fixing the time translation
invariance of the solution. Provided the set of the lattice equations is
solved, the arguments of $f_{n_0}$ and $g_{n_0}$ will be automatically
equal up to discretization errors. In realistic
numerical  calculations we use Eq.~(\ref{eq:31}) for a highly
populated mode with\footnote{The respective physical momentum is
  $k_{n_0} \approx \pi (n_0-1)/L_x$.} 
$n_0 = 2$ or $3$; the equality of phases is then satisfied with accuracy
better than ${|\mathrm{arg}\, f_{n_0} - \mathrm{arg}\, g_{n_0}|
  \lesssim 10^{-3}}$ for all our solutions. 

Let us now discuss the additional equation fixing the time translation
invariance. It is convenient to use different conditions at
energies below and above the barrier height $E_{cb}$. At low energies
we impose the relation~\cite{kuznetsov_tinyakov}
\begin{equation}
\label{eq:29}
\int_0^{L_x} dx \, w(x) \, \mathrm{Re}\, \partial_t \phi \big|_{t_C} = 0
\qquad \qquad \mbox{with} \qquad w(x) = \mathrm{e}^{-4x^2/L_x^2}
\end{equation}
at the point $C$ of the time contour in Fig.~\ref{fig3}a. Equation
(\ref{eq:29}) places the turning point $\partial_t \phi = 0$ of the
solution, if there is one, at a given time $t = t_C$. Even if the
solution does not have turning points, Eq.~\eqref{eq:29} keeps the
singularities of the solution (crossed circles in Fig.~\ref{fig3}a)
away from the time contour. 

At high energies the structure of the semiclassical solutions changes,
and we use different constraint for the time translations. Namely, 
we fix the center--of--mass coordinate $R$ of the 
wave packet at the start of the process, see
Fig.~\ref{fig:lattice}, 
\begin{equation}
\label{eq:30}
\int_0^{L_x} dx \, (r - R) \, \mathrm{Re} \, \rho_E (x) \big|_{t_{-1}}
= 0\;,
\end{equation}
where $\rho_E$ is the energy density entering
Eq.~(\ref{eq:25}): $g^2E = 2\int_0^{L_x} dx \, \rho_E(x)$. In realistic
calculations we take $R \approx 0.7 L_x$ to guarantee that the
initial wave packet is inside the lattice range at $t = t_{-1}$.

Constraints (\ref{eq:29}), (\ref{eq:30}) are discretized in
the standard way using the substitutions (\ref{eq:16}) and
(\ref{eq:17}). We repeat that exchanging one complex equation $f_{n_0}
= \mathrm{e}^{-\theta} g_{n_0}$ for the two real --- 
Eq.~(\ref{eq:31}) and one of Eqs.~(\ref{eq:29}), (\ref{eq:30}) --- 
one keeps the number of equations equal to the number of unknowns.

\subsection{Solving the equations}
\label{sec:numerical-algorithm}
Let us select the appropriate values of the lattice parameters. To this
end we regard the schematic form of the semiclassical solution    
$\phi_{j,\, i}$ in Fig.~\ref{fig:lattice}. 

First of all, the site numbers $N_t$, $N_x$ should be large enough to
ensure small 
discretization errors $O(\Delta t^2)$, $O(\Delta x^2)$. Note, however,
that the characteristic frequencies of the solutions are estimated
as $\omega_k \sim E/N$, see Eqs.~(\ref{eq:26}). They become 
higher at larger energies and smaller multiplicities. In these regions
the solutions are sharp and require fine lattice resolution. On the other
hand, large $N_t$ and $N_x$ are costly for numerical algorithms. In
practical calculations we use lattices $N_t \times N_x = 3200 
\times 150$ and $11000 \times 500$ below and above $E_{cb}$,
respectively. We keep track of the discretization errors and
stop\footnote{See~\cite{Demidov:2015} for the study of the
  high--energy region.} obtaining solutions whenever the relative
accuracies become worse than $1\%$.  

Next, we want to  keep the nonlinear part of the solution at
$x,\, {\rm Re}\;t \sim 0$. Then the initial time $ {\rm Re}\, t_i
\equiv t_{-1}$ should be large negative to ensure linear evolution at
the 
start of the process. We find that\footnote{Recall that in our units
  $m_{-} \approx 1$.} 
$\mathrm{Re}\, t_{-1} = - (6\div 8)$ is large enough: in this case 
the relative contributions of nonlinear interactions at $t\sim t_{-1}$
are smaller than $10^{-3}$. Besides, the spatial
volume  $|x| \leq L_x$ should encompass the entire solution. Since the
initial wave packets propagate inside the lightcone (diagonal lines in  
Fig.~\ref{fig:lattice}), we require $L_x \gtrsim |\mathrm{Re}\,
t_{-1}|$. In 
practice it is convenient to fix\footnote{At low energies we use
  $L_x = 8\div 15$ due to larger sizes of the respective true
  vacuum bubbles.} $L_x = 7$, then choose $t_{-1}$ to keep the 
wave packets at $|x| < L_x$. We checked that our results are
insensitive to the value of $L_x$ at the relative level 
$10^{-4}$. Finally, we explained in Sec.~\ref{sec:2} that the solutions with
$E<E_{cb}$ are real at the real time axis. In   
this case we reduce the part CD of the time contour to two sites
$t_{N_t},\; t_{N_t+1}$ where the condition (\ref{eq:24}) is
imposed. At $E\geq E_{cb}$ the solution becomes real--valued only
asymptotically at $t\to +\infty$, and we are obliged to extend the
contour to large positive times. We choose
$t_{N_t+1} \sim |\mathrm{Re}\, t_{-1}|$. Then the  waves emitted in
the interaction region remain within the lattice range, see
Fig.~\ref{fig:lattice}.

We performed several tests of the numerical procedure, see
Appendix~\ref{App:D} for details. The overall conclusion is that the
linearization and finite--volume effects cause relative errors
smaller than~$10^{-3}$, while the relative
discretization accuracies remain below~$1\%$.

We proceed by describing the numerical procedure to solve the set
of algebraic lattice equations. Our choice of the numerical
method is  limited  because the field equation~(\ref{eq:eq}) is of
hyperbolic and  elliptic types at the Minkowski and Euclidean parts of
the time contour, respectively. In addition, the semiclassical
solution $\phi_s(t,\, x)$ is complex and satisfies the boundary rather
than  initial conditions. The most convenient numerical technique in this
case~\cite{kuznetsov_tinyakov, rst} is based on 
multidimensional Newton--Raphson method. To simplify notations,
let us denote the complete set of lattice field equations (\ref{eq:35})
and boundary conditions (\ref{eq:24}), (\ref{eq:23}), (\ref{eq:31}), 
(\ref{eq:30}) by ${\cal F}_{l,\, k}(\phi) = 0$. In the  
Newton--Raphson method one starts from the approximation
$\phi^{(0)}_{j,\, i}$ to the solution and repeatedly refines it by
finding the correction $\delta \phi \equiv \phi_s - \phi^{(0)}$. The
latter is obtained from the set of linear equations  
\begin{equation}
\label{eq:34}
 {\cal F}_{l,\, k}(\phi^{(0)}) 
+ \sum\limits_{j,\, i} \left.\frac{\partial {\cal
     F}_{l,\, k}}{\partial \phi_{j,\, i}}\right|_{\phi^{(0)}}\; 
 \delta \phi_{j,\, i} = 0
\end{equation}
in the background of $\phi^{(0)}$. After obtaining the correction, one
redefines the approximation $\phi^{(0)} \to \phi^{(0)} + \delta \phi$ and
repeats the procedure. The 
iterations stop once $\delta \phi$ becomes smaller than the
predetermined numerical error.

The Newton--Raphson method converges quadratically~\cite{Press}.
Typically, the acceptable precision $\delta \phi \lesssim 10^{-10}$ is
achieved in $5-6$ interactions. However, this method is extremely
sensitive to the very first approximation 
$\phi^{(0)}$: even slightly incorrect choice  of the latter may cause
divergence of the iterative procedure.

We therefore use a careful strategy for finding the numerical
solutions. First, we obtain the ``simpler'' solution at some
particular values of the parameters $(T,\; \theta) = (T_0,\,
\theta_0)$. The obvious candidate is the
bounce~\cite{Coleman}. It satisfies the semiclassical boundary 
value problem (\ref{eq:Ttheta}) at ${T_0 = +\infty}$, $\theta_0  =
0$ and, on the other hand, can be computed using half--analytic
methods. In the next Section we will discuss an extended 
one--parametric family of ``simpler'' solutions. Second, we
apply the Newton--Raphson method to find the solution at $(T,\,
\theta) = (T_0 + \Delta T,\, \theta_0 + \Delta \theta)$ using the
one at 
$(T_0,\, \theta_0)$ as the first approximation. At sufficiently small
$\Delta T$, $\Delta \theta$ the method has to converge. Third,
starting from the new solution, we change $(T,\, \theta)$ by a small
step,  
again, and obtain yet another solution. We repeat this procedure 
until a complete two--parametric family of the numerical solutions is
found.  

We finish this Section with a remark that the Newton--Raphson
method requires numerical solution of the sparse linear system
(\ref{eq:34}). This is the most time--consuming part of the numerical
procedure. We perform it using two alternative algorithms
described in Appendix~\ref{sec:numerical-alorithm}. Our ``fast'' 
algorithm arrives at the solution in CPU time $t_{CPU} \propto N_t
\cdot N_x^2$ 
operations but has poor stability properties. We find that it works for the
high--energy solutions but accumulates round--off errors at $E\lesssim
E_{cb}$, see explanation in Appendix~\ref{sec:numerical-alorithm}. In
the latter region we use stable (yet slower) algorithm involving $t_{CPU}
\propto N_t \cdot N_x^3$ operations. Both our algorithms are highly
parallelizable and implemented at the multiprocessor cluster.

\section{Numerical results}
\label{sec:3}

\subsection{Periodic instantons}
\label{sec:4}
We start by considering the periodic
instantons~\cite{Khlebnikov:1991th}~--- semiclassical solutions
at $\theta=0$ and arbitrary~$T$. The 
boundary conditions (\ref{000}), (\ref{111}) and (\ref{222}) in this
case imply reality of $\phi_s(t,\, x)$ in the beginning and at the end
of the process i.e.\ along the Minkowski parts AB
and CD of the time contour in Fig.~\ref{fig3}a. Further insight is
gained if we assume that the solutions have turning
points at the corners B and C of the contour, 
\begin{equation}
\label{period}
\partial_t \phi_s(0,\, x) = \partial_t \phi_s(iT,\, x) = 0\;.
\end{equation}
Then the semiclassical evolution along the Euclidean part BC 
also proceeds with real--valued $\phi_s(t,\, x)$. The
solutions satisfying Eq.~(\ref{period}) are $2T$--periodic in
Euclidean time. 

Physical arguments~\cite{Khlebnikov:1991th} suggest that all relevant
periodic instanton solutions satisfy the Ansatz \eqref{period}. Indeed,
consider a somewhat different process: transition between the sectors of 
the false and true vacua at temperature $\beta^{-1}$. Its rate
$\Gamma(\beta)$  is obtained by integrating the
multiparticle probability (\ref{eq:3}) with the Boltzmann exponent,
\begin{equation}
\label{eq:43}
\Gamma(\beta) \sim \int dE\,dN \,\mathrm{e}^{ -\beta E-F_N(E)/g^2} \;;
\end{equation}
where the prefactors are ignored. The integral~(\ref{eq:43}) is
saturated near the saddle point
\begin{equation}
\label{eq:51}
\beta + \partial_E F_N/g^2 \equiv \beta - 2T = 0\;, \qquad  \qquad
\partial_N F_N \equiv - g^2 \theta =  0\;, 
\end{equation}
where we used the Legendre transforms~(\ref{eq:6}) in the first
equalities. One sees that the saddle--point parameters $\theta=0$ and
$T =\beta/2$ in Eq.~(\ref{eq:51}) correspond to a periodic
instanton\footnote{If such solution exists. At high
  temperatures $\beta < 2\pi / |\omega_-|$, where $\omega_-$ is
  defined below, the integral (\ref{eq:43}) is saturated near the
  barrier top $E = E_{cb}$.}. 
Using Eq.~(\ref{s_exp}) at real $\phi_s$, we obtain
the thermal rate
\begin{equation} 
\label{eq:50}
\Gamma \sim \mathrm{e}^{-2\mathrm{Im}\, S[\phi_s]}
\end{equation}
suppressed by the Euclidean action of this solution. One
finds it natural that real Euclidean solutions with period $\beta
= 2T$ describe thermal transitions.

\begin{figure}[t]
\centerline{\includegraphics[width=0.45\textwidth]{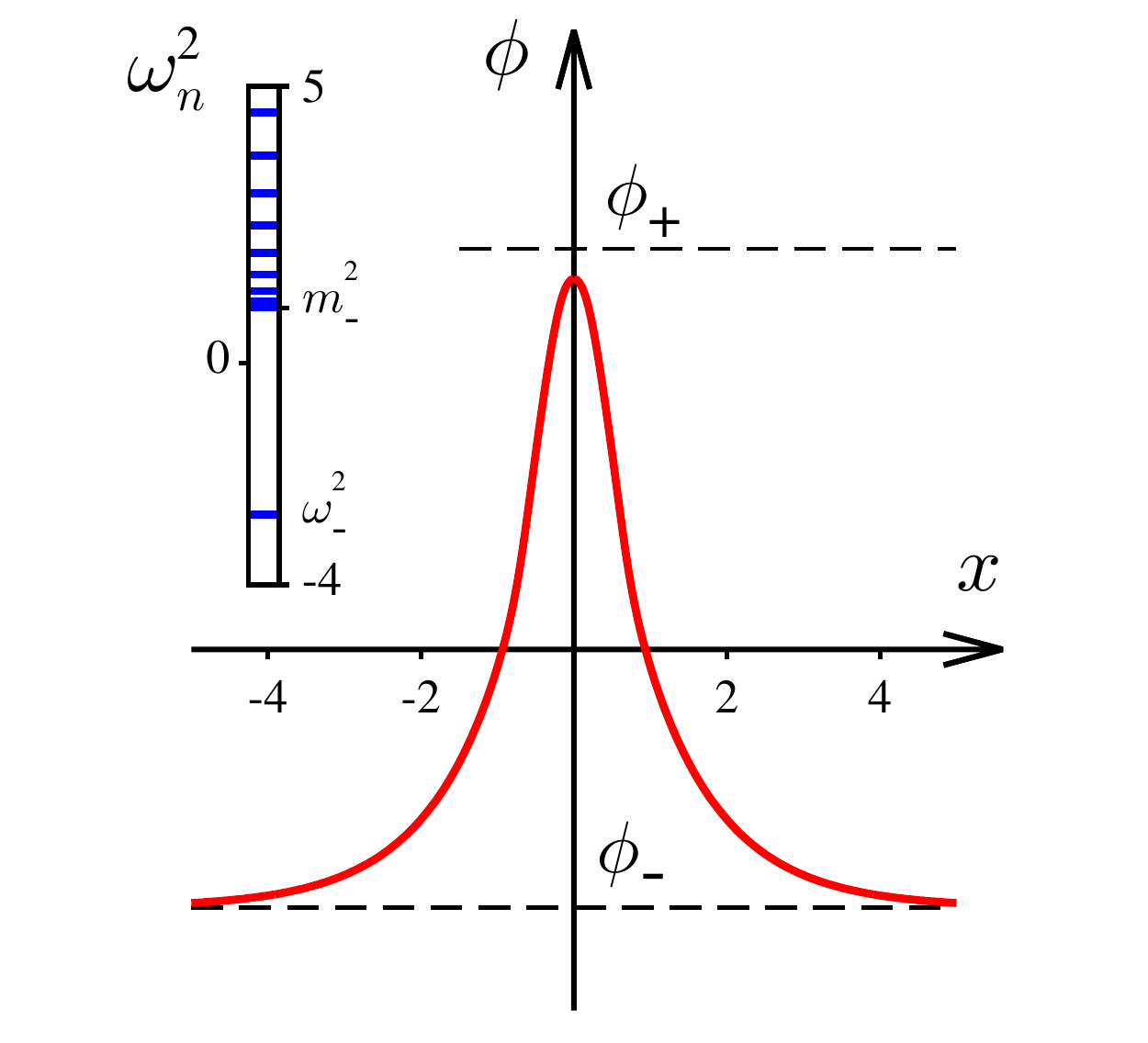}}
\caption{Critical bubble $\phi_{cb}(x)$ and the spectrum
  $\omega_n^2$ of linear perturbations in its background; $\delta \rho
  = 0.4$ and $L_x = 7$.\label{fig:sphaleron}}
\end{figure}
In our study the very first periodic instanton is obtained from the
critical bubble 
$\phi_{cb}(x)$~--- the static true vacuum bubble at the verge of
collapse. In 
Fig.~\ref{fig:sphaleron} we show $\phi_{cb}(x)$ 
at $\delta \rho = 0.4$. Since the critical bubble is unstable, the
spectrum of linear perturbations in its background contains an
exponentially growing mode $\delta \phi_-(x)$ with eigenfrequency
$\omega_-^2 < 0$. This means that the configuration  
\begin{equation}
\label{eq:46}
\phi(t,\, x) = \phi_{cb}(x) + A_- \, \delta \phi_-(x) 
\cosh(|\omega_-| t)
\end{equation}
solves the field equation at small $A_-$. The approximate
solution~\eqref{eq:46} is periodic in Euclidean time with turning
points at $t = 0$ and $t = \pi i/|\omega_-|$. It satisfies the
boundary condition~(\ref{eq:46}) and therefore represents the 
periodic instanton with $T =  \pi/|\omega_-|$.

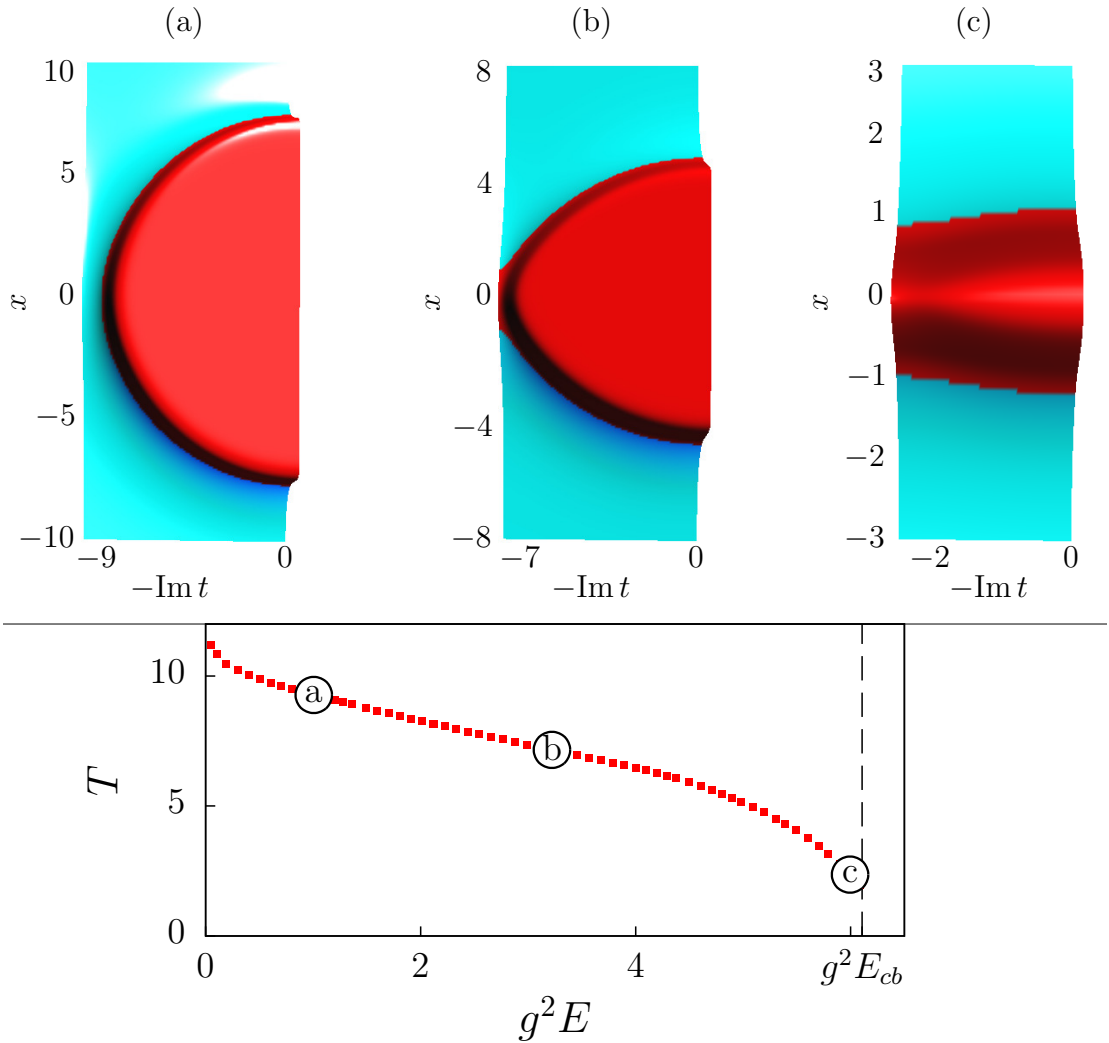
\begin{figure}[t!]
\centerline{\unitlength=0.009\textwidth\begin{picture}(100,90)
\put(0,43){\includegraphics[height=47\unitlength]{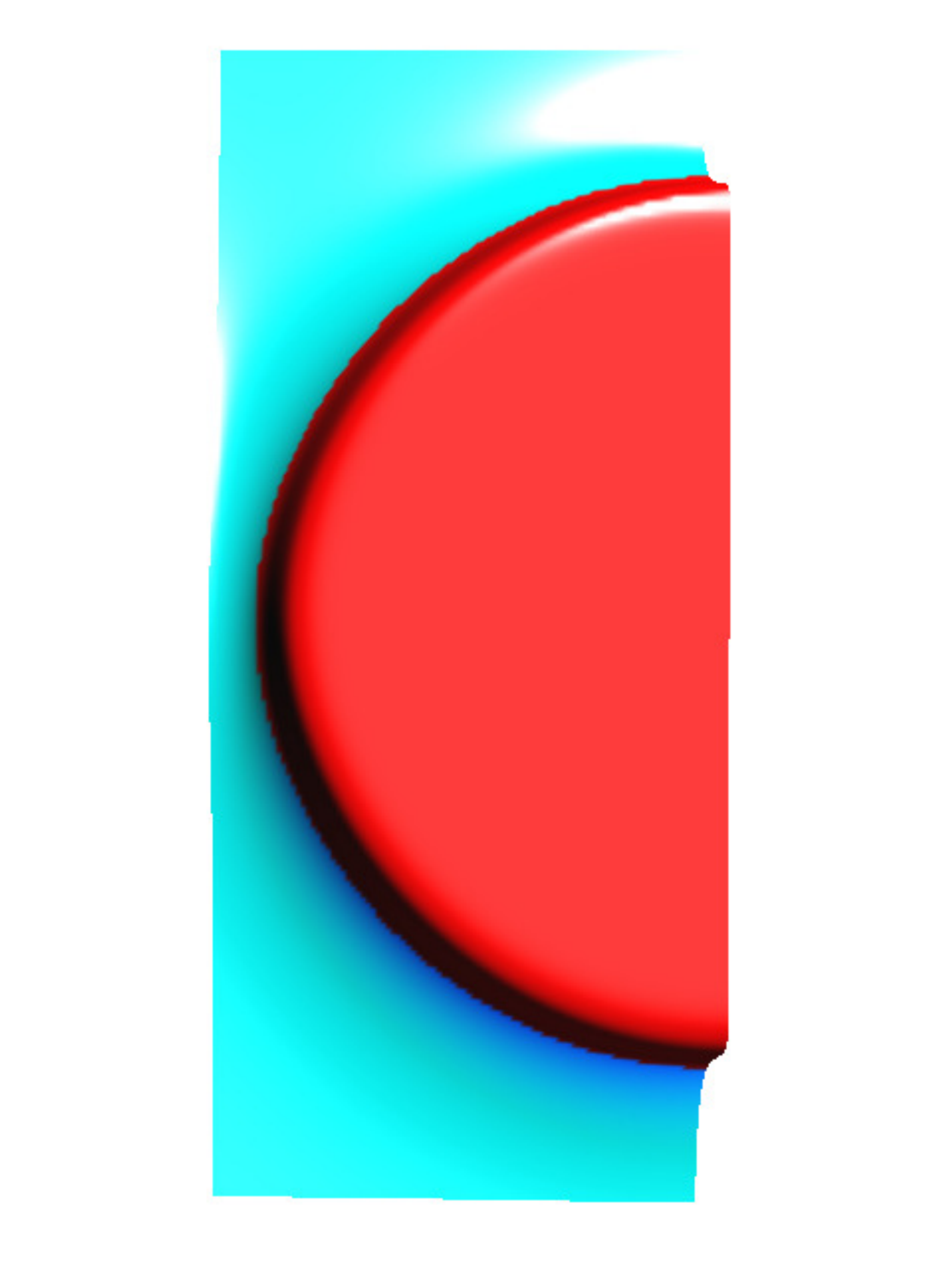}}
\put(5.7,66.6){$0$}
\put(2.1,45){$-10$}
\put(4.3,86.5){$10$}
\put(3.6,55.8){$-5$}
\put(5.8,77.5){$5$}
\put(7.2,43.2){$-9$}
\put(25,43.2){$0$}
\put(1.5,66){\begin{sideways}$x$ \end{sideways}}
\put(12,40.5){$-\mathrm{Im}\, t$}
\put(36.6,41.6){\includegraphics[height=49\unitlength]{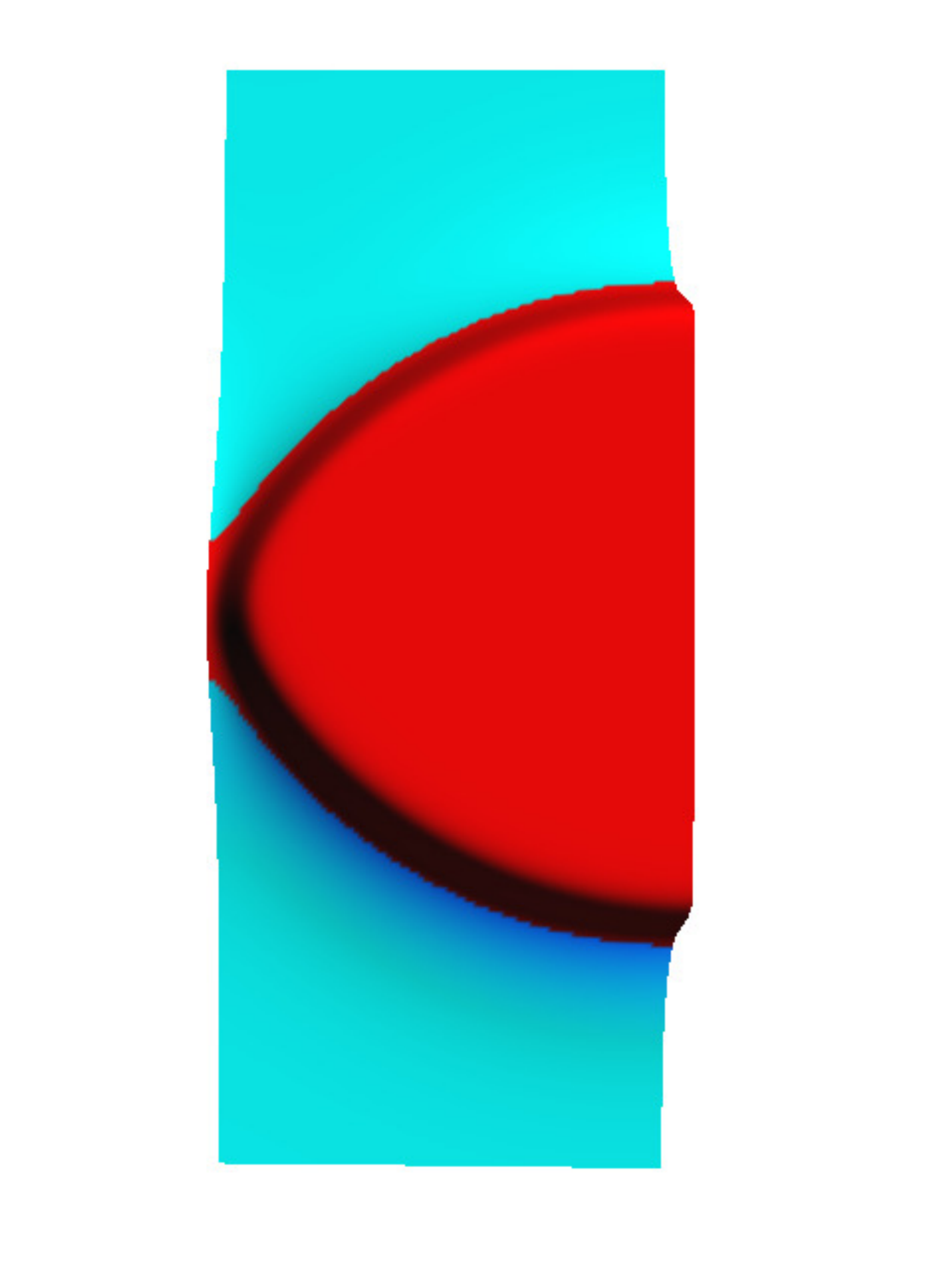}}
\put(42.6,66.6){$0$}
\put(40.4,45){$-8$}
\put(44.7,43.2){$-7$}
\put(61.4,43.2){$0$}
\put(42.6,86.5){$8$}
\put(42.7,76.55){$4$}
\put(40.5,54.9){$-4$}
\put(38.2,66){\begin{sideways}$x$ \end{sideways}}
\put(49.3,40.5){$-\mathrm{Im}\, t$}
\put(68.7,40.6){\includegraphics[height=52\unitlength]{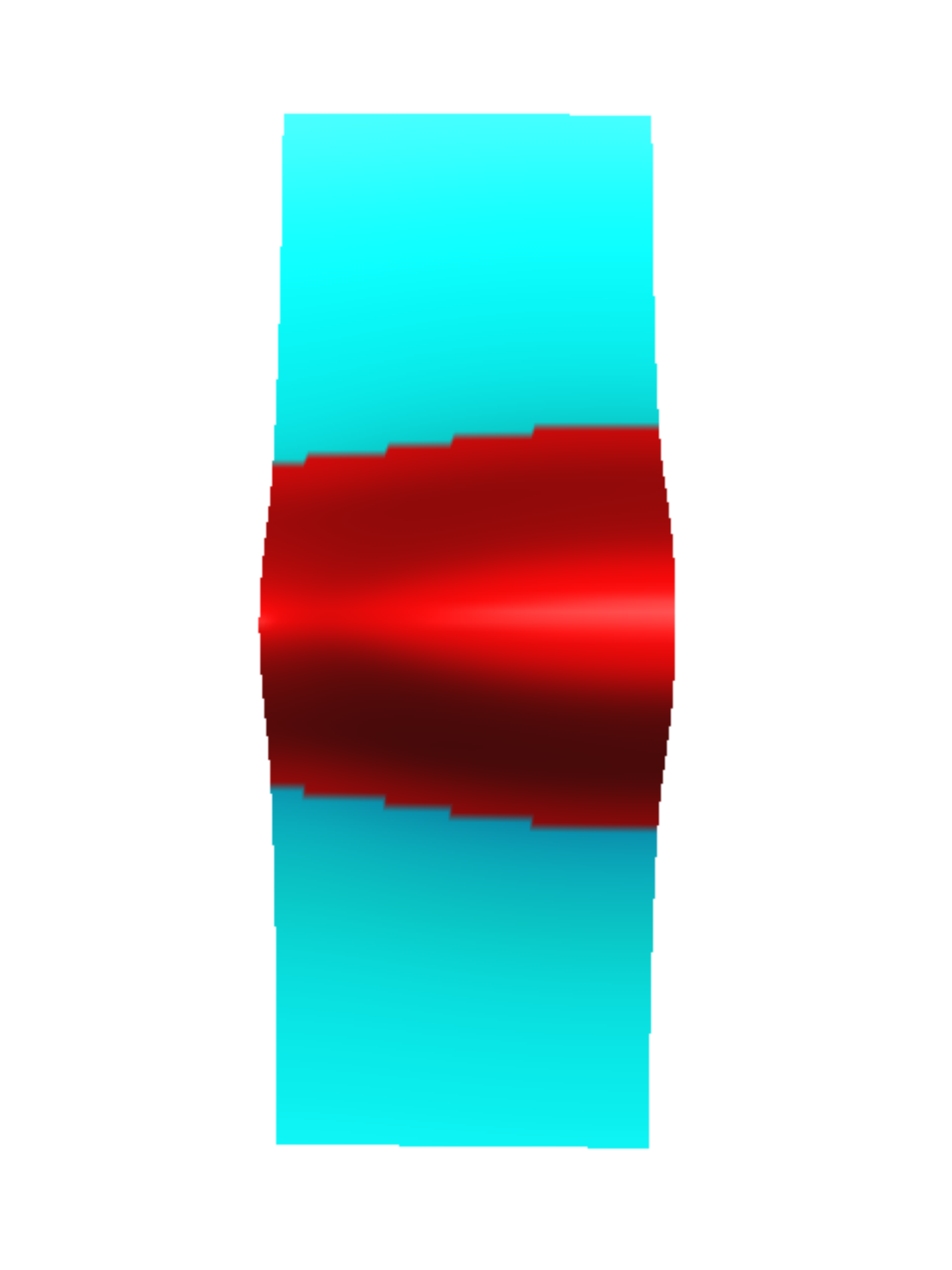}}
\put(77.3,66.6){$0$}
\put(75.2,45){$-3$}
\put(77.3,86.5){$3$}
\put(75.2,52.2){$-2$}
\put(75.4,59.4){$-1$}
\put(77.6,74){$1$}
\put(77.3,81){$2$}
\put(81,43.2){$-2$}
\put(94.6,43.2){$0$}
\put(72.5,66){\begin{sideways}$x$ \end{sideways}}
\put(84.5,40.5){$-\mathrm{Im}\, t$}
\put(0,0){\includegraphics[width=100\unitlength]{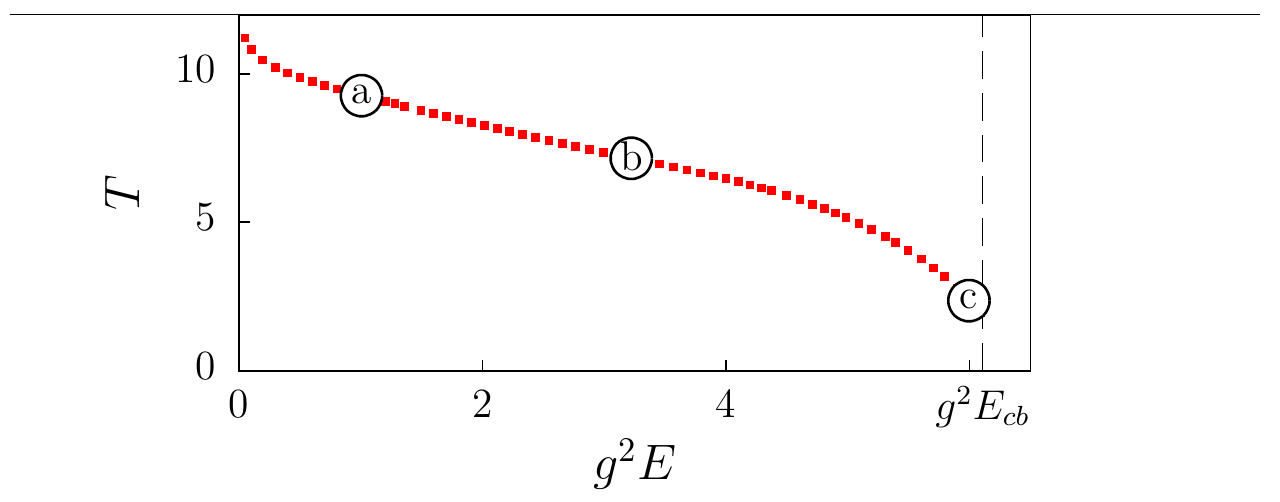}}
\put(85,91){(c)}
\put(51,91){(b)}
\put(15,91){(a)}
\end{picture}}
\caption{(a)---(c) Three--dimensional plots of the   periodic 
  instantons ${\phi_s(t,x) \in \mathbb{R}}$ at different $T$; these
  solutions have $\theta=0$. Only Euclidean parts $t\in [iT,\, 0]$ of
  the solutions 
  are shown. Red/dark and blue/light regions correspond to $\phi_s > 0$ and
  $\phi_s < 0$, respectively. {\it Lower panel:}~Parameters $T(E)$ of the
  periodic instantons. Circles with letters correspond to the 
  solutions~(a)--(c). \label{fig:PI_ET}}   
\end{figure}
To compute numerically the approximate
configuration~(\ref{eq:46}), we solve the static field equation and
find the critical bubble $\phi_{cb}(x)$. The spectrum of its linear
perturbations  is given by the 
matrix--diagonalization routine of Sec.~\ref{sec:lattice-system}, see
the inset in Fig.~\ref{fig:sphaleron}. Picking up the negative mode 
$\delta \phi_-$ with $\omega_-^2 < 0$  and choosing $A_- = 0.3$, we
construct the approximate solution~(\ref{eq:46}). 

Now, we are ready to find the entire family of the
periodic instantons. We numerically solve the field
equation along the Euclidean part BC of the time contour with the
boundary conditions (\ref{period}). We use the Newton--Raphson method
described in the previous Section. The very first solution
(Fig.~\ref{fig:PI_ET}c)  is obtained at $T = \pi/|\omega_-| + \Delta
T$, where $\Delta T$ is small. In this case the
configuration~(\ref{eq:46}) serves as a zeroth--order approximation
for the numerical procedure. Next, 
we increase the parameter $T$ in small steps finding one solution at a
time. The starting configuration at each step is 
provided by the last known solution. Examples of the periodic
instantons are plotted in Figs.~\ref{fig:PI_ET}a--c. Their energies
$E(T)$  are given by Eq.~(\ref{eq:25}) (squares in the lower panel of 
Fig.~\ref{fig:PI_ET}).

One sees a clear distinction between the low-- and high--energy
solutions in Figs.~\ref{fig:PI_ET}a and~\ref{fig:PI_ET}c. While the
latter resembles the critical bubble, the former 
is nearly rotationally--invariant and has large
Euclidean period $T$.  At $E\to 0$ the solutions approach
the bounce thus describing spontaneous decay of the false vacuum. The
periodic instantons are absent in the opposite region $E>E_{cb}$.

Now, consider the limit $\delta \rho \to 0$ of the solutions with
$\theta=0$. In Appendix~\ref{App:C} we remind that 
the sizes of the true vacuum bubbles become infinite 
at small $\delta \rho$ justifying the celebrated thin--wall
approximation~\cite{Coleman, Voloshin:1986zq, voloshin_induced,
  Rubakov:1992gi,Ivlev:1987zz}. The respective solutions
$\phi_s(t,\,x)$ can be obtained analytically. Their action equals
\begin{equation}
\label{eq:45}
2\mathrm{Im}\, S[\phi_s] = \frac{2 g^2 M_S^2}{\delta\rho} \left[
\mathrm{arcsin} \, \frac{T\delta \rho}{g^2 M_S} + \frac{T
  \delta\rho}{g^2 M_S}  \sqrt{1 - \left(\frac{T \delta\rho}{g^2
    M_S}\right)^2}\right] \qquad \mbox{at}\;\; T < g^2 M_S/\delta \rho
\end{equation}
and stays constant at larger periods, see Appendix \ref{App:C}
and~\cite{Rubakov:1992gi}. Note that the typical values of $T$ and 
$2\mathrm{Im}\, S$ in Eq.~(\ref{eq:45}) are proportional to $1/\delta
\rho$ implying infinite suppression $F_N(E)\to +\infty$ 
at $\delta \rho \to 0$. This comes with no surprise, since
transitions between the degenerate vacua are energetically forbidden
below the threshold $E = 2M_S$ of soliton pair creation. On the
other hand, at finite $T$ and $\delta \rho \to 0$ the
rate~(\ref{eq:50}), (\ref{eq:45}) reduces to the Boltzmann probability
\begin{equation}   
\notag
\Gamma \sim \mathrm{e}^{-2M_s \beta} \;,
\end{equation}
of finding the soliton pair in the thermal ensemble at
temperature~${\beta^{-1} =   (2T)^{-1}}$,  
cf.~\cite{Grigoriev:1988bd}.  

\begin{figure}[t]
\centerline{\includegraphics[width=0.49\textwidth]{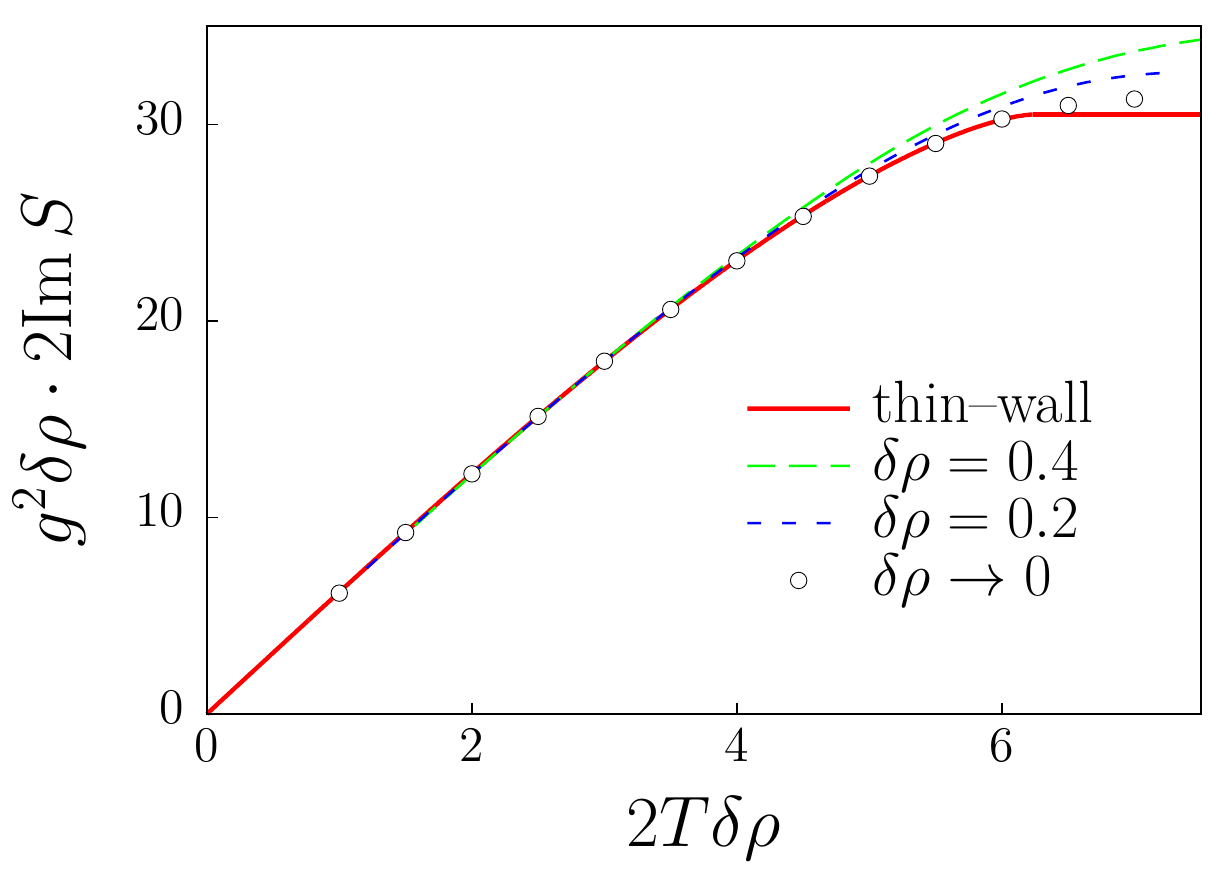}}
\caption{Euclidean action of the periodic instantons at different
  $\delta \rho$ (dashed lines) and their limit $\delta \rho \to 0$
  (empty points) versus the thin--wall result~(\ref{eq:45}) (solid line).\label{fig:thermal}}   
\end{figure}
In Fig.~\ref{fig:thermal} we compare the actions of the periodic
instantons  at different $\delta \rho$ (dashed lines) with
Eq.~(\ref{eq:45}) (solid line). One observes nice agreement which becomes almost
perfect if one extrapolates the numerical results to $\delta \rho = 0$
(empty points in Fig.~\ref{fig:thermal}). 

\subsection{Solutions below $E_{cb}$}
\label{sec:solutions-below-e_cb}
\begin{figure}[t]
\unitlength=0.009\columnwidth
\centerline{\begin{picture}(100,40)(0,-2.5)
\put(5.4,38.5){\includegraphics[width=35\unitlength,angle=-90]{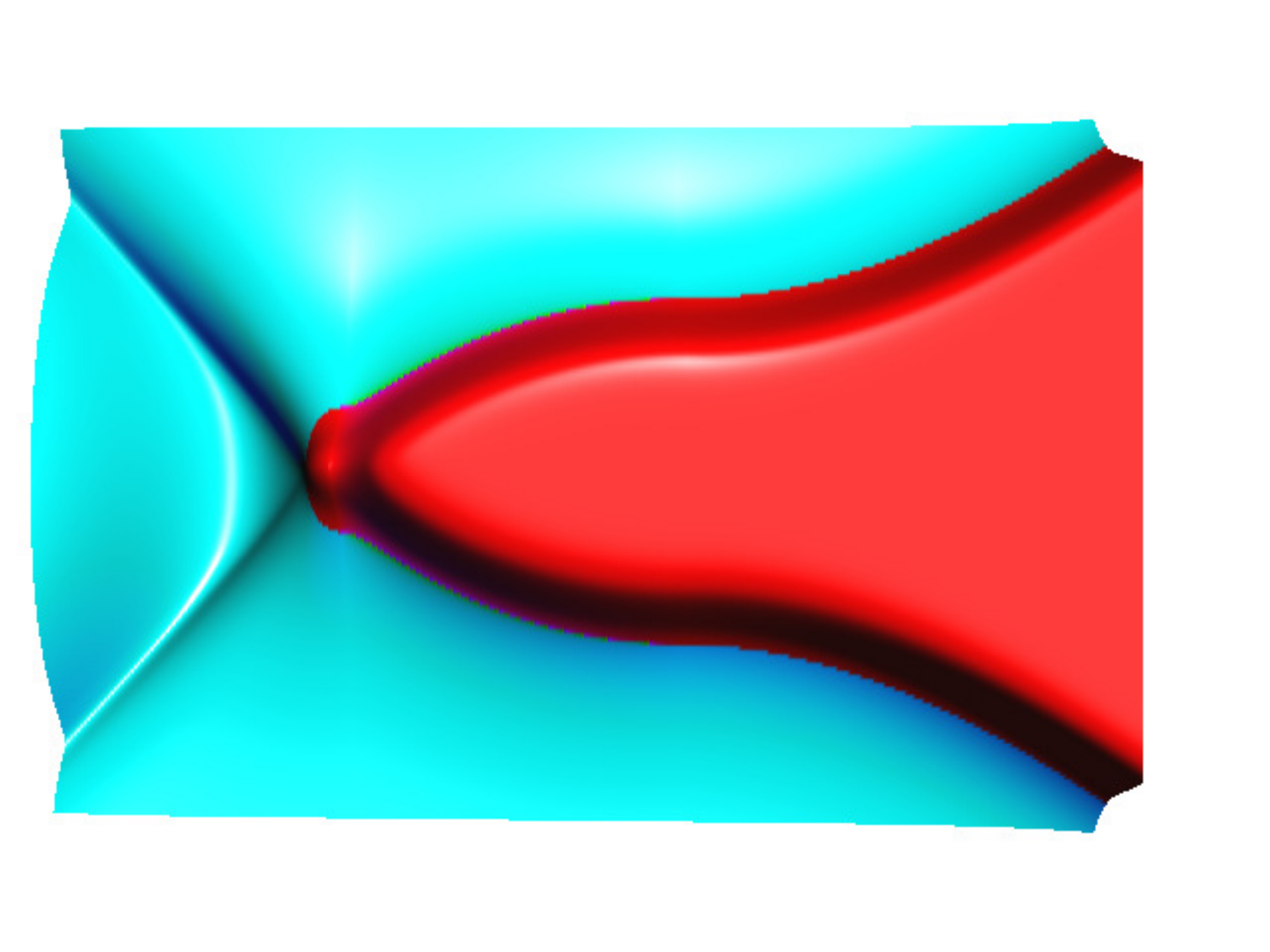}}
\put(21,2){${\rm Re}\, t - {\rm Im}\, t$}
\put(26,-2.5){$(a)$}
\put(4.5,30.8){4}
\put(4.5,21){0}
\put(1.5,21){\begin{sideways}$x$\end{sideways}}
\put(3.9,11){-4}
\put(11.5,5.5){-8}
\put(20.8,5.5){-4}
\put(29.9,5.5){0}
\put(17.5,34.8){B}
\put(7,34.8){A}
\put(29.8,34.8){C}
\put(45,34.8){D}
\put(39.4,5.5){4}
\put(57.7,38.1){\includegraphics[width=35\unitlength,angle=-90]{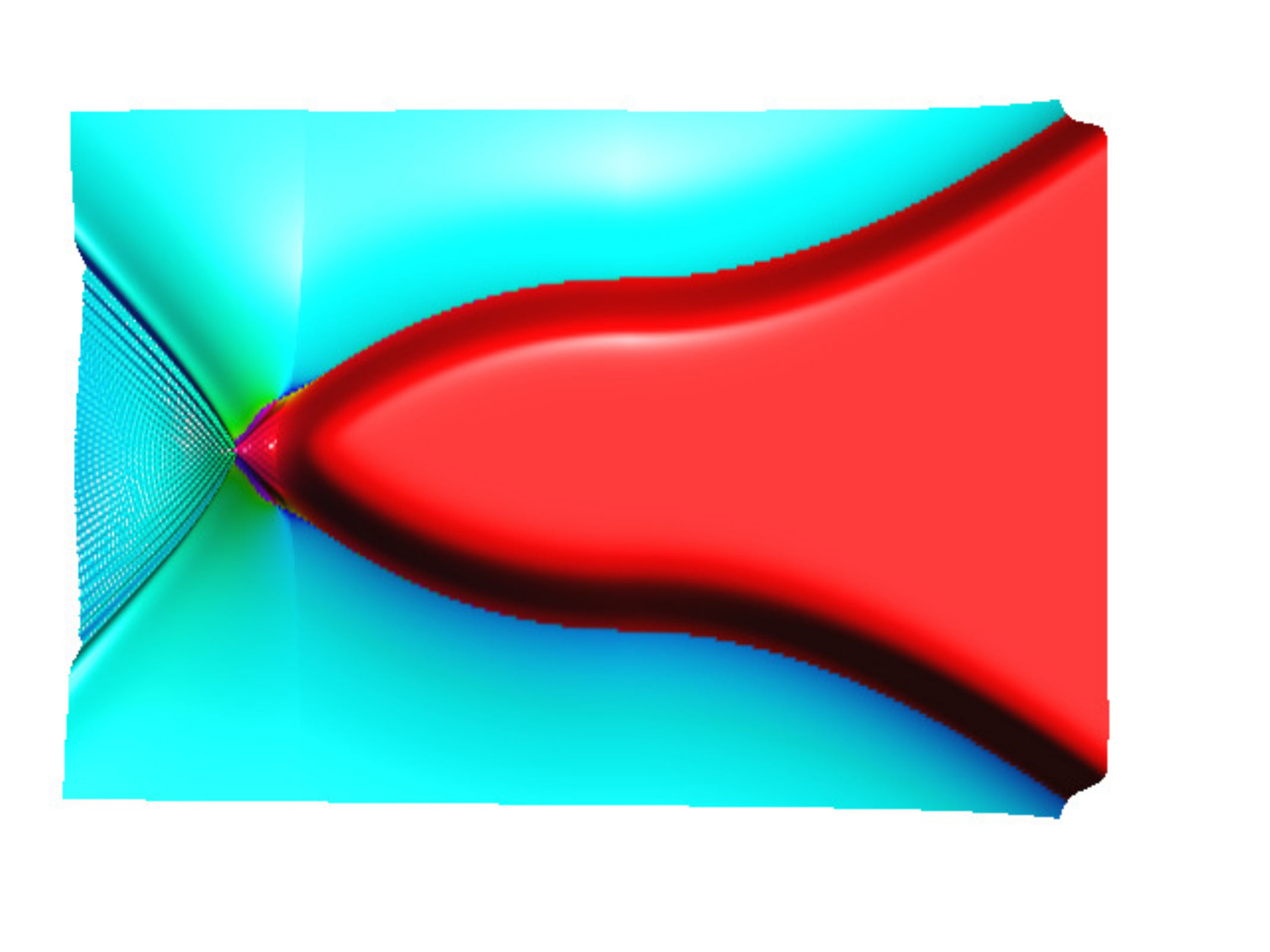}}
\put(57.9,30.8){4}
\put(57.9,20.8){0}
\put(57.1,10.5){-4}
\put(61.8,5.5){-8}
\put(71,5.5){-4}
\put(80.2,5.5){0}
\put(89.5,5.5){4}
\put(68.0,34.8){B}
\put(59.5,34.8){A}
\put(80.0,34.8){C}
\put(96,34.8){D}
\put(55,21){\begin{sideways}$x$\end{sideways}}
\put(73,2){${\rm Re}\, t - {\rm Im}\, t$}
\put(78,-2.5){$(b)$}
\put(49.5,8){\includegraphics[width=7\unitlength,angle=-90]{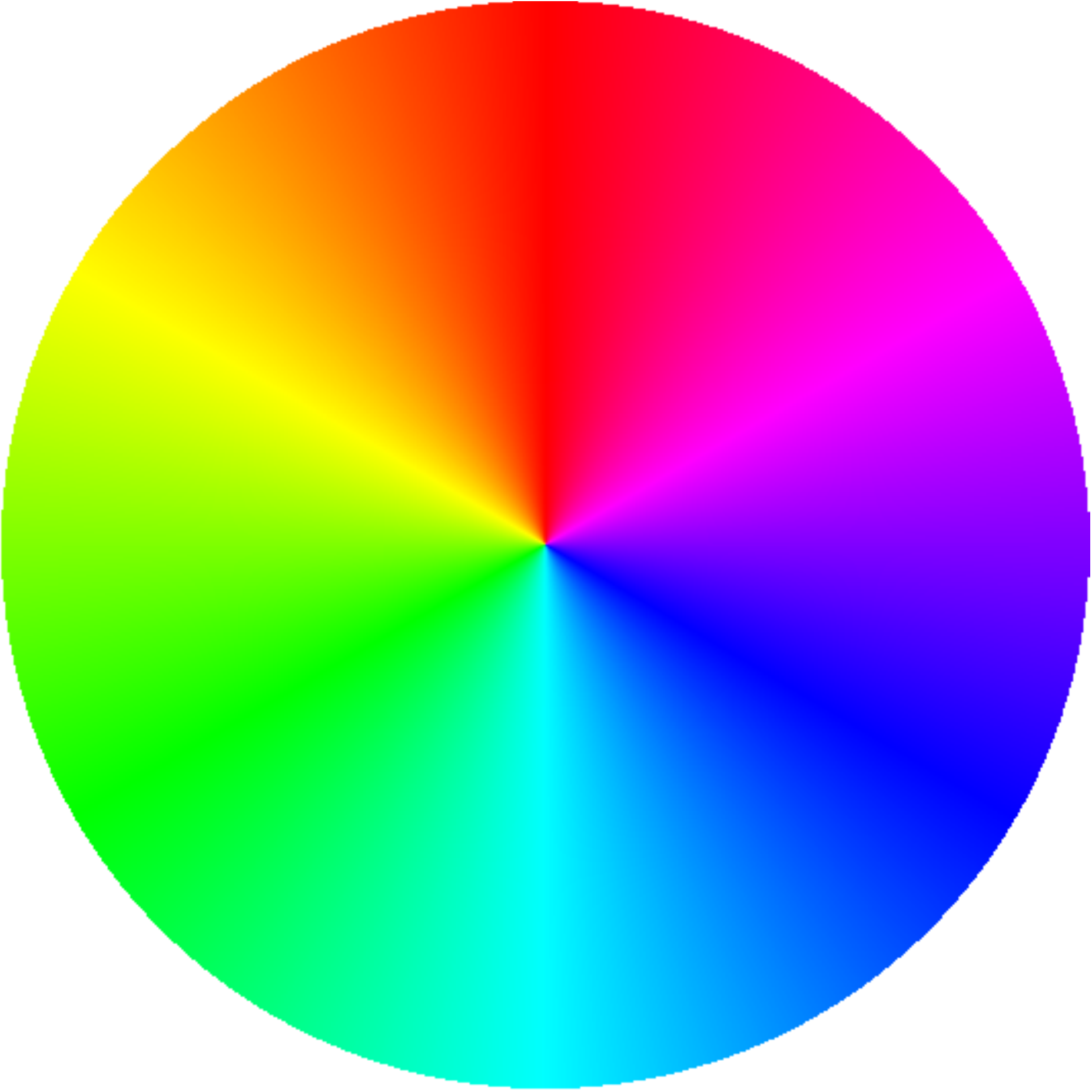}}
\put(49.5,-1.5){$\mathrm{arg}\, \phi_s$}
\put(57,3.5){$0$}
\put(47.3,3.7){$\pi$}
\end{picture}}
\caption{\label{pi_solutions} Solutions at $E< E_{cb}$.
  (a) The periodic instanton with $(T,\, \theta) = (5.2,\, 0)$, 
  $g^2 (E,\, N)\approx(5.0,\,  3.8)$. (b) Real part of the solution
  $\mathrm{Re}\, \phi_s(t,\, x)$ at $(T,\, \theta) = (5.0,\, 0.76)$
  and $(g^2E,\, g^2N) \approx (5.0, \, 2.0)$. Only the central 
  regions of moderately small $|t|$ and $|x| < 5$ are shown. The
  points A, B, C, D of the time contour are marked above each
  solution. Color represents $\mathrm{arg}\,\phi_s$.}
\end{figure}
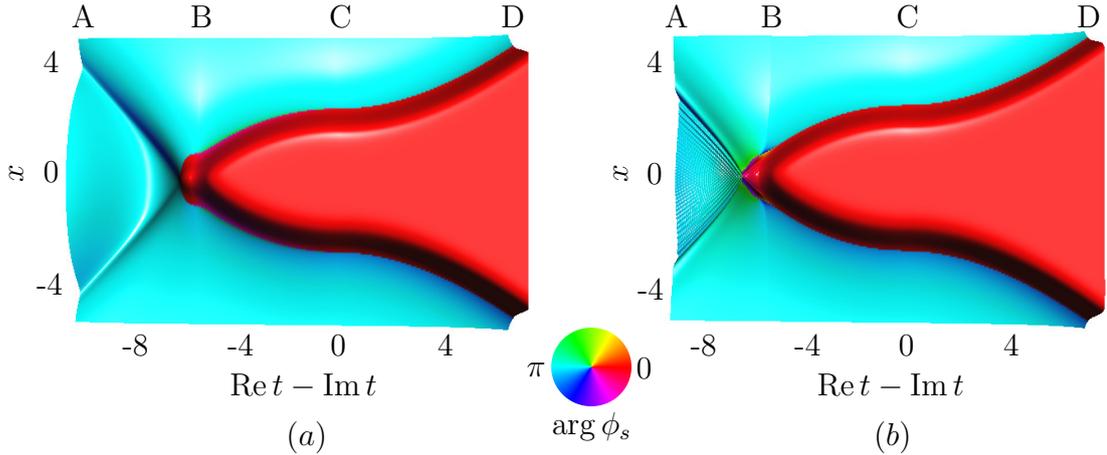

\begin{figure}[ht!]
\centerline{\includegraphics[width=0.65\textwidth]{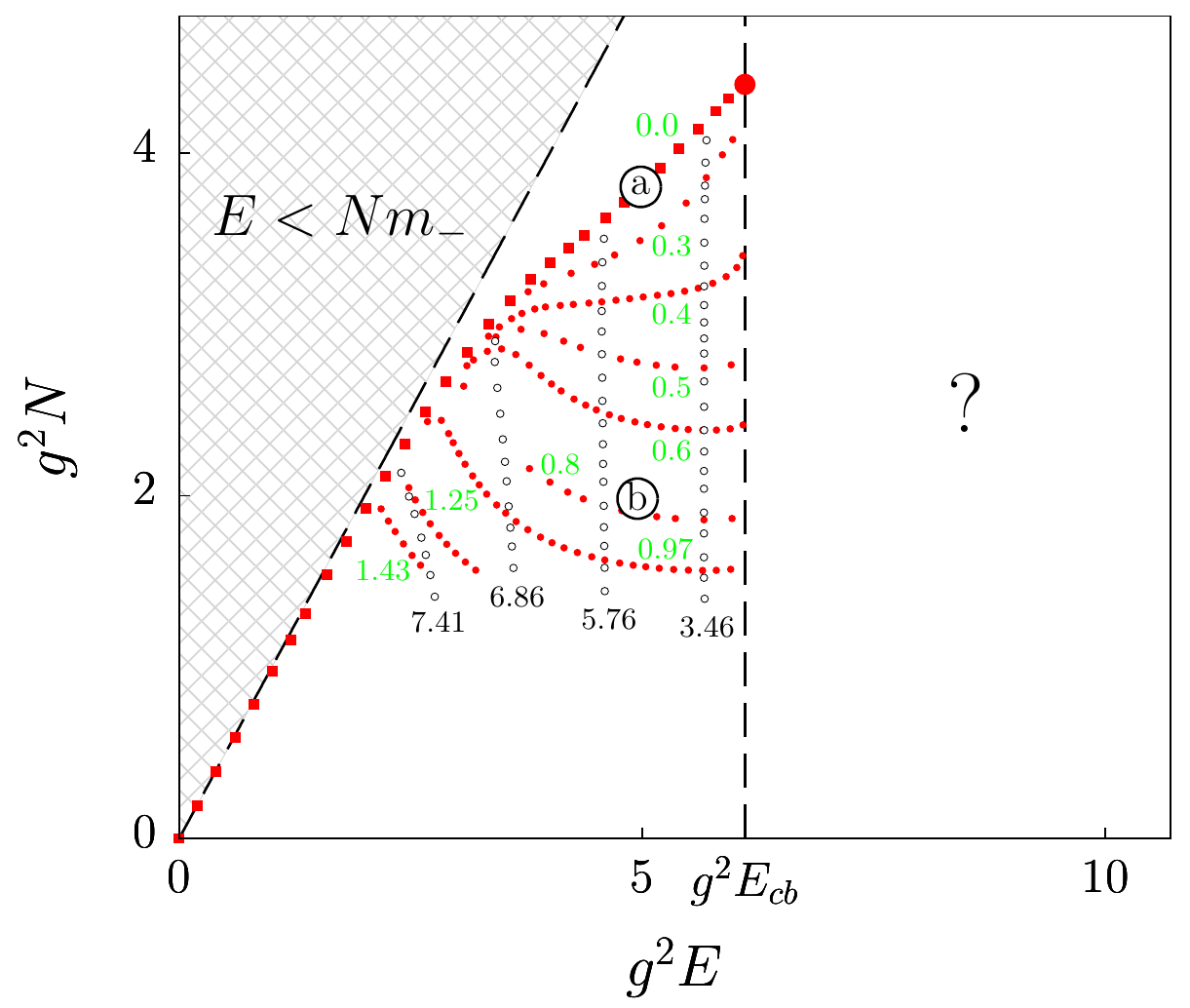}}
\caption{\label{EN} Parameters $(E,\, N)$ of the 
  numerical solutions at $E < E_{cb}$. Empty and filled points are
  situated along the lines $T = \mbox{const}$ and $\theta =
  \mbox{const}$, respectively; the value of the constant parameter is
  written near each line in black for $T$ and in green/gray for 
  $\theta$. Circles with letters represent parameters of the solutions in
  Fig.~\ref{pi_solutions}. Filled squares are the periodic 
  instantons ($\theta=0$). 
}
\end{figure}

Solving the field equation backwards and forwards in time, we continue
the periodic instantons to the parts AB and CD of the time contour. We
thus obtain the complete solutions, see the one in
Fig.~\ref{pi_solutions}a. At large negative times they 
describe wave packets in the false vacuum which collide at 
$t\sim iT$ producing expanding bubbles of the true vacuum. We compute the
in--state quantum numbers $(E,\, N)$ of the solutions by
Eqs.~(\ref{E_N}). The 
line of the periodic instantons is shown with filled squares in the
$(E,\, N)$ plane of Fig.~\ref{EN}.

Starting from the known solutions at $\theta=0$, we find the ones with
positive $\theta$. To this end we increase the value of
$\theta$ in small steps keeping $T = \mbox{const}$.  At each step we
numerically solve\footnote{Recall that the lattice analogs
  (\ref{eq:24}), (\ref{eq:29}) of the final boundary conditions are
  imposed at the point C of the time contour. After solving the
  equations, we continue each 
  solution to the part CD of the contour.} the
boundary value problem~(\ref{eq:Ttheta}). An example of the solution
$\mathrm{Re}\, \phi_s (t,\, x)$ with $\theta>0$ is shown in
Fig~\ref{pi_solutions}b. It is complex and  contains sharp wave 
packets at large negative time, see the color representing 
$\mathrm{arg}\, \phi_s$. 

Evaluating the quantum numbers $(E,\,N )$ of the solutions by
Eqs.~(\ref{E_N}), we mark them with points in the initial data plane
of Fig.~\ref{EN}. One sees that the numerical data cover the region
$E<E_{cb}$ and $N>1.4/g^2$:  at small $N$ the solutions become sharp
and require better lattice resolution. On the other hand, the
high--energy region 
will be explored in the next Section. The suppression exponent
$F_N(E)$ is computed using Eq.~(\ref{s_exp}).

\begin{figure}[t]
\centerline{\includegraphics[width=0.49\textwidth]{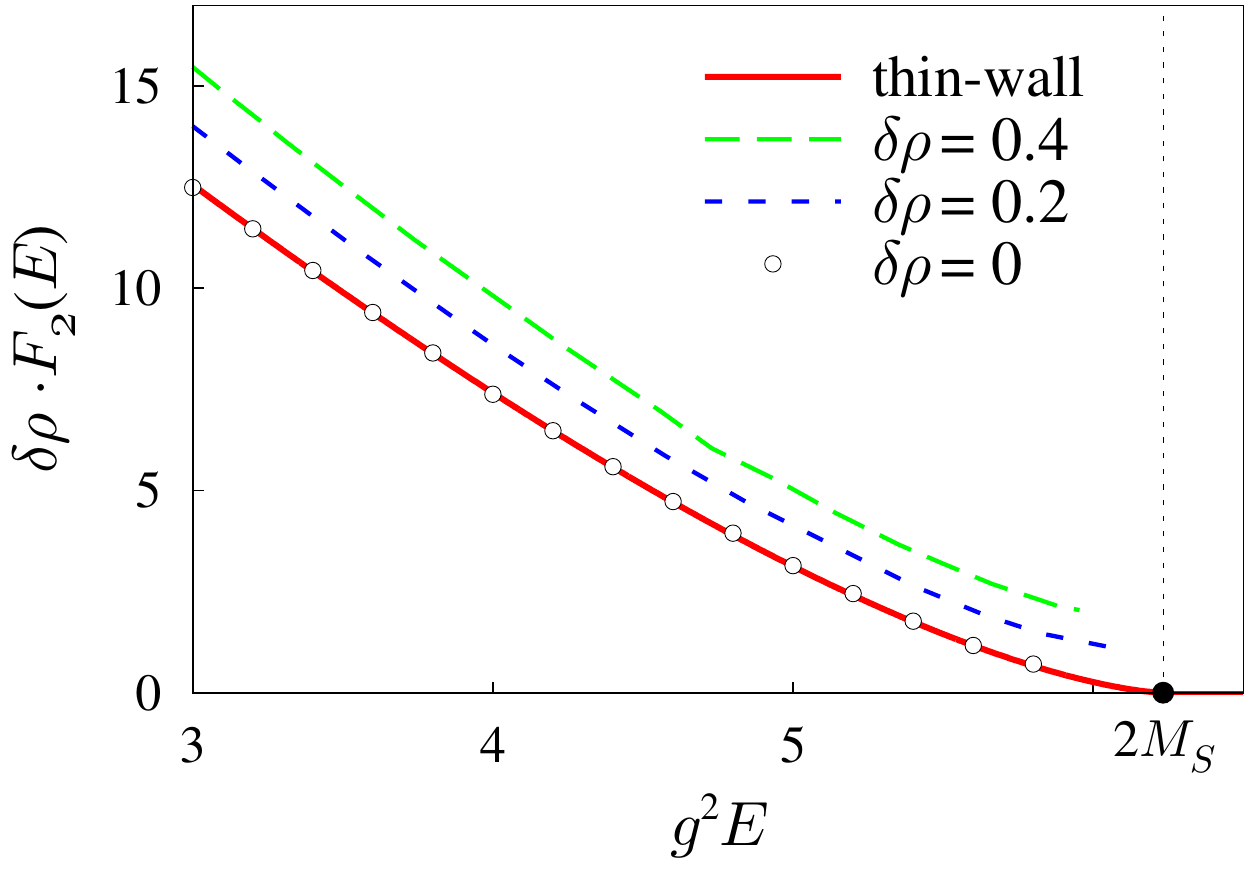}}
\caption{\label{voloshin_pi} The thin--wall prediction~\eqref{eq_4.2}
  (solid line) and numerical results for the
  two--particle exponents $F_2(E)$ at $E<2M_S$ and different $\delta
  \rho$ (dashed lines). Points are obtained by linearly extrapolating  
the numerical data to $\delta \rho=0$. }
\end{figure}

The thin--wall arguments of the previous Section suggest that the
exponent is inversely proportional to  $\delta \rho$ at small
values of the latter. Thus, its Laurent series expansion
starts with the singular term,
\begin{equation}
\label{eq:61}
F_{N}(E) = \frac{F_{N,\, -1}(E)}{\delta \rho} + F_{N,\, 0}(E) + {\cal
  O}(\delta\rho)\;.
\end{equation}
In Appendix \ref{App:C} we deduce
\begin{equation}
\label{eq_4.2}
F_{N,\, -1}(E) = 2 g^4M_{S}^2\left[{\rm
  arccos}\left(\frac{E}{2M_{S}}\right) -
\frac{E}{2M_{S}}\sqrt{1-\frac{E^2}{4M_{S}^2}}\right]
\end{equation}
considering dynamics of the thin--wall bubbles,
see also~\cite{Rubakov:1992gi}. The function (\ref{eq_4.2}) is shown
by solid line in Fig.~\ref{voloshin_pi}. It does not depend on $N$ and 
decreases with energy reaching zero at $E=2M_S$. At higher
energies  Eq.~(\ref{eq_4.2}) does not give any sensible result: 
in Appendix~\ref{App:C} we argue that the thin--wall approximation
breaks down in that region. Presumably, this means that the limit $\delta
\rho \to 0$ of the suppression exponent exists at $E\gtrsim 2M_S$.

Let us compare the numerical results for $F_N(E)$ with the thin--wall
expression \eqref{eq_4.2}. Since the periodic instantons with $\theta
= 0$ have been already studied, we consider the opposite case $\theta
\to +\infty$ or $g^2N \to 0$. In this limit $F_N(E)$ coincides 
with the exponent $F_2(E)$ of transitions from the two--particle
initial states, see the conjecture~(\ref{eq:4}). We extrapolate
the numerical data for 
$F_N(E)$ to $g^2 N=0$ and obtain the dashed lines in
Fig.~\ref{voloshin_pi}; the details of this extrapolation will be discussed
in Sec.~\ref{sec:two-part-proc}. One observes that the numerical
graphs  approach Eq.~(\ref{eq_4.2}) at
smaller $\delta \rho$ and coincide with it  after the additional
extrapolation to $\delta \rho =0$ (empty points). 

\subsection{Going to $E>E_{cb}$}
\label{sec:5}
\begin{figure}[t]
\centerline{
\unitlength=0.01\columnwidth
\begin{picture}(65,38)(3,0)
\put(7,41.4){\includegraphics[width=.4\columnwidth,angle=-90]{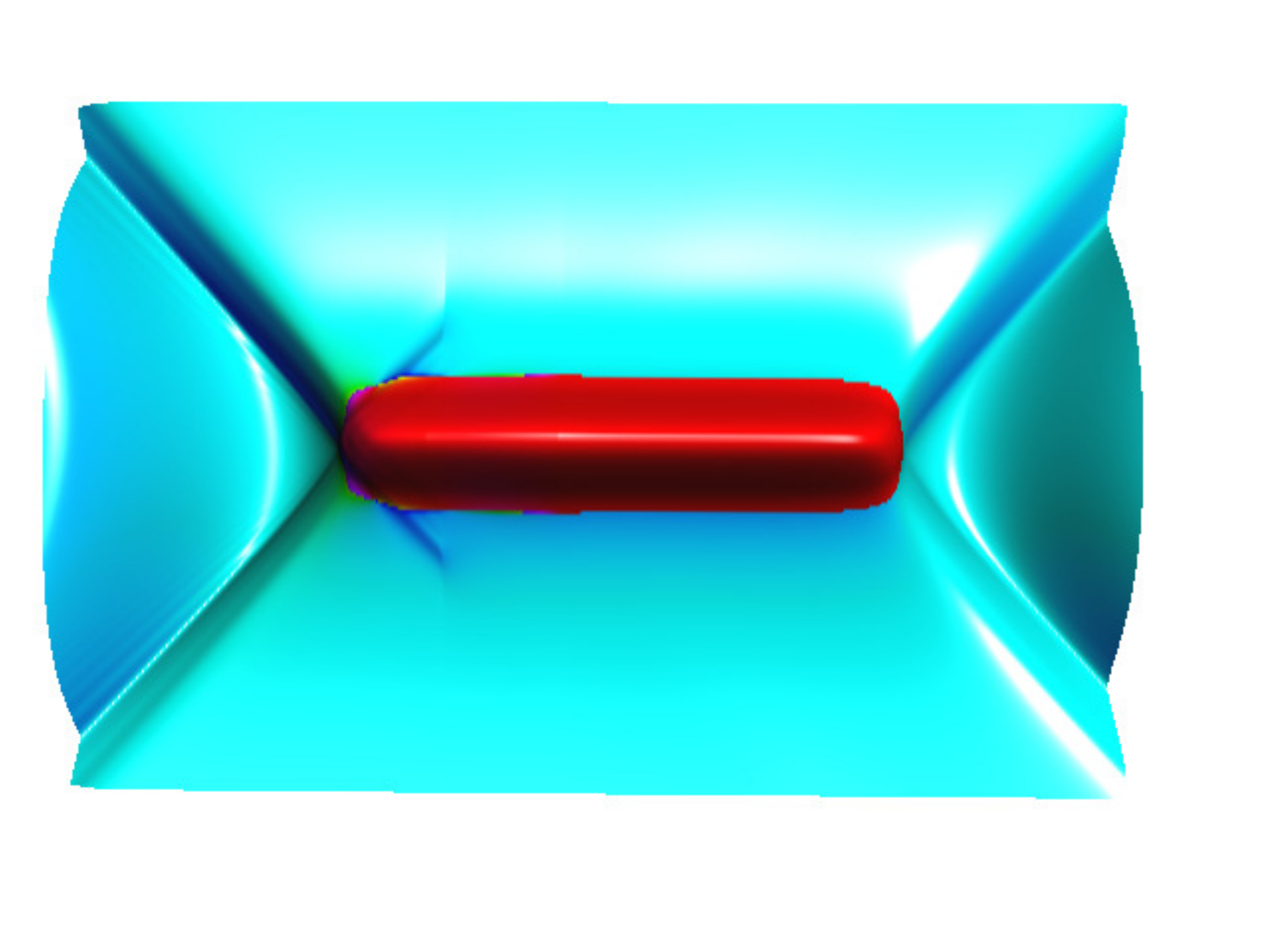}}
\put(6.8,33.7){4}
\put(6.8,28.3){2}
\put(6.8,22.1){0}
\put(3,22.1){\begin{sideways}$x$\end{sideways}}
\put(6.0,16.3){-2}
\put(6.0,10.5){-4}
\put(9.8,5.7){-8}
\put(19.9,5.7){-4}
\put(30.4,5.7){0}
\put(41.0,5.7){4}
\put(51.5,5.7){8}
\put(25.4,2){${\rm Re}\, t - {\rm Im}\,t$}
\put(60,13){\includegraphics[width=0.07\columnwidth,angle=-90]{pie.pdf}}
\put(61,3){$\mathrm{arg}\, \phi_s$}
\put(58.0,8.9){$\pi$}
\put(67.5,8.8){0}
\end{picture}}
\caption{\label{reflected_solution} The unphysical ``reflected''
  solution at $E > E_{cb}$; $(T,\, \theta) = (1.866,\, 0.3)$ and
  $g^2(E,\, N) \approx (6.11,\, 4.1)$.}
\end{figure}
Since the thin--wall bubbles do not describe transitions at
$E>E_{cb}$, one expects to find new properties of the semiclassical
solutions in that region. In realistic calculations we observe
somewhat different effect: at energies above\footnote{More precisely,
  this happens at $E>E_1(N) \approx E_{cb}$. We ignore small
  difference between these two thresholds in  what follows.} $E_{cb}$
our numerical method either diverges or produces unphysical
``reflected'' solutions exemplified in
Fig.~\ref{reflected_solution}. The latter satisfy the semiclassical
equations but approach the false vacuum at $t\to +\infty$. They
apparently cannot describe transitions between the vacua. One
immediately identifies~\cite{Bezrukov:2003yf} the root of the
problems: a condition forcing the solutions to interpolate between the
two vacua is 
not present in the semiclassical boundary value problem
(\ref{eq:Ttheta}). As a consequence, the numerical procedure can
produce unphysical solutions even if the ones with correct properties
exist.  

The problem of fixing the qualitative behavior of the saddle--point
configurations is rather general. We solve it using the
$\epsilon$--regularization technique of Refs.~\cite{Bezrukov:2003yf,
  epsilon, rst_confirmations2}. To this end we recall that in our
numerical method the 
solutions with shrinking bubbles are continuously obtained from the 
physical ones by decreasing the parameter $T$ below some value
$T_*(\theta)$. The ``boundary''  solutions at $T = T_*(\theta)$ are
very special: their  true vacuum bubbles neither shrink  nor expand
but survive to $t\to +\infty$. The main idea of the
$\epsilon$--regularization is to exclude all ``boundary'' 
configurations from the set of accessible semiclassical
solutions. Once this is achieved, one cannot obtain solutions with
shrinking bubbles from the physical ones by continuous
deformations. Then at $E>E_{cb}$ we will find solutions with correct
properties (if they exist).

With this logic in mind, we add~\cite{Bezrukov:2003yf} a small imaginary
term to the classical action,
\begin{equation}
\label{action-epsilon}
S \to S_{\epsilon} = S[\phi] + i\epsilon T_{int}[\phi]\;,
\end{equation}
where the modification parameter $\epsilon$ and functional
$T_{int}$ are positive--definite. Importantly, we require special
properties of $T_{int}[\phi]$: it should take finite
values on any configuration interpolating between the vacua and 
diverge if $\phi(t,\, x)$ contains a static finite--size bubble in the
final state. Then the latter ``boundary'' configurations with static bubbles
represent singularities of the modified action  $S_{\epsilon} \to
+i\infty$ and cannot coincide with its extrema. To the contrary, the 
semiclassical solutions in the modified system extremize
$S_{\epsilon}$ and do not belong to the class of ``boundary'' 
configurations. We conclude that the modified solutions cannot change
qualitative properties in the course of continuous deformations;
finding their continuous family at $E>E_{cb}$ and sending $\epsilon\to
+0$, one arrives at correct high--energy solutions. 

Let us construct the functional $T_{int}[\phi]$ with the desired 
properties. We choose
\begin{equation}
\label{eq:52}
T_{int}[\phi] = \frac1{g^2}\int dt\, dx\; f(x) \, W_{int}\left(\frac{\phi-\phi_{+}}{a}\right)\;,
\end{equation}
where $f(x) = \mathrm{e}^{-x^2/2\sigma_f^2}$ and $W_{int}(u) = u^4\, {\rm
  e}^{-u^2}$; $a = 0.4$. The function $f(x)$ restricts the spatial
integral to the central region $|x| \lesssim \sigma_f$, where 
$\sigma_f = 0.4$ in our numerical calculations. At the same time,
$W_{int}$ is vanishingly small at $\phi \approx \phi_\pm$ and takes 
positive values between the vacua, $\phi_- < \phi < \phi_+
$. Accordingly, the time integral in Eq.~\eqref{eq:52}  diverges 
if the configuration $\phi(t,\, x)$ contains a static finite--size bubble at
$t\to +\infty$. However, if the configuration is  physical and the bubble
expands, the value of the field at $|x|\lesssim \sigma_f$ tends to
$\phi_+$ at large times and the $t$--integral converges. We
conclude that the functional (\ref{eq:52}) discriminates between the
``boundary'' and physical configurations. Note that the modification
(\ref{action-epsilon}), (\ref{eq:52}) simply deforms the scalar potential,
\begin{equation}
\label{eq:54}
V \to V_{\epsilon}(\phi, \, x) = V (\phi) - i\epsilon f(x)
W_{int}\left(\frac{\phi-\phi_+}{a}\right)\;,
\end{equation}
introducing minimal changes of the numerical code.

\begin{figure}[t]
\unitlength=0.009\columnwidth
\centerline{\begin{picture}(105,43)
\put(1.5,44){\includegraphics[width=42\unitlength,angle=-90]{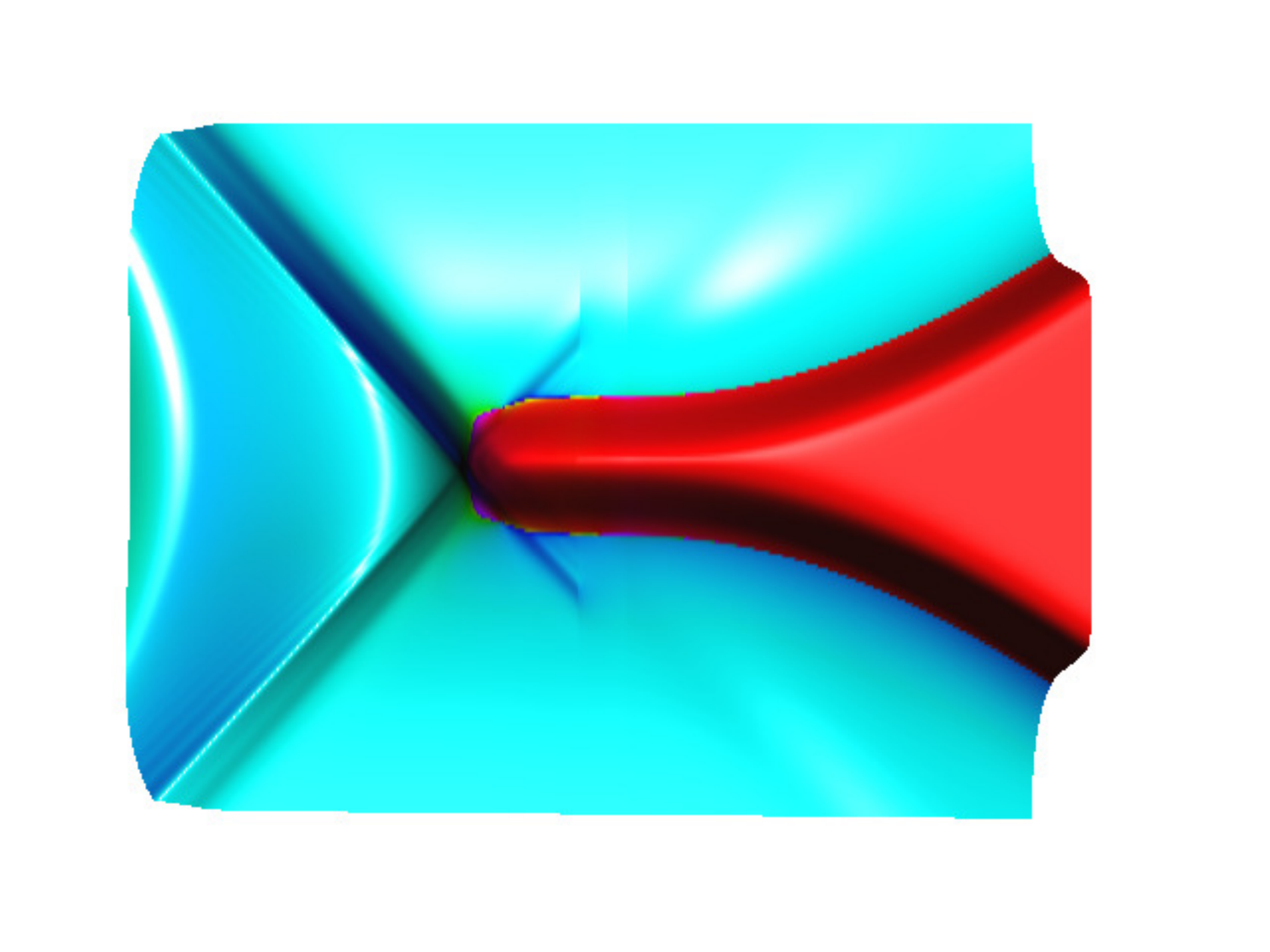}}
\put(23.3,2){${\rm Re}\, t - {\rm Im}\, t$}
\put(4.9,35.7){4}
\put(4.9,29.7){2}
\put(4.9,23.3){0}
\put(1,23.3){\begin{sideways}$x$\end{sideways}}
\put(4.0,17.2){-2}
\put(4.0,10.8){-4}
\put(8.0,5.7){-8}
\put(17.9,5.7){-4}
\put(28.8,5.7){0}
\put(40,5.7){4}
\put(56,44){\includegraphics[width=42\unitlength,angle=-90]{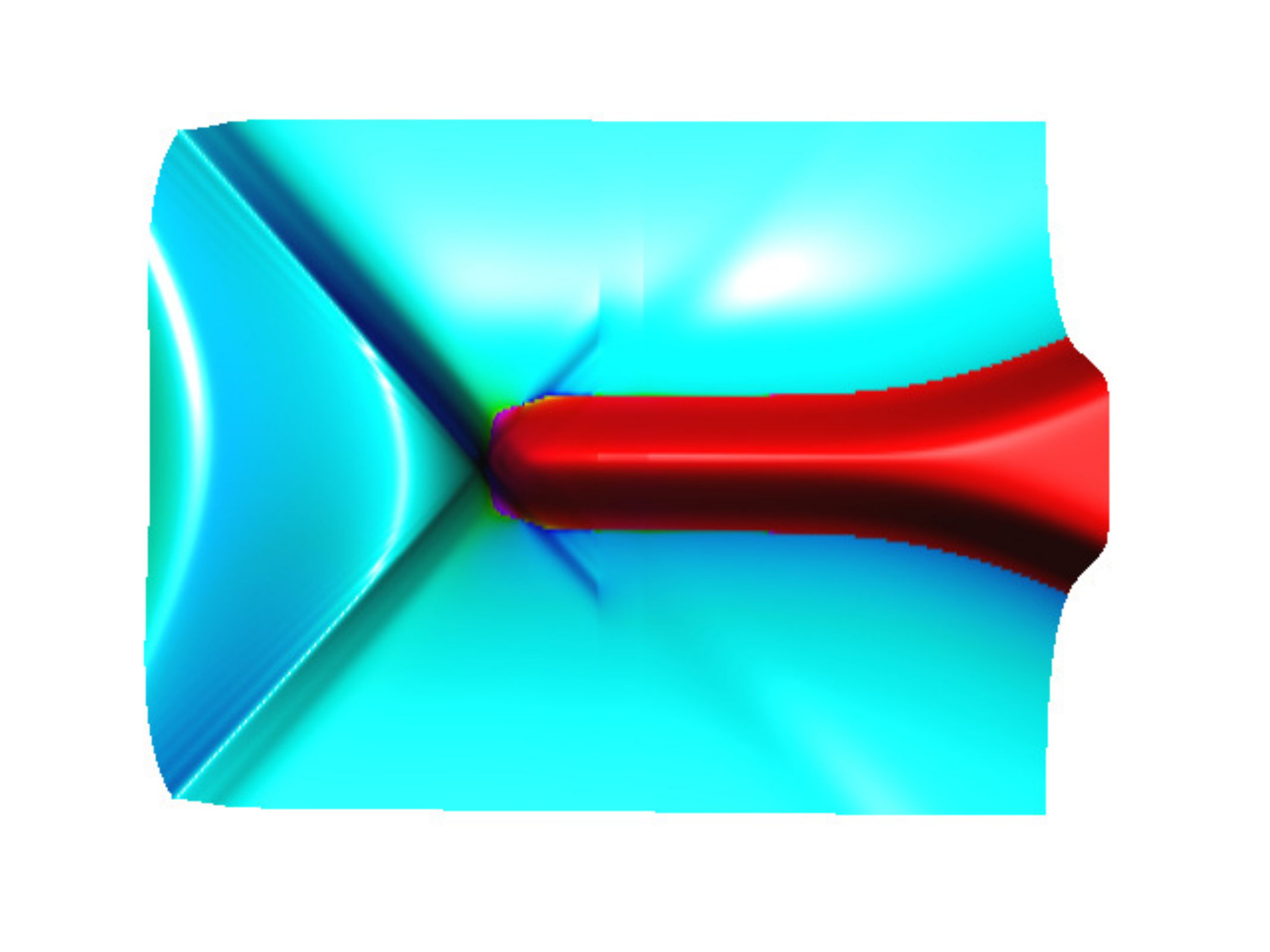}}
\put(60.2,35.7){4}
\put(60.2,29.6){2}
\put(60.2,23.2){0}
\put(56.3,23.3){\begin{sideways}$x$\end{sideways}}
\put(59.4,17){-2}
\put(59.4,10.7){-4}
\put(63.4,5.7){-8}
\put(73.1,5.7){-4}
\put(84.15,5.7){0}
\put(95.3,5.7){4}
\put(79,2){${\rm Re}\, t - {\rm Im}\, t$}
\put(51,10){\includegraphics[width=7\unitlength,angle=-90]{pie.pdf}}
\put(58.5,5.6){0}
\put(51.2,0.7){$\mathrm{arg}\, \phi_s$}
\put(48.8,5.7){$\pi$}
\put(27.2,41){(a)}
\put(82,41){(b)}
\end{picture}}
\caption{\label{fig:boundary_solutions} Modified solutions at
  $\theta = 0.3$, $E \approx 6.11/g^2 > E_{cb}$ and different values
  of the regularization parameter: (a) 
  $\epsilon=5\cdot 10^{-2}$, (b)~$\epsilon =  10^{-3}$. The other
  parameters are (a)~${T = 0.75}$, $g^2N \approx 
  4.14$, (b) $T = 0.7$, $g^2 N \approx 4.13$.}
\end{figure}

\begin{figure}[ht]
\centerline{\includegraphics[width=0.65\textwidth]{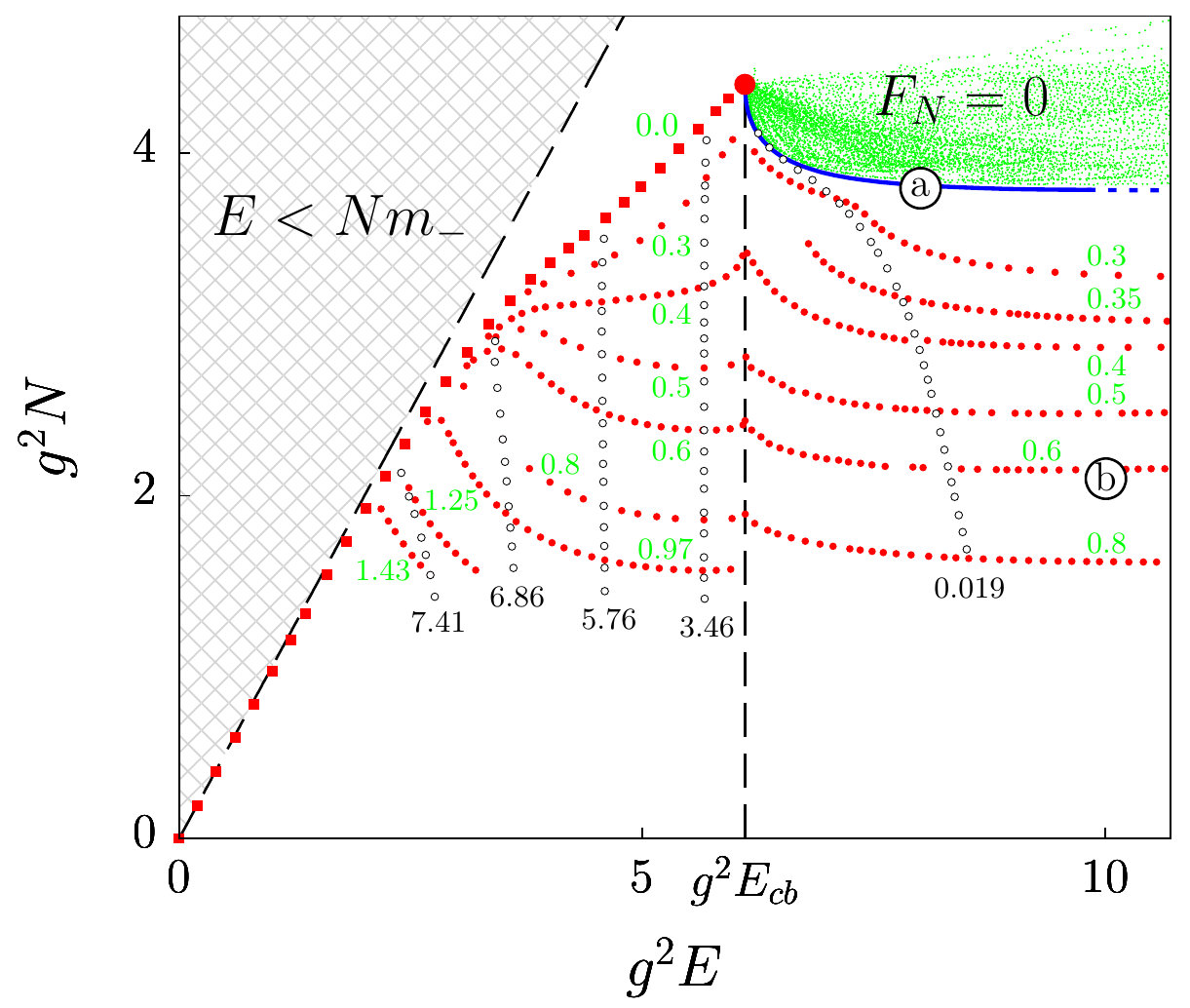}}
\caption{\label{EN1} A complete set of numerical solutions in the
  $(E,\, N)$ plane; $\delta\rho = 0.4$, notations of Fig.~\ref{EN}
  are used. For simplicity the results at $E>E_{cb}$ are obtained at
  nonzero $\epsilon \lesssim 10^{-3}$ using the smaller lattice
  $N_t\times N_x = 3200 \times 150$. Solutions with the highest
  energies $g^2E>11$ and smallest multiplicities $g^2 N < 1.6$ are not
  shown: they need better lattice resolution. Circles with letters
  are the solutions in Figs.~\ref{above_sphaleron_solutions}a,b.}
\end{figure}

We remark that the regularization~(\ref{action-epsilon}) is pretty
general: it was successfully applied in quantum
mechanics~\cite{Bezrukov:2003yf, epsilon, rst_confirmations2,
  Levkov:2007ce, tunneling_time}, field theory~\cite{rst} and 
gravity~\cite{Bezrukov:2015ufa}. Besides, it can be justified by adding
a constraint to the path integral with the Faddeev--Popov  
trick~\cite{epsilon,rst_confirmations2},  cf.~\cite{Affleck:1980mp}.

Following the above recipe, we pick up some physical solution at
$E<E_{cb}$, introduce regularization \eqref{eq:54} with $\epsilon
= 5\cdot 10^{-2}$ and find the modified solution with the same $T$ and
$\theta$. After that we decrease $T$ obtaining the set of modified
solutions at $E>E_{cb}$, see Fig.~\ref{fig:boundary_solutions}a and
cf.\ Fig.~\ref{reflected_solution}. As expected, all these solutions
contain expanding bubbles at $t\to +\infty$ and therefore describe
transitions between the vacua. We finally decrease the modification
parameter to infinitesimally small values ${\epsilon \lesssim
  10^{-3}}$, Fig.~\ref{fig:boundary_solutions}b, and continue to
explore the plane of initial data by changing the parameters $T$ and
$\theta$ in small steps. Eventually, we obtain all possible solutions
at $E>E_{cb}$, see Fig.~\ref{EN1}. 

\begin{figure}[ht]
\unitlength=0.0095\columnwidth
\begin{picture}(102,43)
\put(1.9,44.5){\includegraphics[width=42\unitlength,angle=-90]{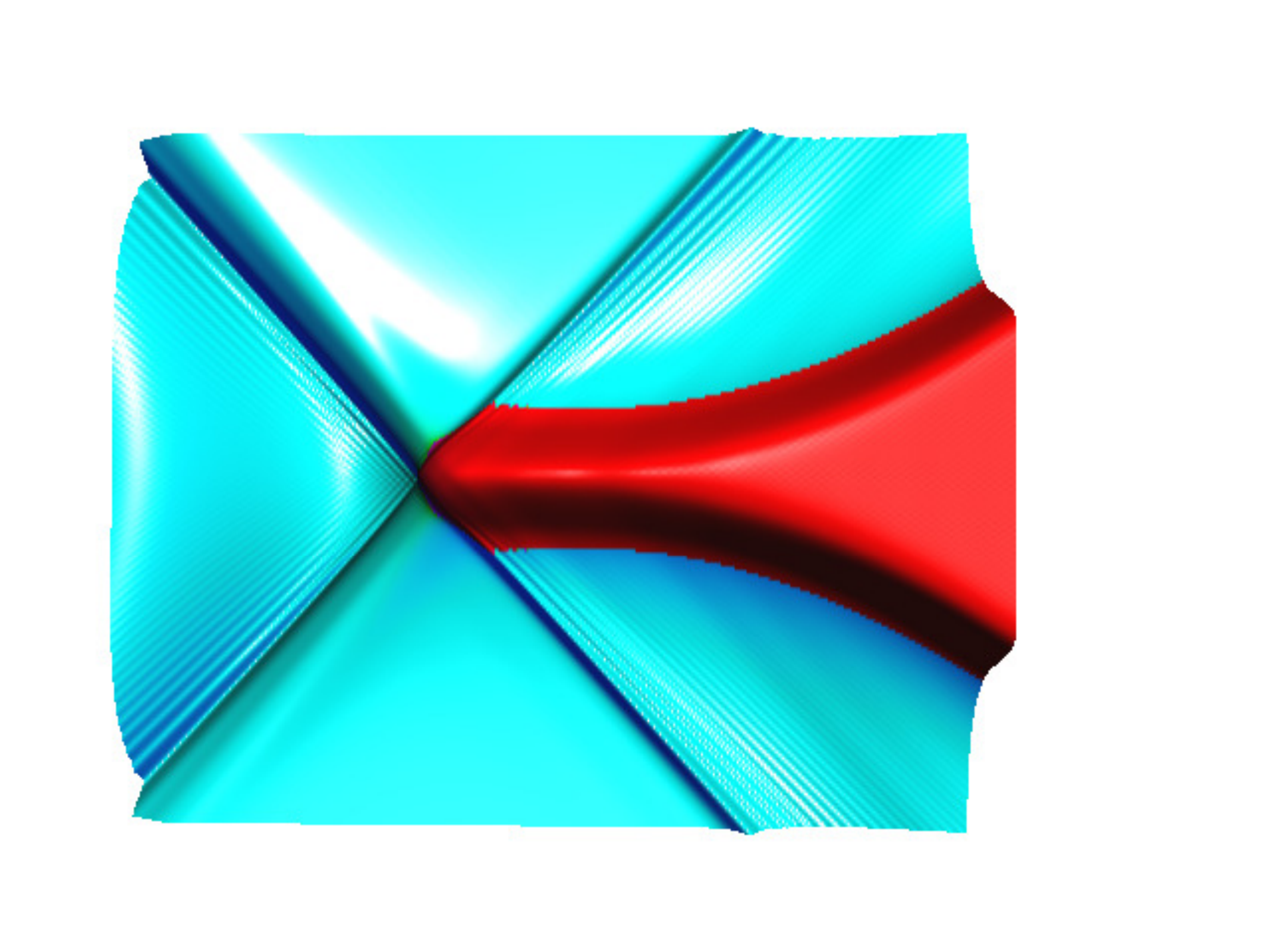}}
\put(21,2){${\rm Re}\, t - {\rm Im}\, t$}
\put(5.1,35.3){4}
\put(5.1,29.6){2}
\put(5.1,23.2){0}
\put(4.3,17){-2}
\put(4.3,11){-4}
\put(10,5.5){-6}
\put(18.4,5.5){-3}
\put(26.4,5.5){0}
\put(34.9,5.5){3}
\put(43.3,5.5){6}
\put(53.3,43.8){\includegraphics[width=42\unitlength,angle=-90]{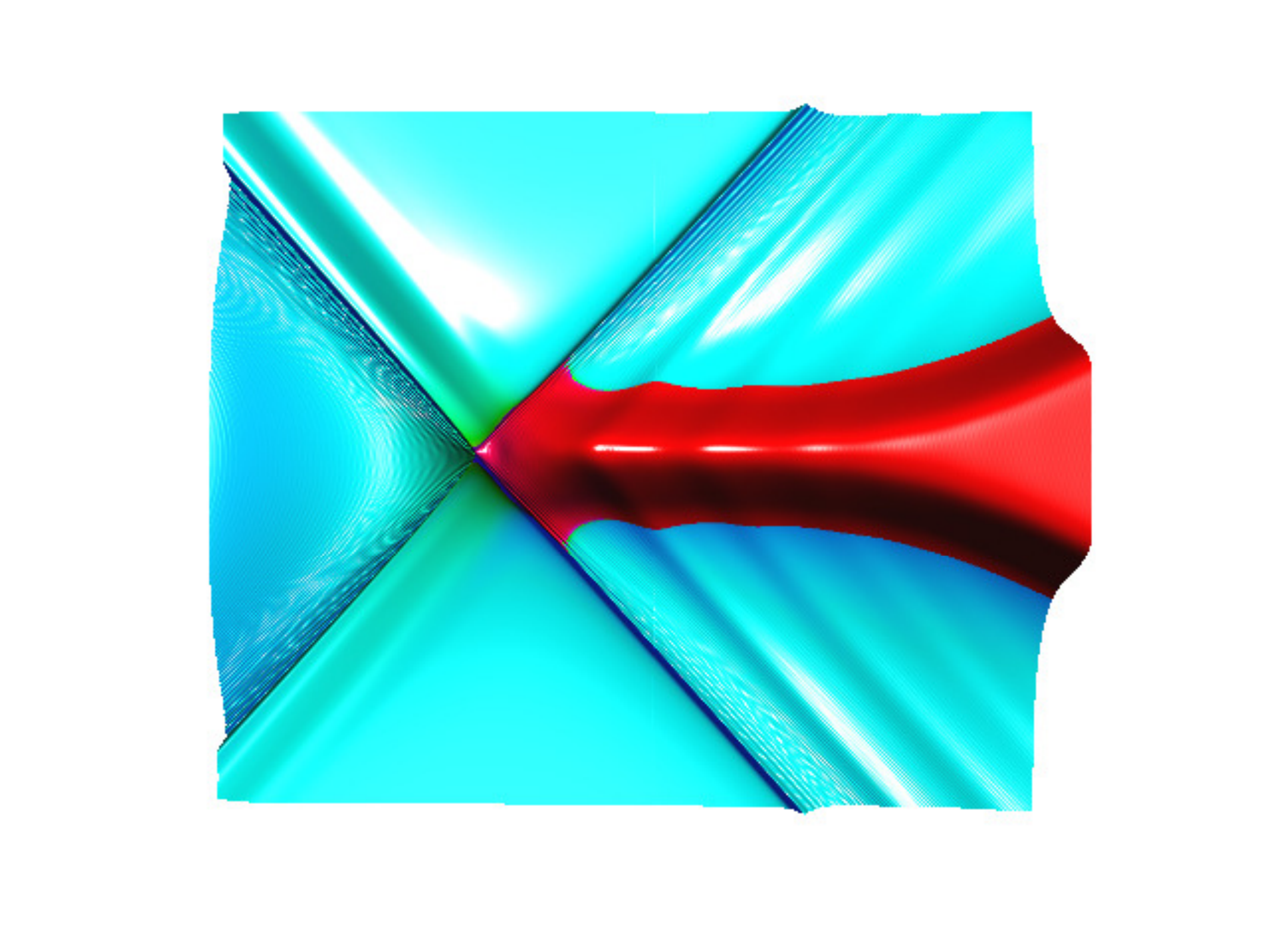}}
\put(60.7,35){4}
\put(60.7,29.6){2}
\put(60.7,23.4){0}
\put(59.9,17.1){-2}
\put(59.9,11.2){-4}
\put(65.5,5.5){-6}
\put(73.7,5.5){-3}
\put(81.9,5.5){0}
\put(90.2,5.5){3}
\put(98.2,5.5){6}
\put(77,2){${\rm Re}\, t - {\rm Im}\, t$}
\put(49.5,10){\includegraphics[width=7\unitlength,angle=-90]{pie.pdf}}
\put(50.5,1){$\mathrm{arg}\, \phi_s$}
\put(47.4,5.6){$\pi$}
\put(57,5.5){0}
\put(79.5,41){(b)}
\put(24,41){(a)}

\end{picture}
\caption{\label{above_sphaleron_solutions} Solutions at $E>E_{cb}$,
  $\epsilon = 5\cdot 10^{-4}$ and $\delta \rho = 0.4$. Their
  parameters are indicated by circles with letters in
  Fig.~\ref{EN1}. Namely, the solution (a) has~$(T,\, \theta) = (1.8\cdot
  10^{-3},\, 0.1)$ and $g^2 (E,\, N) \approx (8.0,\, 3.8)$, in the
  case (b)~$(T,\, \theta) = (8\cdot 10^{-3},\, 0.6)$ and $g^2(E,\, N)
  \approx (10.0,\, 2.1)$.} 
\end{figure}
\begin{figure}[h!]


\centerline{\includegraphics[width=0.47\textwidth]{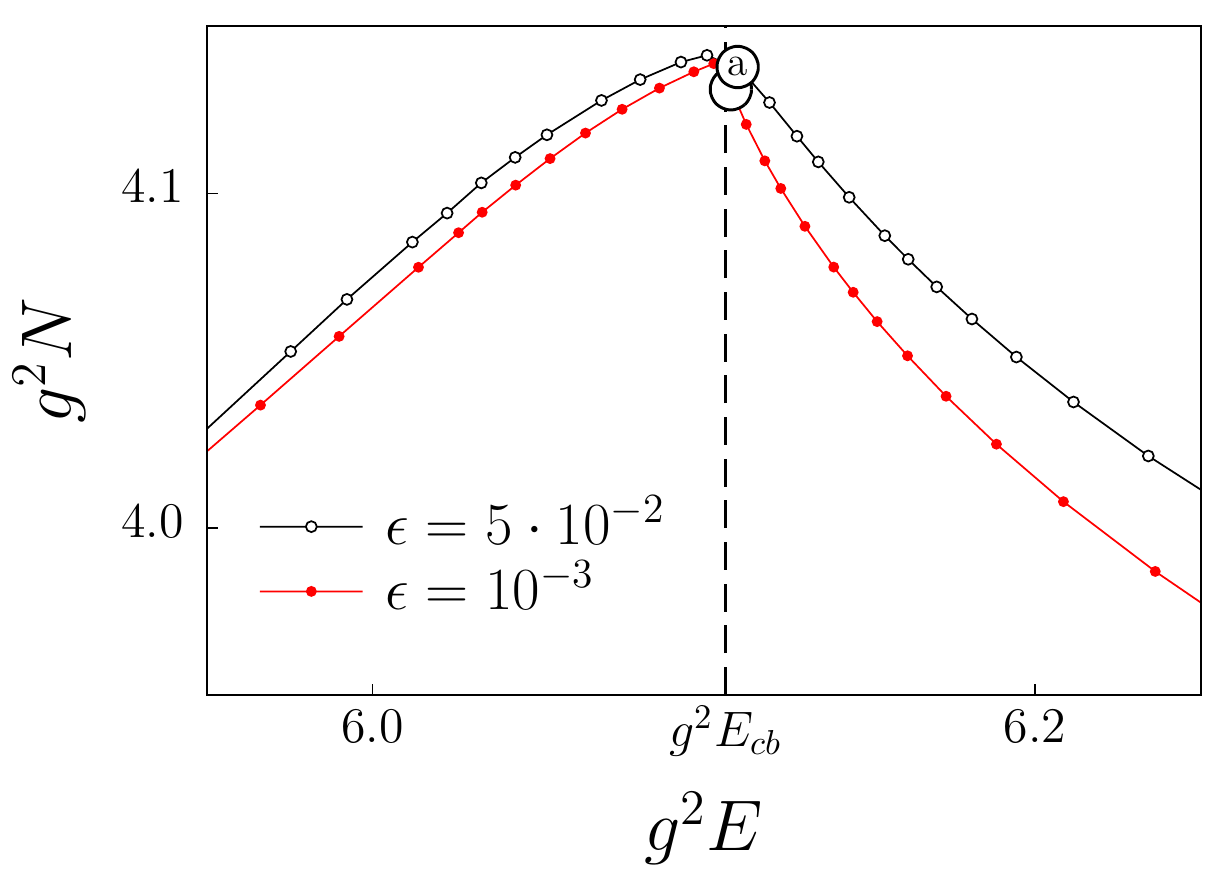}\hspace{3mm}
\includegraphics[width=0.47\textwidth]{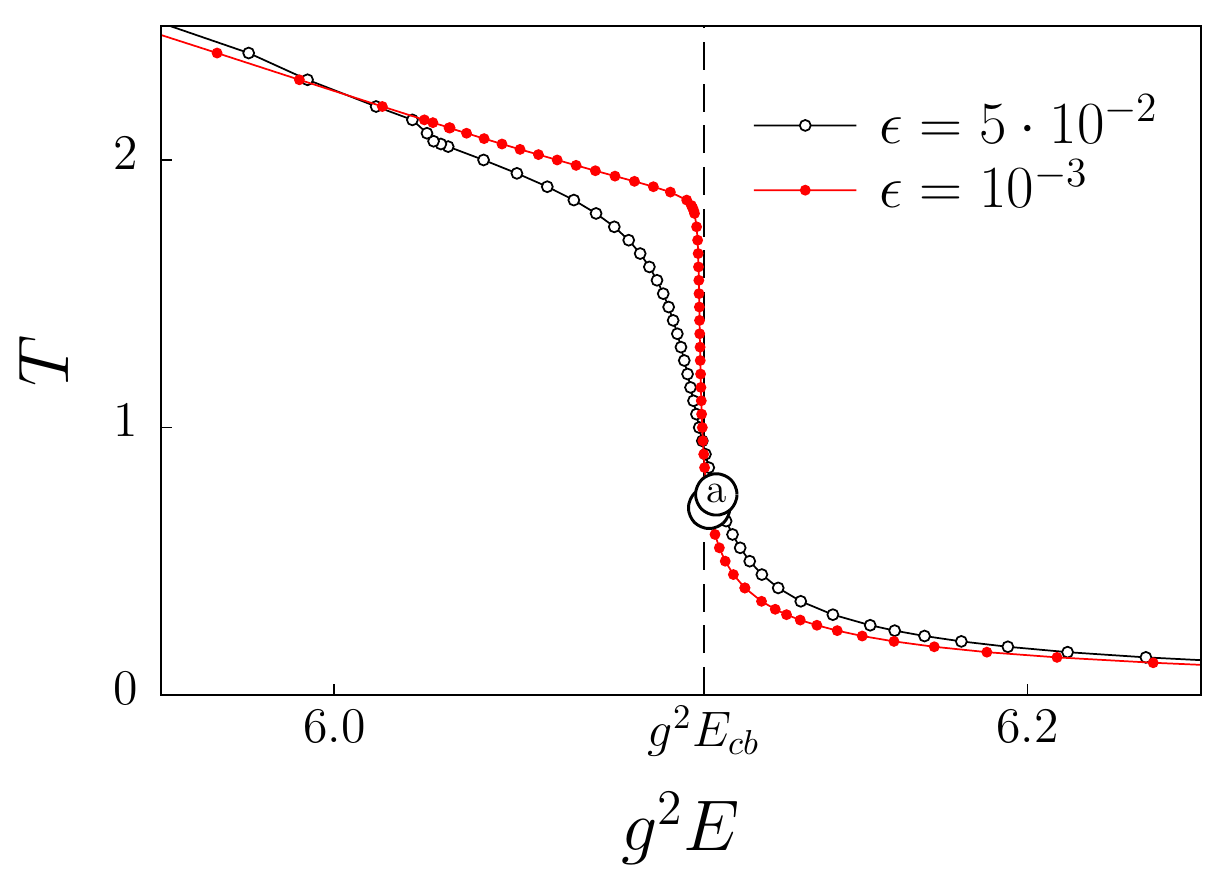}}
\caption{\label{bif_transition_1}Transition to the high--energy region
  $E>E_{cb}$ at $\epsilon>0$. The graphs show the parameters $N$ (left
  panel) and $T$ (right panel) of the modified solutions as functions
  of  energy $E$ along the line $\theta = 0.3$. Large circles mark 
  the solutions in Figs.~\ref{fig:boundary_solutions}a,b.}
\end{figure}
Figure~\ref{above_sphaleron_solutions} displays two solutions at
$E>E_{cb}$; their quantum numbers $(E,\, N)$ are marked  by
circles with letters in Fig.~\ref{EN1}. The initial wave
packets in these solutions are composed of high--frequency modes, their
true vacuum bubbles are small, cf.\ Fig.~\ref{pi_solutions}. Besides,
the respective periods $T$ of Euclidean evolutions are short. This
allows us to use the faster numerical algorithm of
Appendix~\ref{sec:numerical-alorithm}. Recall that the 
solutions become singular in the limits of small $N$ and high $E$; for
our best lattice this bounds the accessible region to $N\gtrsim 1/g^2$ and
$E \lesssim 14/g^2$.

We stress that the transition to the region $E>E_{cb}$ is smooth at
$\epsilon > 0$. In particular, the parameters $N$ and $T$ of the
modified solutions change smoothly along the lines $\theta
= \mbox{const}$, see Fig.~\ref{bif_transition_1}. However, this crossover
becomes sharper at smaller $\epsilon$ suggesting continuous rather
than differentiable changes in the $\epsilon=0$ solutions across
$E_{cb}$. 

\begin{figure}[t!]
\hspace{4.4cm} (a) \hspace{7.4cm} (b)

\vspace{1mm}
\centerline{\begin{minipage}{.47\textwidth}\includegraphics[width=\textwidth]{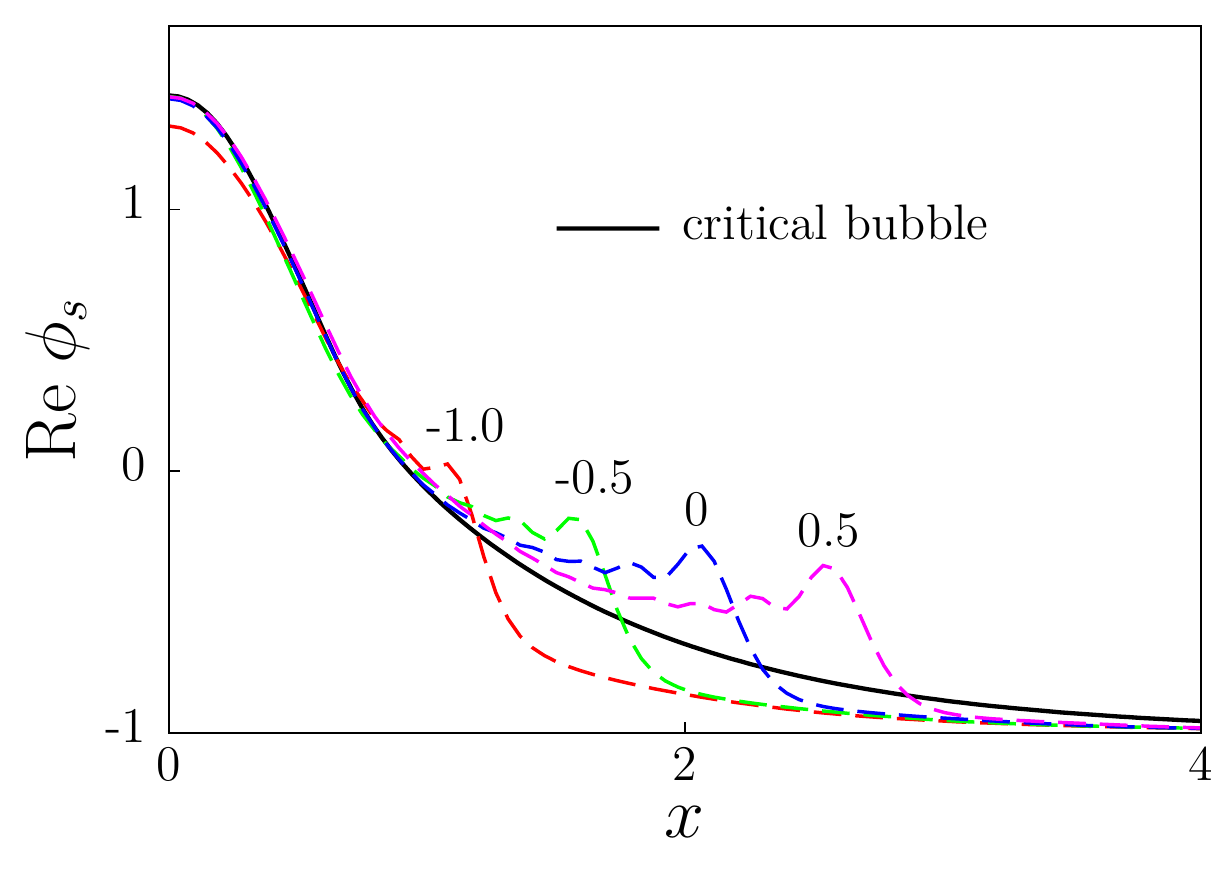}
\end{minipage}\hspace{5mm}\begin{minipage}{.47\textwidth}
\includegraphics[width=\textwidth]{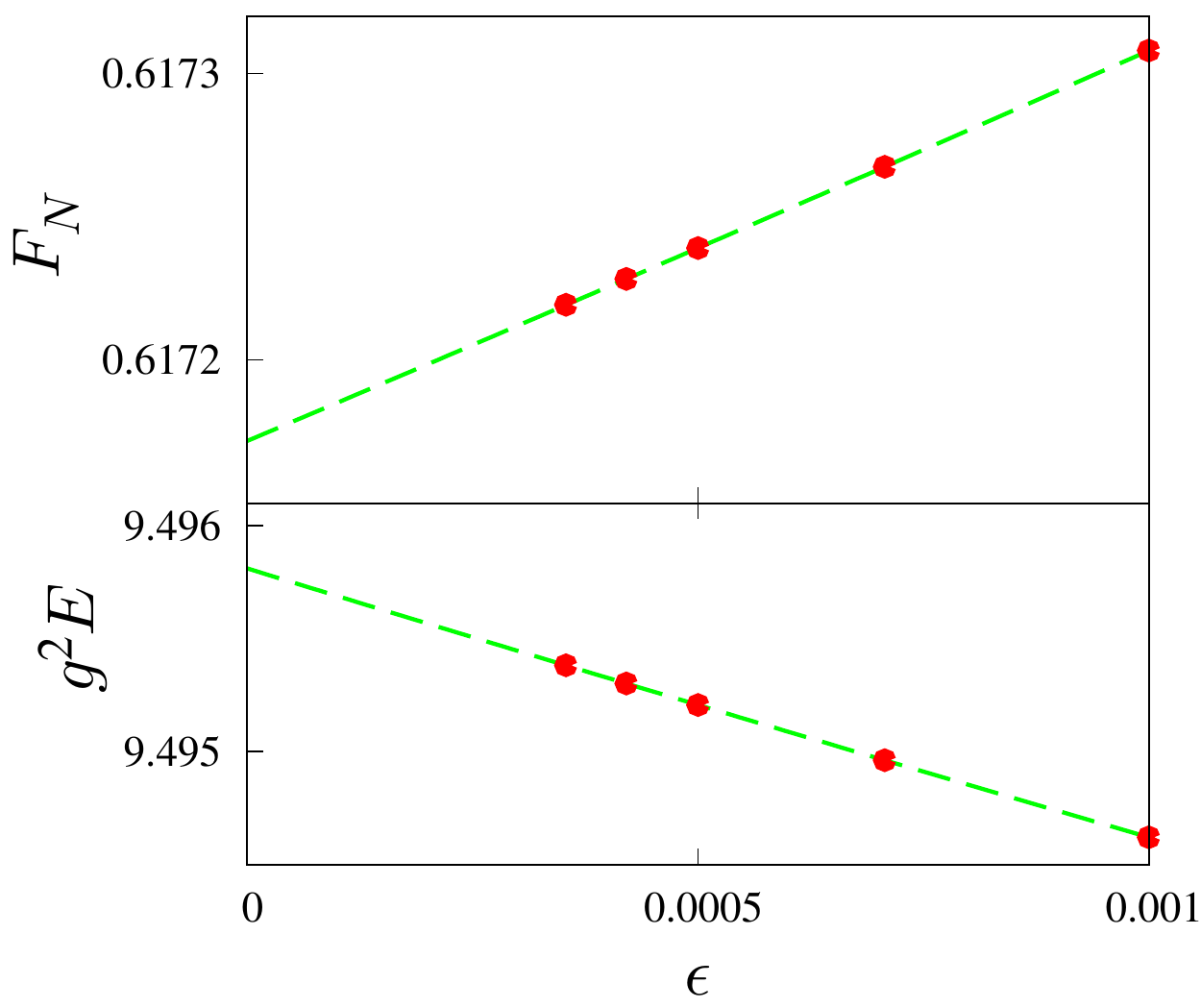}\end{minipage}}
\caption{\label{jump_to_sphaleron} (a) Fixed--time sections of the
  numerical solution $\mathrm{Re}\, \phi_s(t,\, x)$ with $E>E_{cb}$
  and infinitesimally small $\epsilon$. Numbers near the graphs are
  the values of $\mathrm{Re}\, t$. Solid line shows the critical
  bubble $\phi_{cb}(x)$. The parameters of the solution are $(T,\,
  \theta) = (10^{-3},\, 0.1)$ and $g^2(E,\, N) \approx (8.6,\, 3.8)$;
  $\delta \rho = 0.4$.   (b) The limit $\epsilon\to +0$ of the
  suppression exponent $F_N$ and energy $E$ at 
  $(T,\, \theta) = (0.01,\, 0.6)$, $g^2 (E,\, N) \approx (9.5,\, 2.1)$
  and $\delta \rho = 0.4$. Dashed lines are the quadratic fits of
  the data points.}
\end{figure}
\begin{figure}[h!]
\centerline{\includegraphics[width=0.5\textwidth]{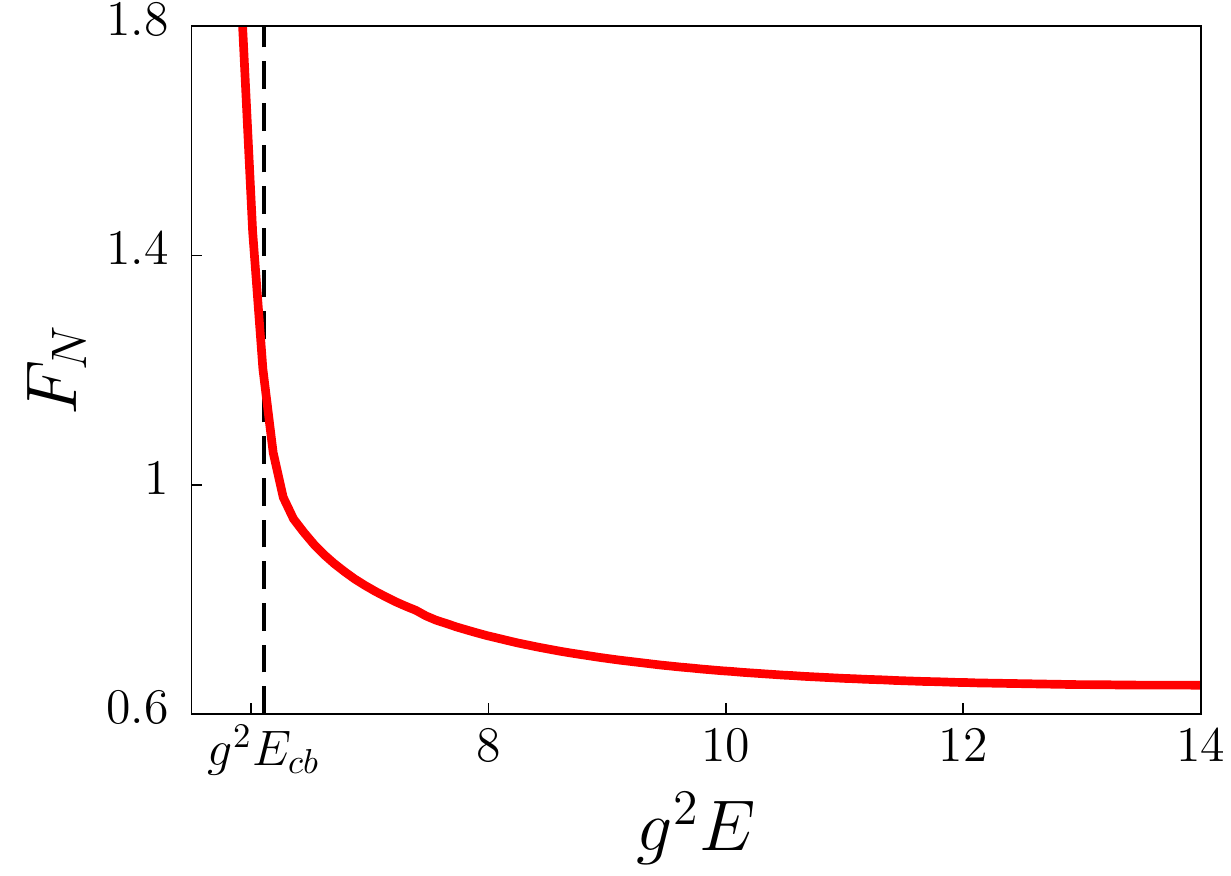}}
\caption{\label{Fgraph} The suppression exponent $F_N(E)$ of
  collision--induced false vacuum decay at high energies and $g^2 N = 
  2$; $\delta \rho = 0.4$. The graph is obtained in the limit
  $\epsilon \to +0$ using  the largest lattice $N_t\times
  N_x = 11000 \times 500$.}
\end{figure}

Now, consider the limit $\epsilon\to +0$ of the semiclassical
solutions. Figure \ref{fig:boundary_solutions} displays two 
solutions with $E>E_{cb}$ at different values of the regularization
parameter $\epsilon$. The true vacuum bubble in the solution with
smaller $\epsilon$ remains static for a notably longer period before it
starts to expand. This suggests that the original solution with
$\epsilon = 0$ contains a static finite--size bubble in the final
state ($t\to +\infty$). The latter property becomes apparent once we plot the $t =
\mbox{const}$ sections of the solution $\mathrm{Re} \,\phi_s(t,\, x)$
with infinitesimally small $\epsilon$, see
Fig.~\ref{jump_to_sphaleron}a, dashed lines. At large times this
solution approaches the critical bubble (solid line):
\begin{equation}
\label{eq:55}
\phi_s(t,\, x) \to \phi_{cb}(x) + \mbox{outgoing waves} \qquad\qquad \mbox{as}
\qquad  t\to +\infty\;.
\end{equation}
Importantly, we find that the asymptotics (\ref{eq:55}) holds
for all original ($\epsilon=0$) solutions above $E_{cb}$, not just the
ones at the boundary of this region. Thus, all
high--energy solutions are unstable and cannot be obtained by direct
numerical method. 

The unusual behavior of the solutions at $E>E_{cb}$ suggests that the
respective processes of false vacuum decay proceed in two
stages. First, the critical bubble is created with exponentially
small probability; the energy excess $E-E_{cb}$ is dropped in the form
of the outgoing waves. Second, the critical bubble, being classically
unstable, decays producing a growing bubble of true vacuum with
probability of order one. Similar tunneling 
mechanisms appear in multidimensional quantum mechanics~\cite{Onishi,
  Bezrukov:2003yf, epsilon, rst_confirmations2, tunneling_time} and
other models of field theory~\cite{rst, rst2}.

Finally, consider the limit $\epsilon \to +0$ of the quantum
numbers $(E,\, N)$ and exponent $F_N$ at fixed $T$ and
$\theta$. We find that these quantities are regular functions of
$\epsilon$: the data points in 
Fig.~\ref{jump_to_sphaleron}b are well fitted with the
quadratic polynomials (dashed lines). We quadratically extrapolate
them to $\epsilon = 0$ and obtain the exponent $F_N(E)$ of
false vacuum decay at high energies. The numerical errors
related to this extrapolation procedure are negligibly small. The
suppression exponent is plotted at $g^2 N=2$ and $\delta  \rho = 0.4$
in Fig.~\ref{Fgraph}. 

\subsection{Classical over--barrier transitions}
\label{sec:class-trans}
\begin{figure}[t]
\centerline{\includegraphics[width=0.6\columnwidth]{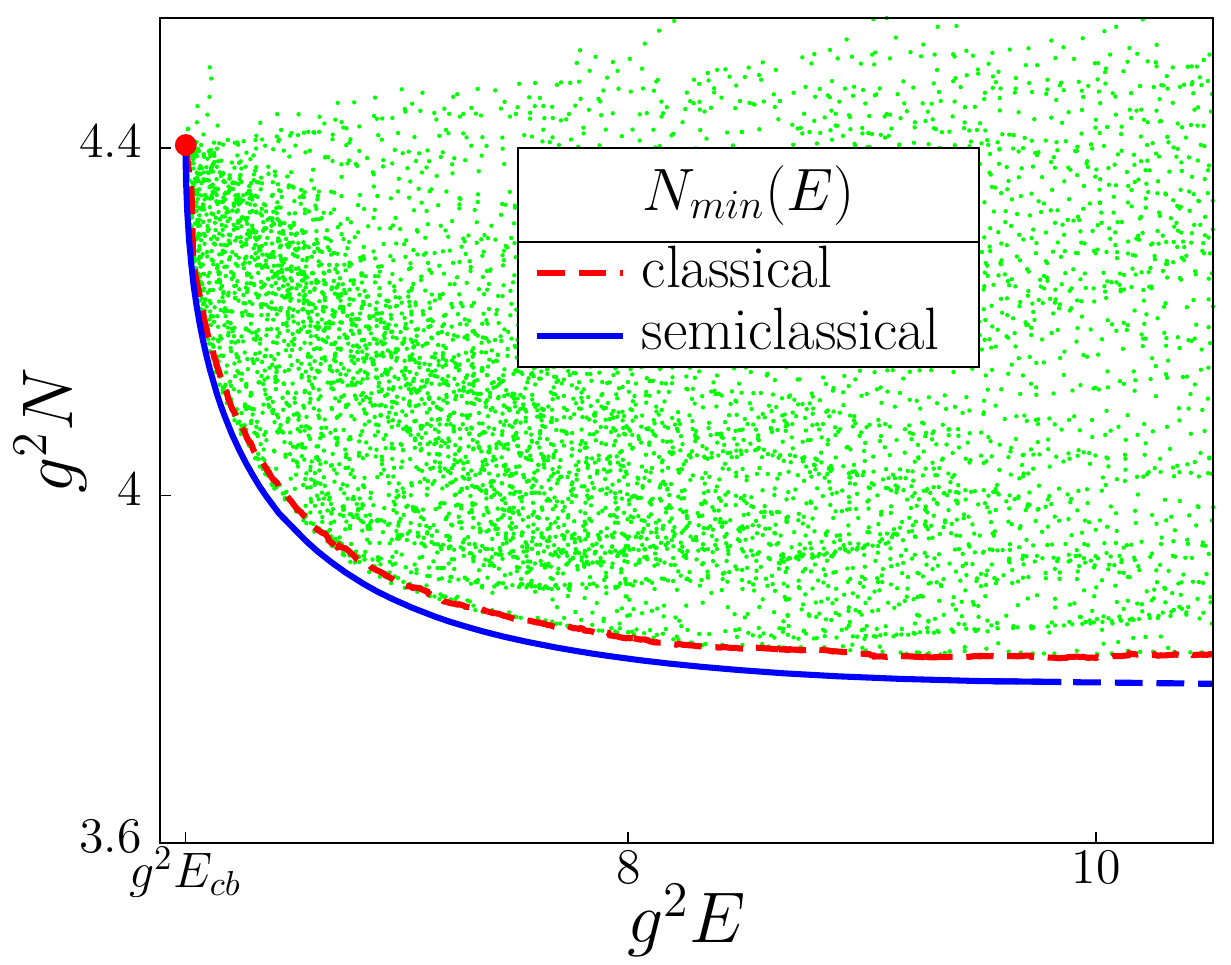}}
\caption{\label{Nmin} Parameters $(E,\, N)$ corresponding to the
  classically allowed transitions at 
  $\delta \rho = 0.4$ (dots) and their lower boundary $N_{min}(E)$
  (dashed line). The solid line is the same boundary extracted from the
  semiclassical results.}
\end{figure}
An important test of our semiclassical method 
involves classically allowed decay of the false vacuum filled with
many particles. These processes proceed with unsuppressed
probabilities, $F_N(E) = 0$, their initial quantum numbers
$(E>E_{cb},\, N)$ occupy the upper right corner in
Fig.~\ref{EN1}. Importantly, the minimal particle number $N =
N_{min}(E)$ required for such transitions can be obtained in two
ways: by studying the real classical solutions and by finding the
region $F_{N}(E) = 0$ from the classically forbidden side. Comparing
the two results, one checks the semiclassical method at
high  energies.
\begin{sloppy}

We studied the classically allowed decay of the multiparticle states
in the false vacuum  
in~\cite{Demidov:2011eu} using the stochastic sampling technique
of~\cite{Rebbi:1996zx}. Namely, we constructed many sets of random
Cauchy data $\{\phi (t_i,\, x),\,  \partial_t \phi(t_i,\, x)\}$ in the  
false vacuum and obtained a classical solution for each set. We
selected the solutions arriving to the true vacuum, i.e.\ those
containing expanding bubbles at $t\to +\infty$. Finally, we calculated
the initial quantum numbers $(E,\, N)$, Eqs.~(\ref{E_N}), for the 
selected solutions and indicated 
them with green/gray dots in Figs.~\ref{EN1} and~\ref{Nmin}. We 
arrived at the region in the $(E,\, N)$ plane corresponding to 
the classically allowed decay of the false vacuum; its lower boundary
(dashed line in Fig~\ref{Nmin}) gives the minimal initial multiplicity
$N_{min}(E)$ of classical over--barrier transitions.

\end{sloppy}
On the other hand, the function $N_{min}(E)$ can be obtained using the
semiclassical results of the previous Section. One notes that any real
solution to the classical field equations satisfies the boundary value problem
(\ref{eq:Ttheta}) at $T =  \theta = 0$ and gives $F_N(E) = 0$. Thus,
at $T,\, \theta \to +0$ the semiclassical solutions become real
and their quantum numbers $(E,\, N)$ approach the boundary $N =
N_{min}(E)$ of classical over--barrier transitions. Besides, since
$F_N(E)$ vanishes at the latter boundary, 
\begin{equation}
\label{eq:53}
\frac{dN_{min}}{dE} \equiv \left.\frac{dN}{dE}\right|_{F_N(E)=0}  = -
\lim_{T,\, \theta\to +0} \left[\frac{2T}{\theta}\right]\;,
\end{equation}
where the Legendre transforms~(\ref{eq:6}) were used in the last
equality. Relation (\ref{eq:53}) implies that the limit $T,\, \theta
\to 0$ should be performed at $\vartheta \equiv \theta/T = \mbox{const}$,
where $\vartheta$ parametrizes the curve $N_{min}(E)$. An exemplary
semiclassical solution with small $T$ and $\theta$ is plotted in
Fig.~\ref{above_sphaleron_solutions}a, its quantum numbers are marked 
by the circle in Fig.~\ref{EN1}. 

In Fig.~\ref{Nmin} we plot the boundary $N_{min}(E)$ extracted from
the semiclassical results (solid line). It almost coincides  with that
obtained from the real classical solutions. This supports our
semiclassical method in the high--energy  region~$E>E_{cb}$.

\subsection{Soliton--antisoliton production: $\delta \rho \to 0$}
\label{sec:solit-antis-prod}
\begin{figure}[t]
\centerline{
\unitlength=0.0088\columnwidth
\begin{picture}(108,43)(3,2)
\put(-0.1,46.7){\includegraphics[width=45\unitlength,angle=-90]{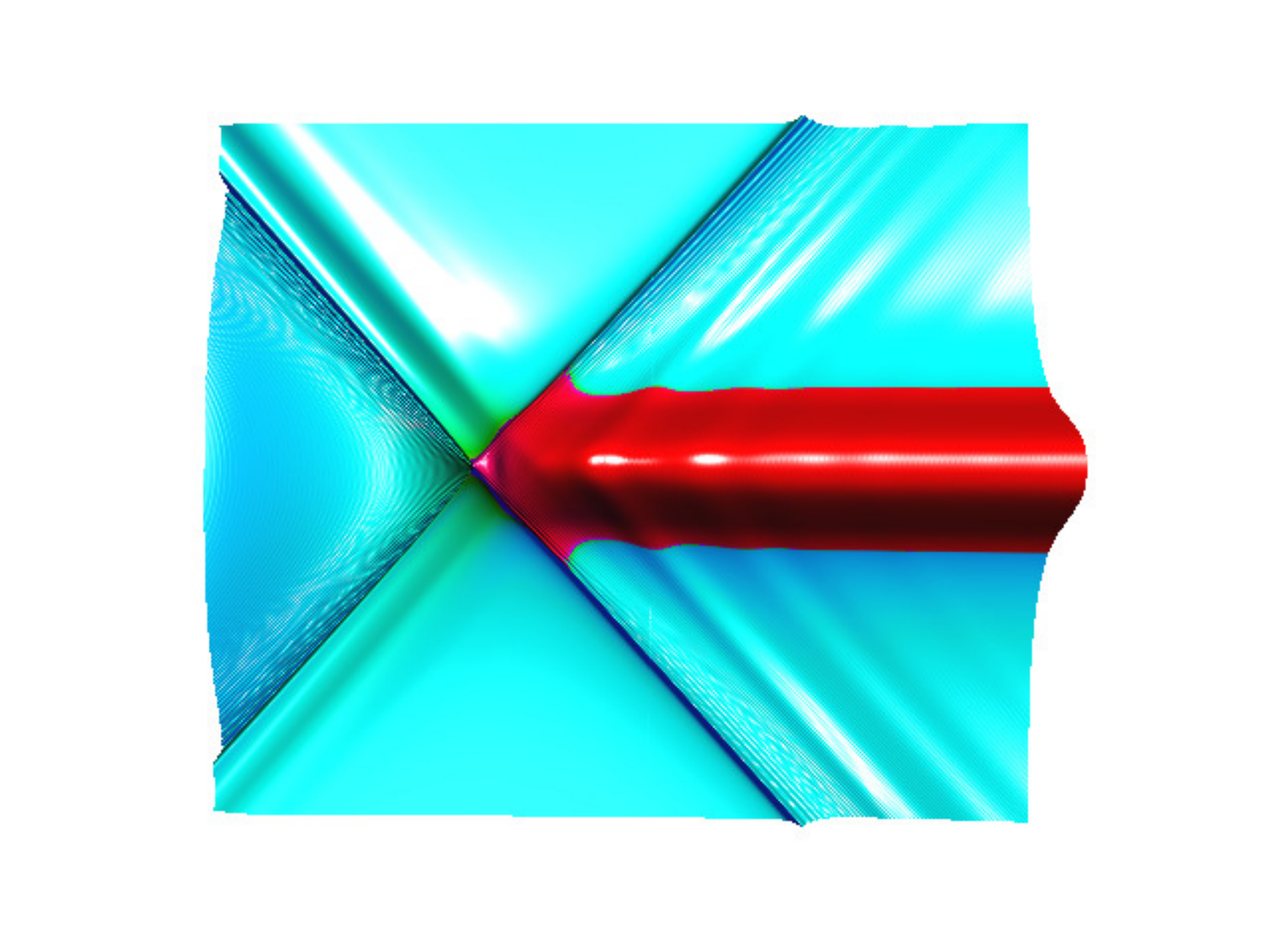}}
\put(6.7,36.7){4}
\put(6.7,30.8){2}
\put(6.7,24.2){0}
\put(3,22.1){\begin{sideways}$x$\end{sideways}}
\put(5.8,17.4){-2}
\put(5.8,11.1){-4}
\put(12.5,5.5){-6}
\put(18.5,5.5){-4}
\put(24.5,5.5){-2}
\put(30.2,5.5){0}
\put(36.1,5.5){2}
\put(42.2,5.5){4}
\put(48,5.5){6}
\put(25.4,2){${\rm Re}\, t - {\rm Im}\,t$}
\put(52,45){\includegraphics[width=0.07\columnwidth,angle=-90]{pie.pdf}}
\put(53,35){$\mathrm{arg}\, \phi_s$}
\put(49.8,40.4){$\pi$}
\put(60.5,40.3){0}
\put(57,1.3){\includegraphics[width=57\unitlength]{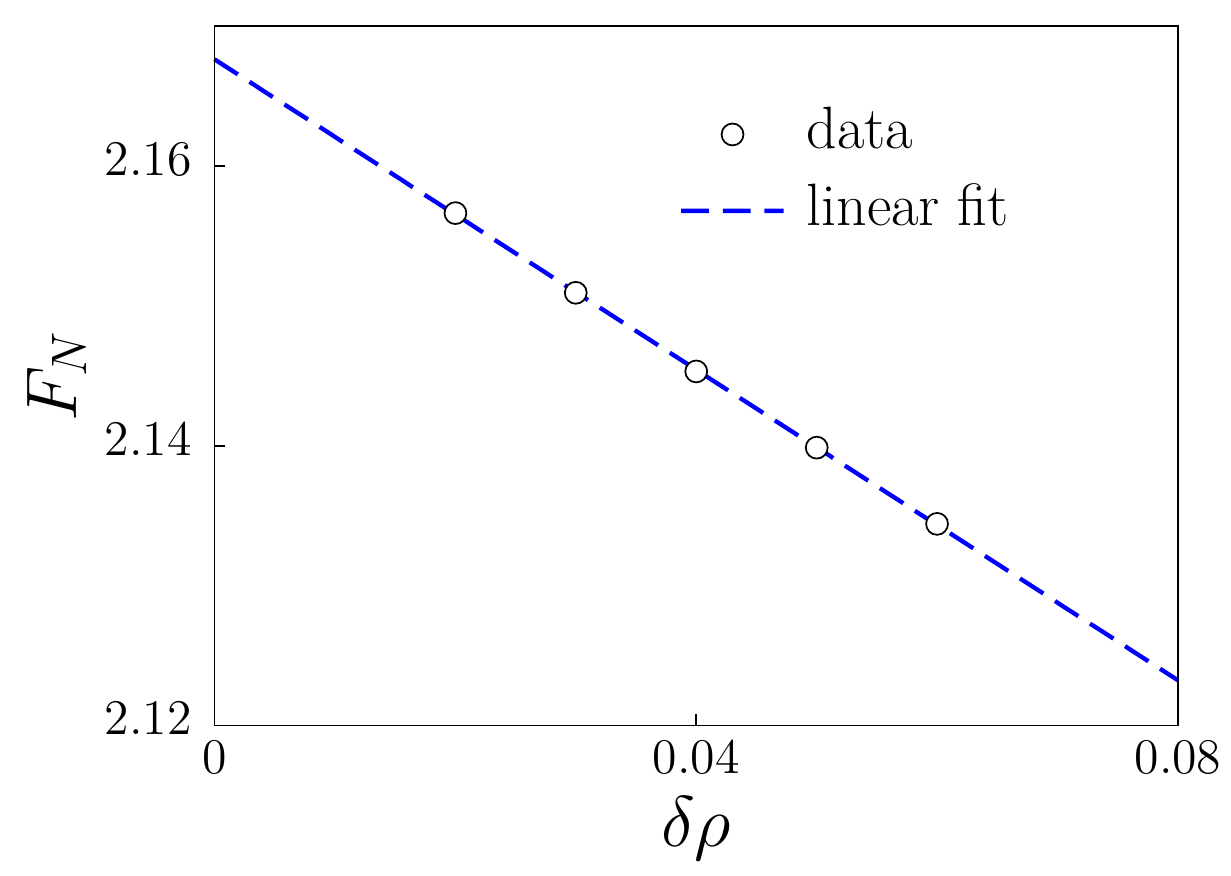}}
\put(27.5,42){(a)}
\put(86,42){(b)}
\end{picture}}
\caption{\label{drhoto0} (a) The semiclassical solution at $\delta \rho =
  0.03$; its parameters $(E,\, N)$ are the same as in
  Fig.~\ref{above_sphaleron_solutions}b. (b) Linear extrapolation of the    
  suppression exponent to $\delta \rho = 0$ at ${g^2 (E,\, N) = (9,\,
  2)}$.}
\end{figure}
So far we have considered the collision--induced decay of the false
vacuum. In this Section we send ${\delta \rho \to +0}$ and extract the
exponent $F_N(E)$ of the soliton--antisoliton pair production in
particle  collisions.

Keeping $E>E_{cb}$ and $\epsilon>0$, we decrease the energy density of
the true vacuum to ${\delta\rho \sim 0.02 \div 0.06}$. We arrive at 
the semiclassical solutions exemplified in
Fig.~\ref{drhoto0}a. Their true vacuum bubbles 
expand at small constant velocities without any apparent
acceleration. Sending, in addition,  $\epsilon \to +0$, we obtain the
solutions arriving at the static critical bubble
$\phi_{cb}(x)$. Recall that the spacial size of $\phi_{cb}$ is
logarithmically large at small~$\delta \rho$. 
\begin{figure}[t]
\centerline{\includegraphics[width=0.6\textwidth]{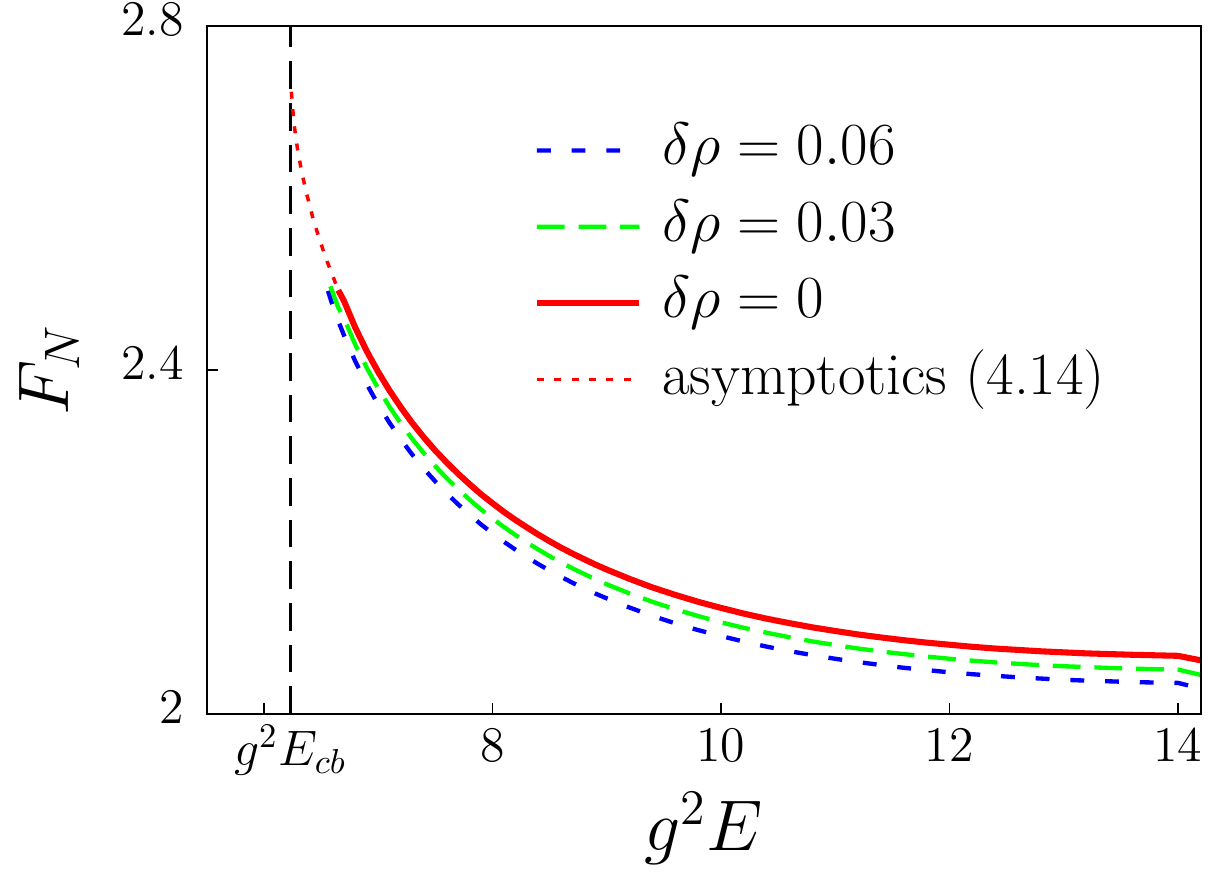}}
\caption{\label{Fgraphdrho0} Suppression exponent $F_N(E)$ at $g^2 N = 2$
  and different $\delta \rho$. Solid line is obtained by linearly
  extrapolating the data to $\delta \rho = 0$.}
\end{figure}

As expected, we find that the semiclassical exponent $F_N(E)$ is not
very sensitive to $\delta\rho$ in the high--energy region
$E>E_{cb}$. Moreover, at small $\delta \rho$ its dependence is linear,
see Fig.~\ref{drhoto0}b. This means that the singular term in the 
``thin--wall'' expansion (\ref{eq:61}) is absent and the limit $\delta
\rho \to 0$ of the exponent exists. Linearly extrapolating $F_N(E)$ to
$\delta \rho = 0$ at fixed $E$ and $N$ (dashed line in
Fig.~\ref{drhoto0}b), we obtain the exponent of soliton--antisoliton
production in particle collisions. 

In Fig.~\ref{Fgraphdrho0} we plot the semiclassical exponent at $g^2 N
= 2$: dashed lines represent the numerical data at different values of
\ $\delta \rho$, solid line is the result of extrapolation to 
$\delta \rho = 0$. The results for $F_N(E)$ at $\delta \rho = 0$ are
also shown in 
Fig.~\ref{EN_schema}b at different values of $g^2 N$. One observes
that the exponent decreases with  energy approaching constant at $E\gg 
2M_S$. In~\cite{Demidov:2015} we argued that this is the expected
behavior for the collision--induced tunneling at~$E\to +\infty$. 

Note that numerical extrapolation to $\delta
\rho = 0$ is harder to perform near the threshold $E_{cb}=2M_S$
because the asymptotic expansion of $F_N$ in $\delta \rho$ has 
different form at smaller energies. However, the thin--wall arguments
of Appendix~\ref{App:C} suggest near--threshold behavior
\begin{equation}
\label{eq:49}
F_{N}(E) \approx c_1(N) + c_2(N) |E - 2M_S|^{1/2} \qquad\qquad \mbox{at}
\qquad E \approx 2M_S\;,
\end{equation}
where $c_1$ and $c_2$ are unknown functions of $N$. We find that the
numerical results in Figs.~\ref{Fgraphdrho0} and~\ref{EN_schema}b are
consistent with the asymptotics~(\ref{eq:49}) (dotted lines in both
figures). 

\subsection{Two--particle processes}
\label{sec:two-part-proc}
Now, consider creation of the soliton--antisoliton pairs in the
two--particle collisions ($N=2$). These processes are natural from the
viewpoint of collider physics but cannot be described by direct
semiclassical methods. We compute their leading exponent $F_2(E)$
using the Rubakov--Son--Tinyakov conjecture~(\ref{eq:4}),
i.e.\ evaluating the limit $g^2 N \to 0$ of the  multiparticle
exponent $F_N(E)$. Recall that the semiclassical solutions develop a
singularity in this limit. Thus, we cannot address them directly. In
what follows we extrapolate the numerical results to $g^2 N \to 0$ using an
educated guess on the behavior of the multiparticle exponent.

\begin{figure}[ht!]
\hspace{4.3cm}(a) \hspace{7.5cm} (b)

\centerline{\includegraphics[width=0.47\textwidth]{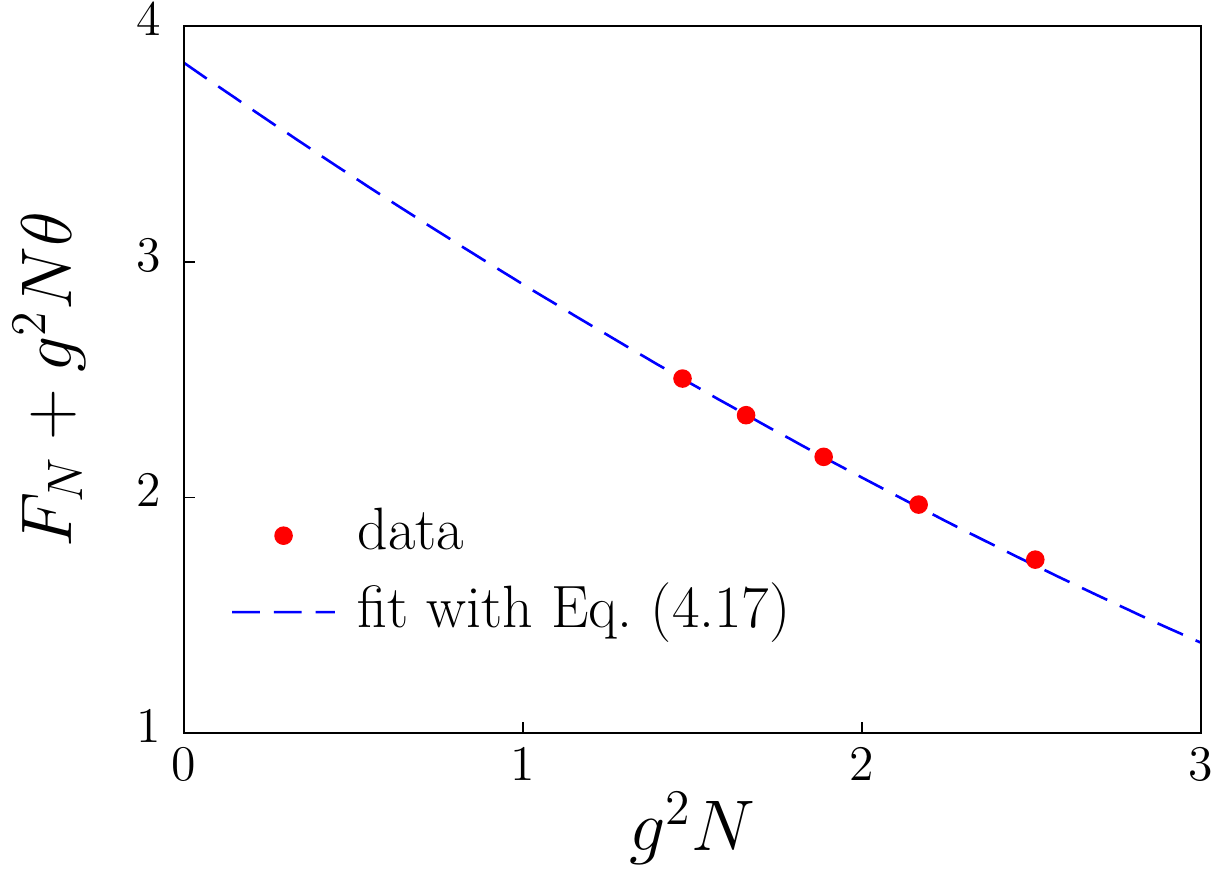}\hspace{3mm}
\includegraphics[width=0.47\textwidth]{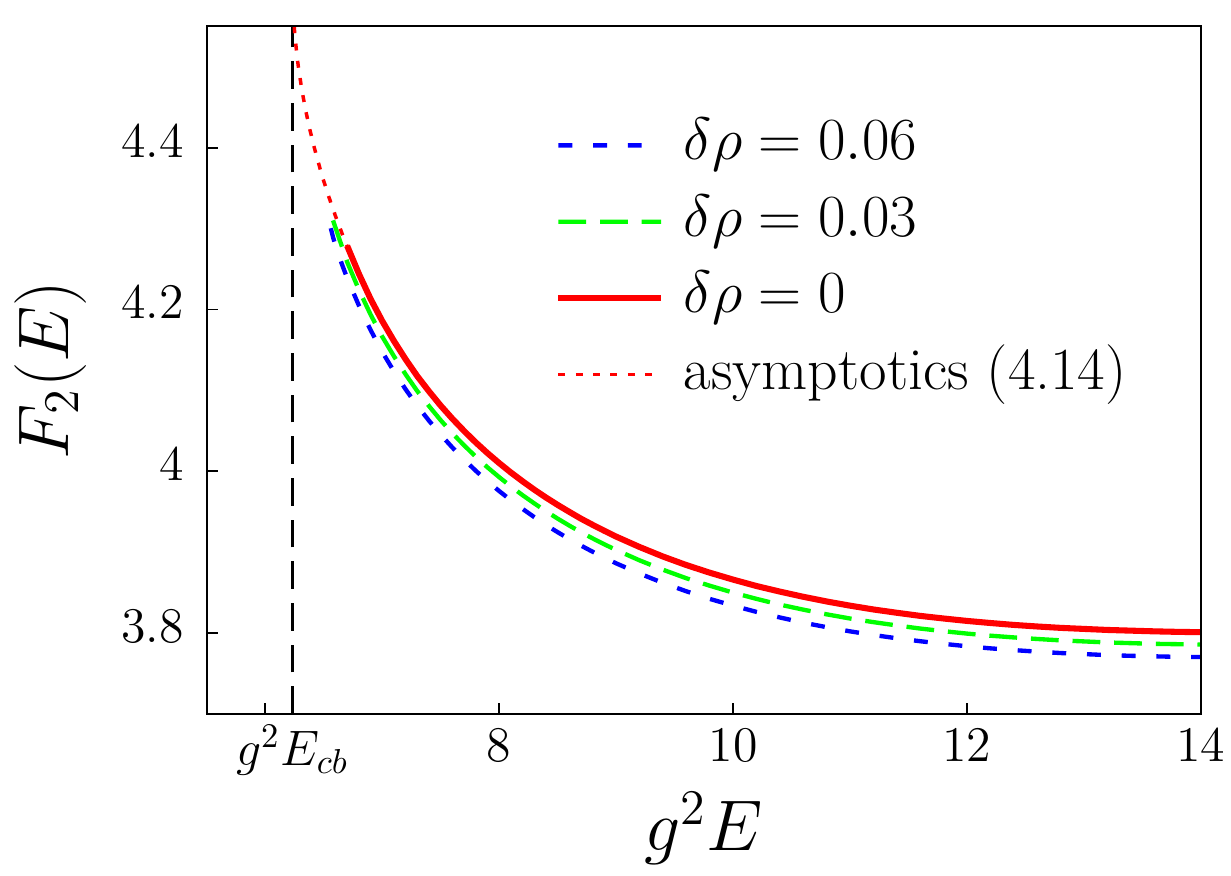}}
\caption{\label{N_extra} (a) Extrapolating the suppression exponent to
  $g^2 N = 0$ at $g^2 E = 10$ and $\delta \rho = 0.06$. (b) Two--particle exponent
  $F_2(E)$ at different $\delta \rho$ (dashed lines) and the result of
  its linear extrapolation to $\delta \rho = 0$ (solid line). } 
\end{figure}

One finds that vanishingly small initial particle number is achieved
at $\theta \to +\infty$: in this case the initial condition
(\ref{222}) reduces to the Feynman one, and the initial state of the
process becomes semiclassically indistinguishable from the
vacuum. Besides, since the combination $\mathrm{e}^{-\theta}$
enters the semiclassical equations, one expects regular
expansion
\begin{equation}
\label{eq:57}
g^2 N = n_1 \cdot \mathrm{e}^{-\theta} + n_2 \cdot
\mathrm{e}^{-2\theta} + \dots \qquad 
\Rightarrow\qquad \theta = - \log (g^2 N) + \theta_0 + \theta_1
\, (g^2 N) +  \dots\;, 
\end{equation}
where $n_i$ and $\theta_i$ are the energy--dependent Taylor
coefficients. Now, we integrate the Legendre transform (\ref{eq:6}) at
fixed energy, 
\begin{equation}
\label{eq:58}
F_{N}=F_2 - \int\limits_{0}^{N}g^2 \theta  \, dN = F_2 - g^2 N \theta
 - \int\limits_{\theta(N)}^{+\infty} g^2 N \, d\theta,
\end{equation}
where $F_N \to F_2$ and $g^2\theta N \to 0$ at $g^2 N \to 0$; besides, we
integrated by parts in the second equality. Substituting the expected
behavior (\ref{eq:57}) into Eq.~(\ref{eq:58}), we obtain the
asymptotic form 
\begin{equation}
\label{eq:59}
F_N(E) + g^2 N \theta \approx F_2(E) - g^2 N + \frac{\theta_1(E)}{2} \cdot (g^2
N)^2 + \dots \;,
\end{equation}
of the exponent at small $g^2N$. This behavior, if confirmed,
automatically implies that the limit $g^2 N \to 0$ of the
multiparticle exponent 
exists.

In Fig.~\ref{N_extra}a we compare the numerical data for the quantity $F_N
+ g^2 N\theta$ (points) with the expectation \eqref{eq:59} (dashed
line). One observes a consistent fit involving two parameters, $F_2$
and~$\theta_1$. Changing the number of data points in the fit,
we learn that the result for $F_2(E)$ is stable with relative
precision of order  1\%. 

The final numerical graphs of the two--particle exponent $F_2(E)$ are
shown in Fig.~\ref{N_extra}b: dashed and solid lines are obtained at
$\delta \rho > 0$ and $\delta \rho \to 0$, respectively. The last
graph corresponding to soliton--antisoliton production is repeated in 
Fig.~\ref{EN_schema}b.

\section{Concluding remarks}
\label{sec:6}
In this paper we developed a new semiclassical method to calculate the
probability of the topological soliton--antisoliton pair production in
particle collisions. Our main idea was previously reported
in~\cite{Demidov:2011dk}, here we presented the details, confirmations
and generalizations of the method.

In spite of all technicalities, the essence of our approach is simple:
it is based on expectation that the semiclassical solutions describing
classically forbidden transitions continuously depend on parameters of
the transitions.\footnote{In general the continuity may be 
  violated due to Stokes phenomenon~\cite{Berry, shudo_stokes, Levkov:2007a}. We 
  demonstrated, however, that the latter is not relevant for the
  processes considered in this paper.} With this idea in mind, we
introduced a small external field ${\cal E}$ coupled to the topological
charges of the soliton and  antisoliton. We started from the
well--known solutions describing spontaneous Schwinger creation of the  
soliton pairs in the field~${\cal E}$. Then we added colliding particles
to the initial state of the process and obtained the solutions for the
collision--induced Schwinger processes. Finally, we gradually
increased the collision energy $E$ above the threshold $2M_S$ of
soliton pair production and switched off the field, ${\cal E}
\to 0$. In this way we continuously arrived at the solutions describing
soliton--antisoliton production in particle collisions.

We illustrated the above semiclassical method with the explicit numerical
calculations in the ${(1+1)}$--dimensio\-nal 
model of a scalar field. In particular, we demonstrated that the above--mentioned 
semiclassical solutions continuously depend on the parameters $E$ and
${\delta \rho \propto |\mathbf{\cal E}|}$ of the induced Schwinger process. We
showed that these solutions reproduce the thin--wall
results at low energies and correctly describe classical over--barrier
transitions at $E>2M_S$ and high initial multiplicities. Finally, we
computed the semiclassical suppression exponents of the soliton--antisoliton pair
production in the $N$--particle and two--particle collisions, see
Fig.~\ref{EN_schema}b. 

We believe the semiclassical approach of this paper will be useful in
other models / setups of particle and condensed matter physics.

\paragraph*{Acknowledgements.}
We thank V.~Rubakov~\cite{Rubakov60} for
criticism and F.~Bezrukov,  D.~Gorbunov, M.~Liba\-nov, E.~Nugaev,
S.~Troitsky, S.~Sibiryakov for discussions. This work is supported by
the RSCF grant 14-22-00161. Numerical calculations have been performed
on the Computational cluster of the Theoretical division of INR RAS.
We thank G.~Rubtsov for supporting its stability. 

\appendix
\section{Deriving the equations}
\label{sec:derivation}
In this Appendix we review the semiclassical method
of~\cite{Rubakov:1992ec}, see also~\cite{Rubakov_Shaposhnikov,
  Rebbi:1996qt}. To this  
end we introduce path integral for the probability~(\ref{eq:3})
and evaluate it in the saddle--point approximation. We obtain 
equations for the saddle--point configuration $\phi_s(t,\, x)$ and
expression for the suppression exponent $F_N(E)$.

Semi--inclusive initial states in Eq.~(\ref{eq:3}) have fixed energy
$E$ and multiplicity $N$. The projector $\hat{P}_{E,N}$ on the 
subspace of these states simplifies~\cite{Rubakov:1992ec} in the
coherent basis~\cite{Faddeev_Slavnov}, 
\begin{equation}
\label{eq:8}
\langle a|\hat{P}_{E,\, N}|b\rangle = -\int_{i\infty}^{-i\infty} \frac{dT
  d\theta}{2\pi^2}\; \exp\left\{2ET + N\theta + \int dk\, \gamma_k
\,a_k^*b_k \right\}\;, \qquad\gamma_k = {\rm e}^{-2\omega_k T -
  \theta}\;.
\end{equation}
where $|a\rangle$, $|b\rangle$ are the eigenstates of the
false--vacuum annihilation operators $\hat{a}_k$ with
eigenvalues $a_k$, $b_k$ and normalization $\langle a | b\rangle  =
\exp\left(\int dk \, a_k^* b_k\right)$. Integrals over $T$ and
$\theta$ run along the imaginary
axes. Expression (\ref{eq:8}) can be proven~\cite{Rebbi:1996qt} by
acting on the Fock states $\hat{a}_{k_1}^\dag \dots\hat{a}_{k_n}^\dag
|0\rangle$ in the coherent--state representation. We transform
Eq.~(\ref{eq:8}) into the configuration representation using 
\[
\langle \phi_i | a\rangle =
  \exp \left\{\frac12 \int dk \left(-a_k a_{-k} -
    \omega_k \phi_i (k)\phi_i(-k)/g^2 +
    2\sqrt{2\omega_k} \, a_k \phi_i(k)/g  \right)\right\}\;,
\]
where $\phi_i(k)$ stands for the spatial Fourier transform of
$\phi_i(x)$. Here and below the prefactors are omitted.
We obtain the matrix elements
\begin{equation}
\label{eq:11}
\langle \phi_i| \hat{P}_{E,\, N} | \phi_i'\rangle =
\int_{i\infty}^{-i\infty} dT d\theta \; \mathrm{e}^{2ET +
  N\theta + B[\phi_i,\, \phi'_i]}\;;
\end{equation}
denoting with
\begin{equation}
\label{eq:10}
B =  \int \, \frac{\omega_k \, dk}{2g^2 (\gamma_k^2 - 1)}\left[(\gamma_k^2+1)
   \left(\phi_i(k)\phi_i(-k) + \phi_i^{\prime}(k)\phi_i^{\prime}(-k)\right) -
   4\gamma_k \phi_i(k)\phi_i^{\prime}(-k)\right]\;,
\end{equation}
the quadratic functional of $\phi_i$ and $\phi_i'$.

Next, we write the probability~(\ref{eq:3}) in the path integral form, 
\begin{align}
\notag
{\cal P}_N & = \int {\cal D} \phi_i \; {\cal D} \phi'_i\; {\cal D}
\phi_f \;\langle \phi_f | \hat{U}(t_f,\, t_i)|\phi_i \rangle \; \langle
 \phi_i |\hat{P}_{E,\, N} | \phi'_i \rangle \; \langle \phi'_i |
\hat{U}^\dag(t_f,\, t_i)| \phi_f \rangle \\ 
\label{eq:9}
& = \int_{i\infty}^{-i\infty} dT d\theta\int {\cal D} \phi_i\,
      {\cal D} \phi_i'\, {\cal D} \phi_f \int {\cal 
  D}\phi\big|_{\phi_i}^{\phi_f} \; {\cal D}\phi'
\big|_{\phi_i'}^{\phi_f} \; \mathrm{e}^{iS[\phi] - iS[\phi'] +
  2ET + N\theta + B[\phi_i,\, \phi'_i]}\;.
\end{align}
In the first line of Eq.~(\ref{eq:9}) we integrate over all initial
and final--state configurations $\phi_i$, $\phi_i'$ and $\phi_f$
projecting onto the relevant subspace of initial states with 
$\hat{P}_{E,\, N}$. In the second line the path integrals for the
evolution operators $\hat{U}$ and $\hat{U}^\dag$ were introduced. We
assume that the configurations $\phi$ and $\phi'$ describe false vacuum decay:
they are close to the false vacuum at   $t=t_i \to   -\infty$ and
contain expanding bubbles as $t = t_f \to +\infty$. 

The integral (\ref{eq:9}) can be evaluated in the saddle--point
approximation at large $S$ and $B$ i.e.\ at $g^2 \ll 1$, see
Eqs.~(\ref{eq:1}) and (\ref{eq:10}). In this case the dominant
contribution to the integral comes from the vicinity of the saddle--point
configuration $\{\phi_s(t,\, x),\; \phi'_s(t,\, x),\; T,\; \theta\}$
which extremizes the integrand in Eq.~(\ref{eq:9}). This means that 
the saddle--point fields $\phi_s$ and  $\phi'_s$
satisfy the classical field equations $\delta S/\delta \phi = \delta
S/\delta \phi' =  0$. Boundary conditions for the latter equations are
obtained by varying the 
integrand with respect to $\phi_i \equiv \phi(t_i,\,x)$, $\phi_i'$ and
$\phi_f$, 
\begin{align}
\label{eq:BC_1}
&i\dot{\phi}_i(k) + \omega_k\phi_i(k) = \gamma_k \left(i\dot{\phi}^{\prime}_i(k) +
  \omega_k\phi_i^{\prime}(k)\right)\;,\\
  \notag
&i\dot{\phi}_i'(k) - \omega_k\phi_i'(k) = \gamma_k\left(i\dot{\phi}_i(k) -
  \omega_k\phi_i(k)\right)\;,\\
  \label{eq:13}
& \dot{\phi}_f =  \dot{\phi}_f'\;, \qquad \phi_f = \phi_f'\;,
\end{align}
where the dots denote the time derivatives coming from the 
variations of the action, e.g.\ $\delta S/\delta \phi_f =
\dot{\phi}_f/g^2$. Finally, differentiation with respect to $T$ and
$\theta$ gives equations
\begin{equation}
\label{eq:15}
2E = -\frac{\partial B}{\partial T} \;, \qquad \qquad 
N = - \frac{\partial B}{\partial \theta} \;,
\end{equation}
where $B$ is defined in Eq.~(\ref{eq:10}).

Two observations greatly simplify the semiclassical equations. First,
$\phi'$ originates from the complex conjugate amplitude suggesting
\begin{equation}
\label{eq:14}
\phi_s'(t,x) = [\phi_s(t,\, x)]^*\;.
\end{equation}
This Ansatz is compatible with Eqs.~(\ref{eq:BC_1}) if the
saddle--point values of $T$ and $\theta$ are real, and we assume that
as well. Second, in the infinite past the solution $\phi_s(t,\,x)$
describes free waves~\eqref{111} in the false vacuum. Hence, 
\begin{equation}
\label{eq:5}
\phi_{s,\,i}(k) = \frac{1}{\sqrt{2\omega_k}} \left[ f_{|k|}
\,\mathrm{e}^{-i\omega_k (t_i-iT)} + g_{|k|}^* \,\mathrm{e}^{i\omega_k (t_i -
  iT)} \right]\;.
\end{equation}
At this point of calculation the initial time $t_i$ is real; it will
be continued to complex values in Sec.~\ref{sec:2}. Substituting 
Eqs.~(\ref{eq:14}),~(\ref{eq:5}), into Eqs.~(\ref{eq:13}),
(\ref{eq:BC_1}),    
(\ref{eq:15}), one obtains the boundary
conditions~\eqref{000}, \eqref{222} and expressions \eqref{E_N}
relating $(T,\, \theta)$ to $(E,\, N)$.

We finally substitute the saddle--point configuration into the
integrand of Eq.~(\ref{eq:9}) and obtain the leading suppression exponent 
\begin{equation}
\label{eq:7}
F_N(E) =  - g^2 (2ET+N\theta) 
 + 2g^2\mathrm{Im}\, S[\phi_s] - g^2B[\phi_{s,\,i},\, \phi_{s,\,i}^*]\;.
\end{equation}
Rewriting the last term with help of Eq.~(\ref{eq:5})  and the boundary
condition~\eqref{222}, we arrive at the standard
expression~\eqref{s_exp}. 

We finish derivation with two remarks. First, the solution $\phi_s$
and all equations will be considered along the complex--time contour
in Fig.~\ref{fig3} corresponding to tunneling. In this case the
initial time $t_i$ is taken complex because the exponent (\ref{s_exp})
does not depend on it anyway. Second, the functional
\eqref{eq:7} depends on $E$, $N$ explicitly  and via the saddle--point
configuration $\{\phi_s,\, T,\,  \theta\}$. However, since we have
already extremized this functional with respect to $\phi$, $T$ and
$\theta$, its $E$ and $N$--derivatives are given by
Eqs.~(\ref{eq:6}).

\section{Lattice formulation}
\label{App:B}
In this Appendix we list the discretized action and finite--difference
semiclassical equations~(\ref{eq:Ttheta}), see~\cite{kuznetsov_tinyakov, rst} for similar approaches.

Our lattice $\{t_j,\, x_i\}$ is defined in
Sec.~\ref{sec:lattice-system}. 
Performing the substitutions~(\ref{eq:16}) and (\ref{eq:17}) in the action~(\ref{eq:1}), we obtain,
\begin{equation}
\label{eq:28}
g^2 S =
\sum_{\substack{i=0 \\[.5ex] j=-1}}^{\substack{ N_t\\[.5ex] N_{x}-1}}
\Delta \bar{x}_{i}\, \frac{(\phi_{j+1,\, i}-\phi_{j,\, i})^2}{\Delta t_{j}}
-\sum_{\substack{i=0 \\[.5ex] j=0}}^{\substack{N_t+1 \\[.5ex] N_{x}-2}}
\Delta\bar{t}_j\, \frac{(\phi_{j,\, i+1}-\phi_{j,\, i})^2}{\Delta x}
 -2\sum_{\substack{i=0 \\[.5ex] j=0}}^{\substack{N_t+1 \\[.5ex] N_x-1}}
  \Delta \bar{t}_j \, \Delta \bar{x}_i\, V(\phi_{j,\, i}).
\end{equation}
Recall that the lattice field equations are the derivatives of this
expression at the internal points $0 \leq j \leq N_t$ of the time
contour. Assuming for simplicity $1 \leq i \leq N_x-2$, we obtain, 
\begin{equation}
\label{eq:36}
{\cal F}_{j,\, i} \equiv \frac{\phi_{j+1,\, i} - \phi_{j,\, i}}{\Delta t_j \Delta \bar{t}_{j}} - 
\frac{\phi_{j,\, i} - \phi_{j-1,\, i}}{\Delta t_{j-1}\Delta \bar{t}_{j}}
- \frac{\phi_{j,\, i+1} + \phi_{j,\, i-1} - 2 \phi_{j,\, i}}{\Delta
  x^2} + V' (\phi_{j,\, i}) = 0\;,
\end{equation}
where $V'$  is the derivative of the scalar potential.
Equations at the spatial boundaries, i.e.\ at $i=0$ and $i=N_x-1$, are
derived in similar manner.

We assume that the evolution is linear in the beginning of the
process. In this case the potential reduces to $V\approx m_-^2 (\phi -
\phi_-)^2/2$. Substituting it into the above action, we obtain
Eq.~(\ref{eq:20}) with discrete time and
\begin{equation}
\notag
H_{i,\, i'} = \left( m_-^2  + \frac{2}{\Delta x^2} \right)\delta_{i,\, i'}  
  - \frac{\delta_{i+1,\, i'} + \delta_{i,\, i'+1}}{
    \Delta x \sqrt{\Delta \bar{x}_i \Delta \bar{x}_{i'}} }\;.
\end{equation}
We directly compute the real eigenvectors $\xi_i^{(n)}$ and 
eigenvalues $\omega_n^2$ of this symmetric matrix by QR
decomposition. The coefficients $f_n$ and $g_n$ of the linear solution
(\ref{eq:21}) are then given by Eqs.~\eqref{eq:22}, where the time
derivative~\eqref{eq:19} simplifies,
\begin{equation}
\label{eq:18}
\sum_i \, \xi_i^{(n)} \, \partial_t \psi_i \big|_{t_{-1}} =
\sum_i \xi_i^{(n)} \left\{\frac{\psi_i(t_0) - \psi_i(t_{-1})}{\Delta t_{-1}} 
+ \Delta\bar{t}_{-1} \omega_n^2 \psi_{i}(t_{-1}) \right\}\;.
\end{equation}
Using explicit expressions for $f_n$ and $g_n$, we rewrite the initial
condition $f_n = \mathrm{e}^{-\theta} g_n$ in the matrix form,
\begin{align}
\notag
& \sum_i \xi_i^{(n)} \left\{ (1-\mathrm{e}^{-\theta}) \, \omega_n
  \,\mathrm{Re} \, \psi_i  - (1+\mathrm{e}^{-\theta})\,  
  \mathrm{Im}\, \partial_t \psi_i\right\}_{t_{-1}} = 0\;,\\
\notag
& \sum_i \xi_i^{(n)} \left\{ (1+\mathrm{e}^{-\theta}) \, \omega_n
  \,\mathrm{Im} \, \psi_i  + (1-\mathrm{e}^{-\theta})\,  
  \mathrm{Re}\, \partial_t \psi_i\right\}_{t_{-1}} = 0\;,
\end{align}
where the discrete operator $\partial_t$ is defined in
Eq.~(\ref{eq:18}). One reads off the matrices $M_R$ and $M_I$ in 
Eq.~(\ref{eq:23}) from the above equations. 

We directly discretize the nonlinear energy~\eqref{eq:25}, 
\begin{multline}
\label{eq:60}
g^2 E_{j+1/2} = \sum_{i=1}^{N_{x}-1}\Delta \bar{x}_{i}\, 
\frac{(\phi_{j+1,\, i}-\phi_{j,\, i})^2}{\Delta t_{j}^2}
\\+ \left\{\sum_{i=0}^{N_{x}-2}\frac{(\phi_{j,\, i+1} - 
  \phi_{j,\, i})^2}{2\Delta x}  + \sum_{i=0}^{N_x-1}\Delta \bar{x}_i\, V(\phi_{j,\, i})
+ (j\to j+1) \right\}\;.
\end{multline}
It conserves, i.e.\ does not depend on $j$ up to $O(\Delta t^2)$
numerical errors. Somewhat different discretization procedure is
natural for the energy of the linear system~(\ref{eq:20}) at $t =
t_{-1}$. We start from the expression
\begin{equation}
\label{eq:27}
g^2 E = \sum_{i=0}^{N_{x}-1} (\partial_t \psi_i)^2\Big|_{t_{-1}} +
\sum_{i,\, i'=0}^{N_{x}-1} \psi_i H_{i,\, i'} \psi_{i'}\Big|_{t_{-1}}
\end{equation}
in continuous time and discrete space.
Substituting the solution (\ref{eq:21}), we obtain the first of
Eqs.~(\ref{eq:26}). The respective formula for the initial multiplicity
is then easily deduced from Eqs.~(\ref{E_N}). 

We finish this Appendix with the lattice expression for the
suppression exponent\footnote{The last integral in this
  expression is $O(\Delta t)$ and $O(\Delta x^2)$ accurate. 
  In practice the first--order correction in $\Delta t$ is negligible
  because $|\Delta t| \ll \Delta x$. All other lattice expressions in
  our study
  are second--order both in space and time.}
\begin{equation}
\label{F_E_discrete}
F_N(E) =  - g^2 (2ET + N\theta)  + 2 g^2 {\rm Im}\; S + {\rm 
  Im}\sum_{i=0}^{N_x-1}\Delta \bar{x}_{i} \, (\phi_{-1,\,
  i}+\phi_{0,\, i} -
2\phi_-) \, 
\frac{\phi_{0,\, i}-\phi_{-1,\,i}}{\Delta t_{-1}}\;,
\end{equation}
where the action is given by Eq.~(\ref{eq:28}).

\section{Tests of the numerical procedure}
\label{App:D}
We subdued the lattice solutions to a number of consistency tests
which support the numerical methods of Sec.~\ref{sec:numerical-method} 
and allow us to estimate the numerical errors.

For a start, we checked sensitivity of the results to the spatial
cutoff $L_x$. To this end we increased the cutoff from $L_x=7$ to $14$
at a fixed lattice spacing $\Delta x = L_x/(N_x-1)$. The 
integral quantities $E$, $N$ and $F_N$ stayed
independent of $L_x$ up to relative errors of 
order~$10^{-4}$.

\begin{figure}[t]
\hspace{4cm} (a) \hspace{7.9cm}(b)

\vspace{1mm}
\centerline{\includegraphics[width=0.47\textwidth]{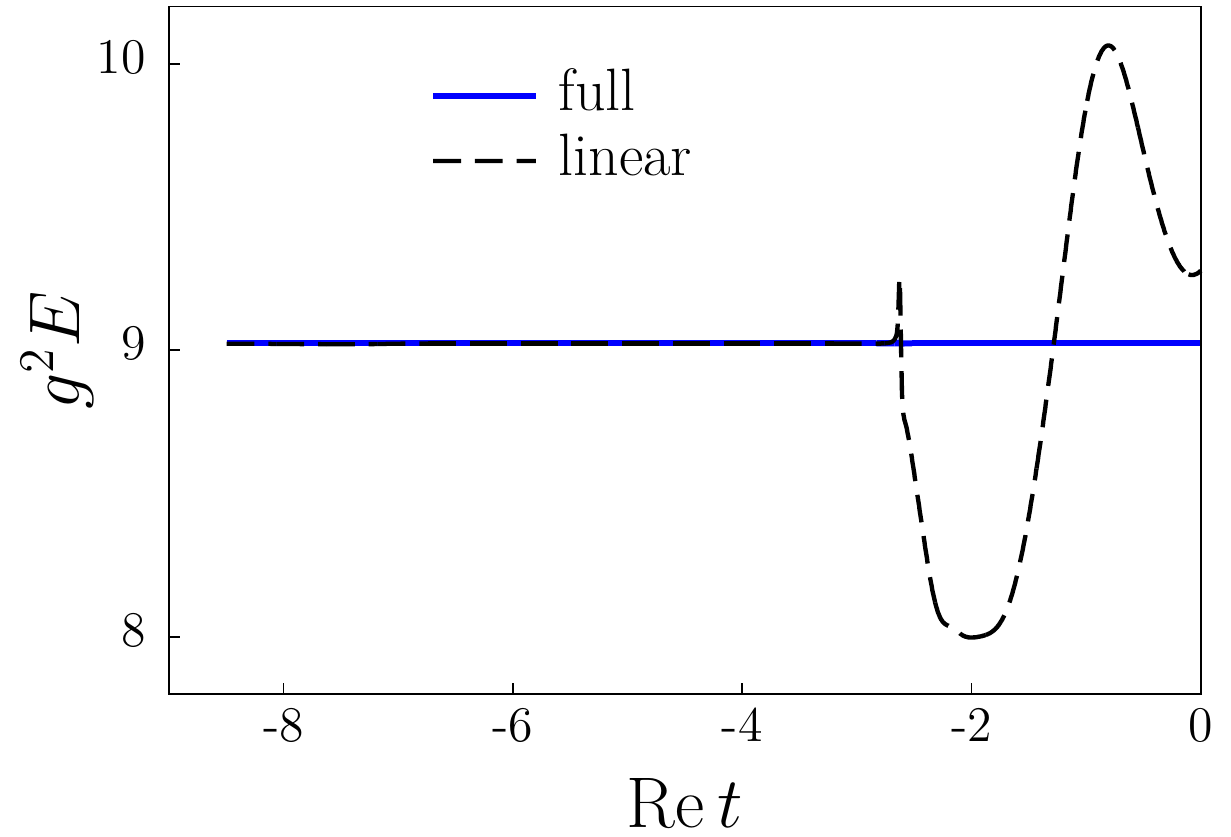}\hspace{5mm}
\includegraphics[width=0.46\textwidth]{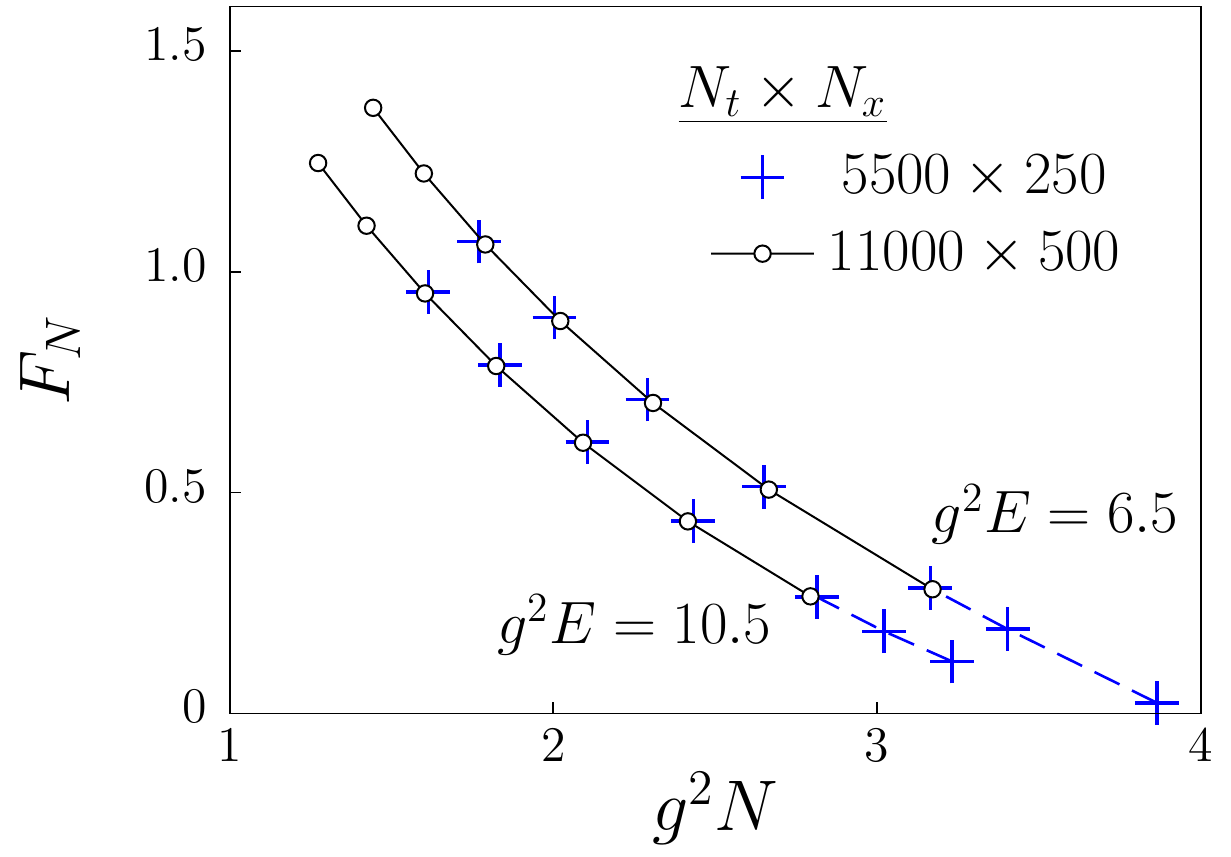}}
\caption{(a) Full and linear energies of the 
  semiclassical solution with $(E,\, N) \approx
  (9.0,\, 2.8)/g^2$ at $\mathrm{Re}\, t<0$. (b) The exponent $F_{N}$ as a function of $N$ at
  different lattice spacings. The space--time box in both figures is $L_x =
  7$, $\mathrm{Re}\, (t_{N_t+1} - t_{-1})  \approx 15$.\label{fig:linearE}}
\end{figure}

Another source of numerical errors is related to the finite extent 
$t_{-1} \leq t \leq t_{N_t+1}$ of the temporal lattice. Recall that
the semiclassical solutions should describe free waves in the false
vacuum at $t\sim t_{-1}$. This property is conceptually important, it
was used in the derivation of the initial condition (\ref{222}). We
estimate the effect of nonlinear interactions in the beginning of the
process by evaluating the energy of the linearized system given by
Eqs.~(\ref{eq:26}), (\ref{eq:22}) at different time sites $t_j$ and
comparing it with the full conserved energy~(\ref{eq:25}), see
Fig.~\ref{fig:linearE}a.  The graphs stay close at  
$t\sim t_{-1} \approx -8.5$, separating in the nonlinear region
$t\gtrsim -3$. The relative error due to nonlinear interactions in
the beginning of the process is estimated as
$|E_{\mathrm{full}}-E_{\mathrm{linear}}|/E_{\mathrm{full}} \lesssim
10^{-3}$. One concludes that the semiclassical evolution at $t\sim
t_{-1}$ is, indeed, free thanks to the clever choice of the scalar
potential in Sec.~\ref{sec:choosing-potential}.

\begin{figure}[t]
\hspace{5cm} (a) \hspace{6.8cm} (b)

\centerline{\includegraphics[width=0.45\textwidth]{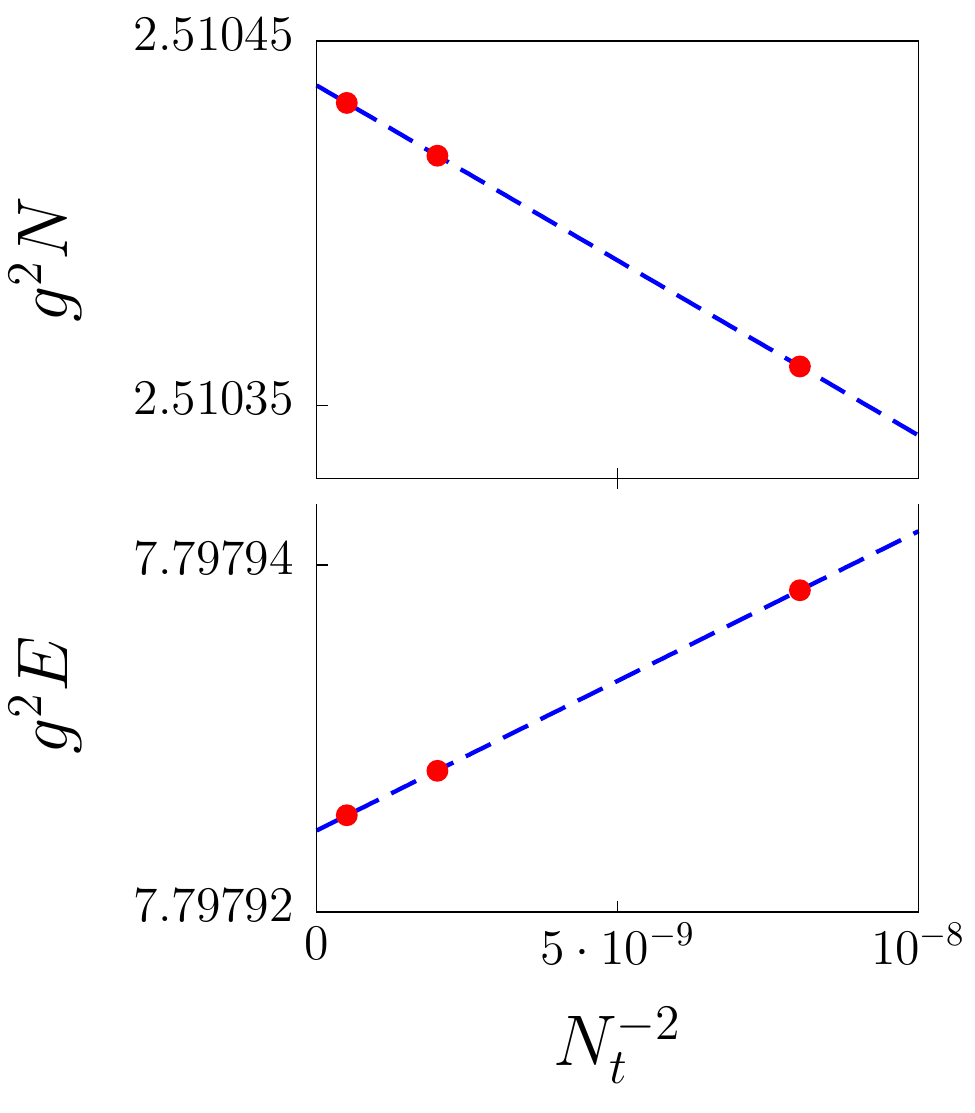}  
\includegraphics[width=0.45\textwidth]{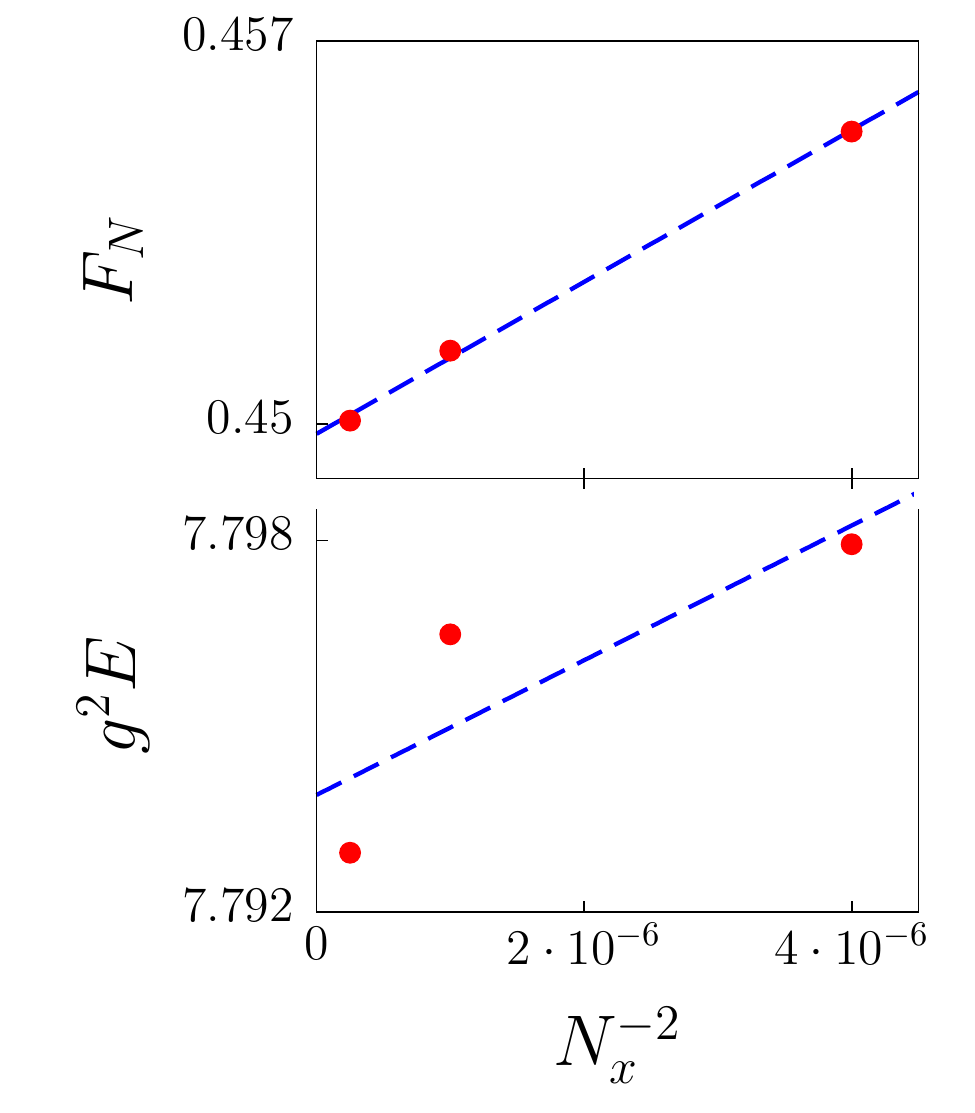}}
\caption{The integral quantities $E$, $N$ and $F_N$ as functions on
  the (a) time and (b)~spatial lattice spacings at $(T,\,
  \theta) = (0.026,\, 0.5)$ and $(E,\, N) \approx (7.8,\,
  2.5)/g^2$; $E>E_{cb}$. The dashed lines are the linear fits of the
  data points. We use the space--time box from Fig.~\ref{fig:linearE}.\label{fig:dxdt}}  
\end{figure}

\begin{figure}[ht!]
\centerline{\includegraphics[width=0.5\textwidth]{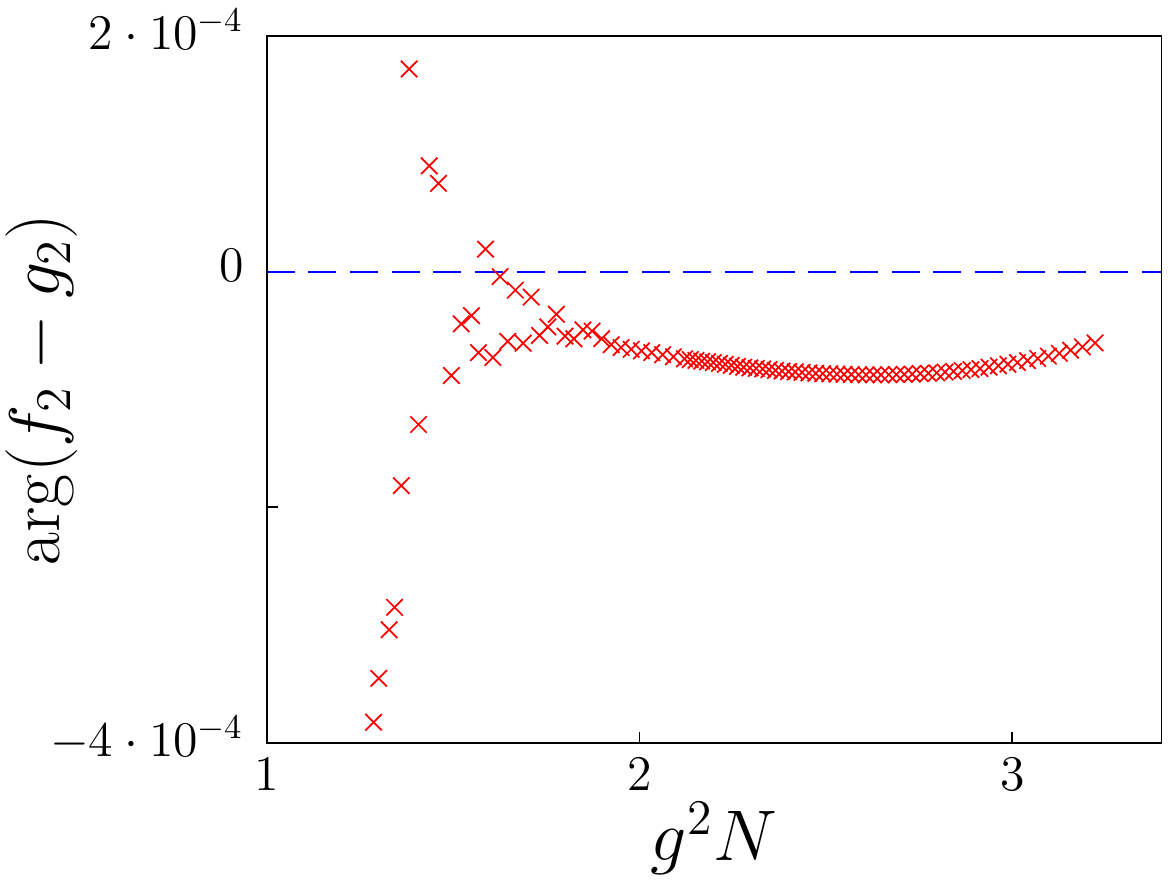}}
\caption{The discrepancy in the initial condition $\mathrm{arg}\,
  f_2 = \mathrm{arg}\, g_2$ at $T = 0.026$ and different~$\theta$; $E>E_{cb}$. The
  standard lattice size and the space--time box from
  Fig.~\ref{fig:linearE} are used.\label{fig:zmode}}
\end{figure}

Let us turn to discretization effects which should be 
$O(\Delta t^2)$ and $O(\Delta x^2)$ small in the second--order
finite--difference scheme. First of all, one observes that
conservation of the full energy in Fig.~\ref{fig:linearE}a is violated 
at the level of ${\Delta   E_{\mathrm{full}}/E_{\mathrm{full}}
  \lesssim 10^{-3}}$. This is the effect of time discretization because
energy conservation is restored at $\Delta t\to 0$ and finite
$\Delta x$. Second, we directly estimate the finite--difference errors
comparing numerical results at different lattices, see
Figs.~\ref{fig:linearE}b and~\ref{fig:dxdt}. Recall that our
reference--point lattices are different at energies below and above
$E_{cb}$: $N_t \times N_x = 3200 
\times 150$ and $11000\times 500$, respectively. In the former case   
the relative errors in $E$, $N$ and $F_N$ stay below $1\%$  
reaching maximum at the smallest~$N$. Fine lattice resolution at high
energies gives errors well below $1\%$ at $E\sim E_{cb} \approx 6.1$
and high~$N$. The errors grow, however, to $1\%$ at $E\gtrsim 14/g^2$
and/or small multiplicities. In what follows we exclude the results with
$g^2 E > 14$ and $g^2 N < 1$ due to improperly high discretization
errors.

To conclude, we keep the finite--difference effects below the relative
level of $1\%$. Figure~\ref{fig:dxdt} shows that most of the integral
quantities linearly depend on $\Delta x^2 \propto N_x^{-2}$ 
and $\Delta t^2 \propto  N_t^{-2}$, like they should in the
second--order discretization scheme. The only exception is the energy
$E$ (lower panel in Fig.~\ref{fig:dxdt}b) which receives small
nonpolynomial correction from adiabatic high--frequency waves in
the solution. This effect is negligible in the region
$E\lesssim 14/g^2$ which we consider (see, however, the study of the
high--energy solutions in~\cite{Demidov:2015}). 

In Sec.~\ref{sec:discretization} we traded the initial condition
$\mathrm{arg}\, f_{n_0} = \mathrm{arg}\, g_{n_0}$ at a given $n_0$
for the artificial constraint fixing the time--translation
invariance. We argued that the omitted condition will be
automatically satisfied once the other lattice equations are solved
correctly. In Fig.~\ref{fig:zmode} we demonstrate that
this is, indeed, the case: at $n_0 = 2$ the related numerical error
is smaller than~$10^{-3}$.   

\begin{figure}[t]
\hspace{4.5cm}(a) \hspace{7.3cm} (b)

\centerline{\includegraphics[width=0.45\textwidth]{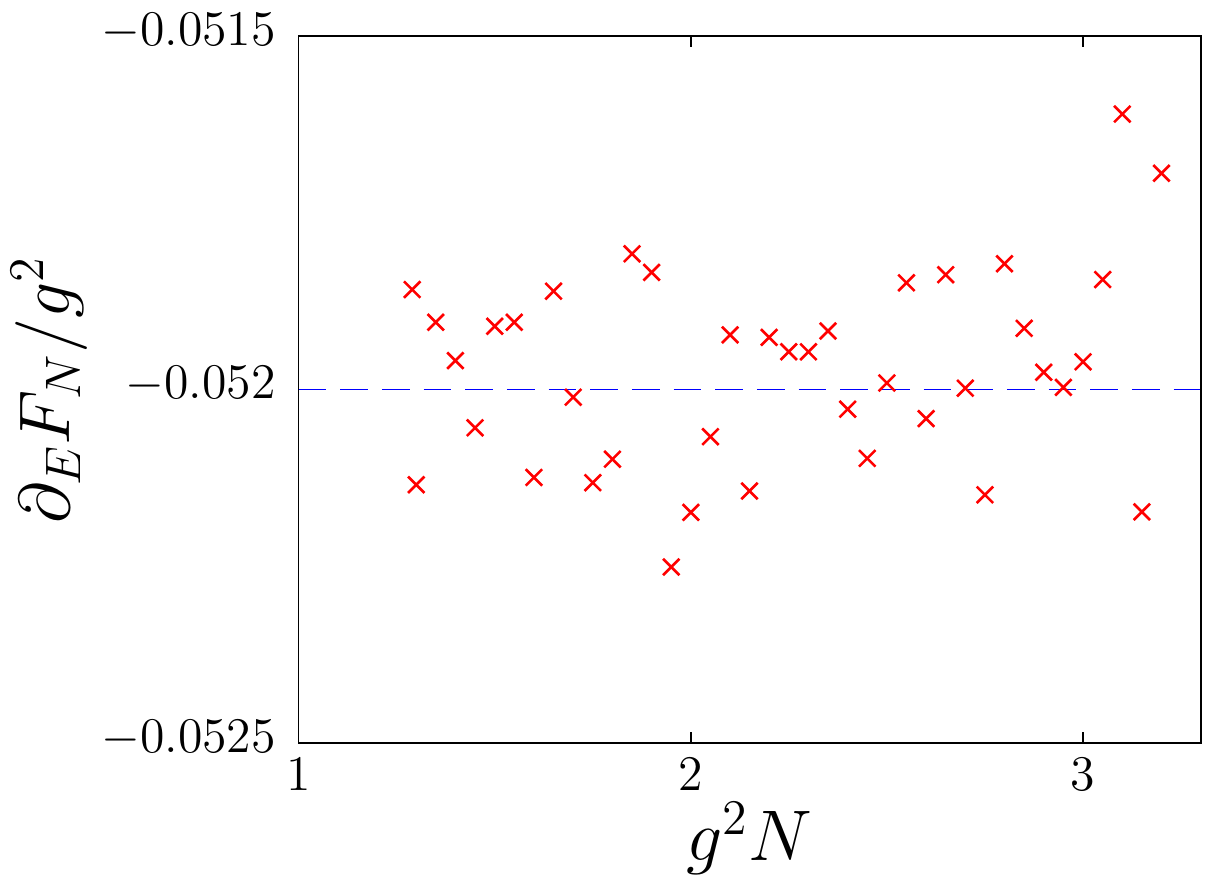} \hspace{5mm}
\includegraphics[width=0.45\textwidth]{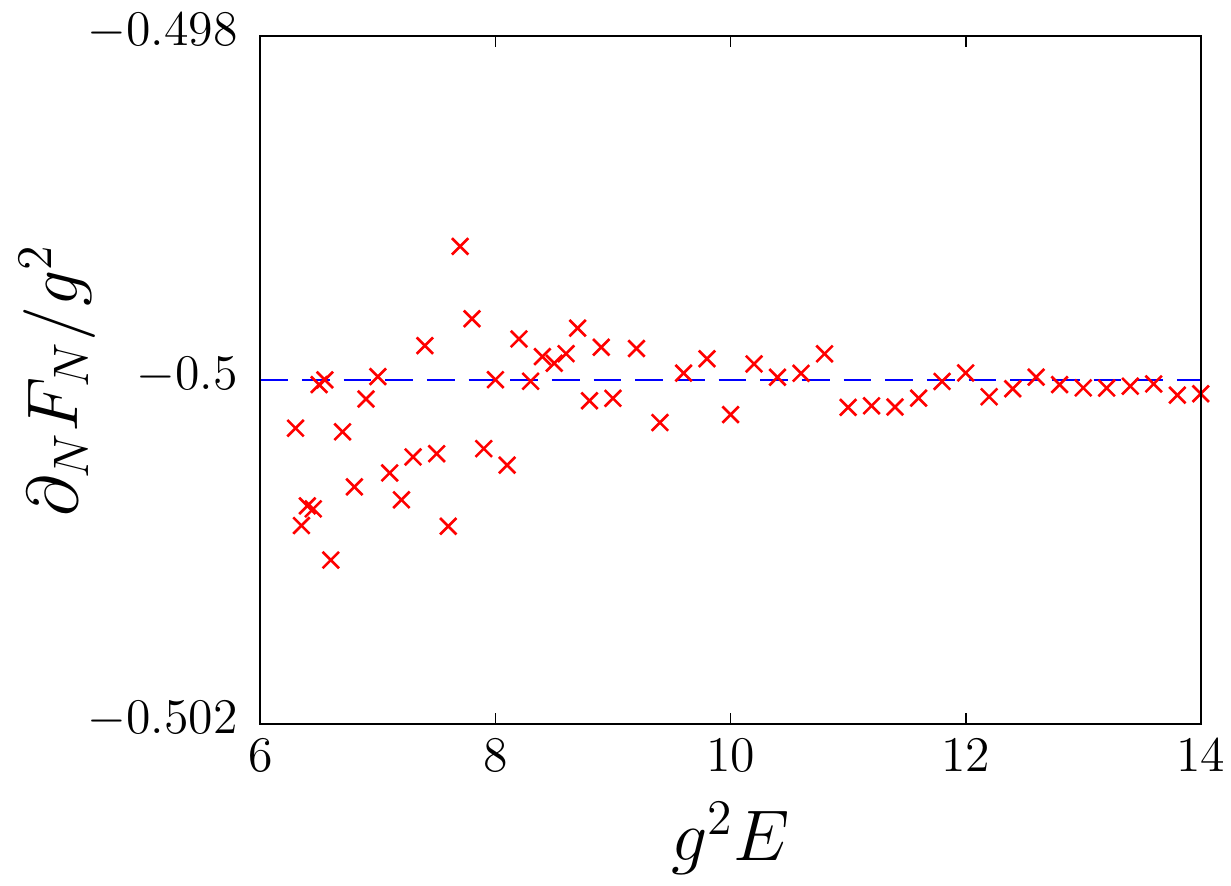}}
\caption{\label{dSdTdtheta} Partial derivatives of the suppression
  exponent with respect to the (a) energy~$E$ and (b) initial particle
  number $N$. The graphs are plotted along the lines (a) $T = 0.026$
  and (b) $\theta = 0.5$. }
\end{figure}

A good piece of our qualitative and quantitative results
is based on the Legendre transforms (\ref{eq:6}) which should
hold with acceptable numerical precision. In Fig.~\ref{dSdTdtheta} we
plot the partial derivatives $(\partial_E F_N,\, \partial_N F_N)/g^2$ of
the suppression exponent and compare them to $- 2 T$ and
$-\theta$ (dashed horizontal lines). One observes that the Legendre
transforms hold with absolute precision $\Delta T,\, \Delta \theta
\lesssim 10^{-3}$.  

To summarize, the relative finite--volume and discretization effects
in our solutions are smaller than $10^{-3}$ and $10^{-2}$,
respectively. The other numerical errors are vanishingly small. 

\section{Solving the linear system}
\label{sec:numerical-alorithm}
Each iteration of the Newton--Raphson method involves solution to the 
system of $({N_t+3})\times N_x \gtrsim 10^6$ complex linear
equations~\eqref{eq:34}. This is the major CPU time--consuming part of
the numerical procedure. We decrease its computational cost using the
sparse structure of the lattice field equations.

In what follows we suppress the spatial indices working with the
$N_x$--dimensional vector field $\phi_j = (\phi_{j,\, 0} ,\,   
\dots ,\, \phi_{j,\,  N_x-1})$. Recall that the linear system
(\ref{eq:34}) is obtained by substituting ${\phi_j = \phi^{(0)}_j +
  \delta \phi_j}$ into the lattice equations and expanding them to the
linear order in~$\delta\phi_j$. In particular, the discrete field
equations (\ref{eq:36}) take the form, 
\begin{equation}
\label{eq:37}
C_j \cdot \delta \phi_j =  L_j \cdot \delta \phi_{j-1} + R_j\cdot
\delta \phi_{j+1} - {\cal F}_j\;,
\end{equation}
where ${\cal F}_j$ is their left--hand side at
$\phi = \phi^{(0)}$, while $C_j$, $L_j$ and $R_j$ are the $N_x\times
N_x$ coefficient matrices which can be deduced from
Eqs.~(\ref{eq:36}), e.g.\ 
$(C_j)_{i,\, i'} = \partial {\cal F}_{j,\, i} /
\partial \phi_{j,\, i'} \big|_{\phi^{(0)}}$. Note that Eq.~(\ref{eq:37}) is sparse: it relates $\delta
\phi_{j-1}$, $\delta \phi_{j}$ and $\delta \phi_{j+1}$. Besides, the
matrices $L_j$ and $R_j$ are diagonal, while $C_j$ is
three--diagonal. We will use both these facts while solving the  
linear system.

Our ``stable'' algorithm~\cite{Rebbi_QM} eliminates equations
from the set (\ref{eq:37}). Namely, suppose we express $\delta
\phi_{j}$ from the $j$-th equation, 
\begin{equation}
\label{eq:40}
\delta \phi_j =  C_j^{-1} \left(L_j \cdot \delta \phi_{j-1} + 
R_j\cdot \delta \phi_{j+1} - {\cal F}_j\right)\;,
\end{equation}
 and substitute it into the
neighbouring {$(j\mp 1)$--th} equations. The latter will keep the form 
(\ref{eq:37}) albeit with new coefficient matrices. In particular, the
$(j-1)$--th equation will relate $\delta \phi_{j-2}$, $\delta
\phi_{j-1}$ and $\delta \phi_{j+1}$ with
\begin{align}
\notag
&C_{j-1}' = C_{j-1} - R_{j-1} \, C_{j}^{-1}\,  L_j \;, &&
L_{j-1}' = L_{j-1}\;,\\
\label{eq:38}
&R_{j-1}' = R_{j-1}\,  C_{j}^{-1} \, R_{j} \;, &&
{\cal F}_{j-1}' = {\cal F}_{j-1} + R_{j-1}\, C_{j}^{-1} \,{\cal
  F}_j\;. 
\end{align}
Repeatedly using Eqs.~(\ref{eq:38}), we eliminate all field 
equations except for the very first and last ones at $j = 0,\,
N_t$.  To make this procedure more stable, we first
apply~(\ref{eq:38}) to the equations with odd $j$, then eliminate   
odd equations from the remaining set, etc. Once we are left with
the equations at $j = 0$ and $N_t$ relating $\delta \phi_{-1}$, $\delta
\phi_0$, $\delta \phi_{N_t}$, $\delta \phi_{N_t+1}$, we add the 
linearized boundary conditions  and solve the
resulting linear system by the direct method of LU
decomposition. The complete solution $\{\delta \phi_j\}$ is restored
from the boundary values of $\delta \phi$ and Eqs.~(\ref{eq:40}).   

The above elimination process is apparently ineffective. Indeed,
the substitutions (\ref{eq:38}) do not preserve the sparse form of the
coefficient matrices thus requiring general matrix multiplications and 
inversions at the second stage of the elimination process. One concludes that
$c N_t N_x^3$ operations with $c\sim 1$ are needed for obtaining the
solution. Note, however, that the disadvantages of this ``slow''
algorithm are compensated by its exceptional stability properties. We
exploit it at $E < E_{cb}$ where the faster procedure accumulates 
round--off errors and causes divergence of the Newton--Raphson
iterations.

Our ``fast'' method benefits from the sparse form of the coefficient
matrices in Eq.~(\ref{eq:37}). It is based on Cauchy problem for
the lattice field equation. Consider the $N_x \times (2N_x+1)$ matrix
$\Xi_j$ satisfying the analog of Eq.~(\ref{eq:37}),
\begin{equation}
\label{eq:39}
C_j \cdot \Xi_j =  L_j \cdot \Xi_{j-1} + R_j\cdot
\Xi_{j+1} - \left(\mathbf{0}, \, \mathbf{0},\,  {\cal F}_j\right)\;,
\end{equation}
where the last matrix term in the right--hand side contains the vector
${\cal F}_j$ in the last column; the bold~$\mathbf{0}$ and
$\boldsymbol{1}$ denote zero and unit $N_x\times N_x$ matrices. We
solve the Cauchy problem for the ``propagator'' $\Xi$ with the initial
conditions 
\begin{equation}
\label{eq:41}
\Xi_{-1} = (\boldsymbol{1},\, \boldsymbol{0},\, 0)\;, \qquad\qquad 
\Xi_{0} = (\boldsymbol{0},\, \boldsymbol{1},\, 0)\;.
\end{equation}
This procedure involves multiplications of $\Xi_j$ by the diagonal
and three--diagonal matrices $L_j$, $R_j^{-1}$ and $C_j$, i.e.\ $c N_t
N_x^2$ operations, $c\sim 1$.

Given the propagator $\Xi$, we introduce a convenient representation
of the solution,
\begin{equation}
\label{eq:42}
\delta \phi_j = \Xi_j \cdot \begin{pmatrix} \delta\phi_{-1}
  \\ \delta\phi_0 \\ 1\end{pmatrix}\;,
\end{equation}
where the $(2N_x+1)$--vector in the right--hand side is composed from
the boundary fields $\delta \phi_{-1}$ and $\delta \phi_{0}$.  One can
check that the vector (\ref{eq:42}) 
satisfies Eq.~(\ref{eq:37}). Equation \eqref{eq:42} relates, in
particular, $\delta \phi_{N_t}$, $\delta \phi_{N_t+1}$ to $\delta
\phi_{-1}$ and $\delta \phi_{0}$. Adding these two relations to the
set of boundary conditions, we obtain a closed linear system for
$\delta \phi_{-1}$, $\delta \phi_{0}$, $\delta \phi_{N_t}$ and $\delta  
\phi_{N_t+1}$. We solve the latter using the LU decomposition
method. Given $\delta \phi_{-1}$ and $\delta \phi_0$, we
Cauchy--evolve Eq.~(\ref{eq:37}) determining the solution $\{\delta 
\phi_j\}$. 

\begin{figure}[t]
\hspace{4.3cm}(a) \hspace{7.8cm}(b)

\centerline{\includegraphics[width=0.45\textwidth]{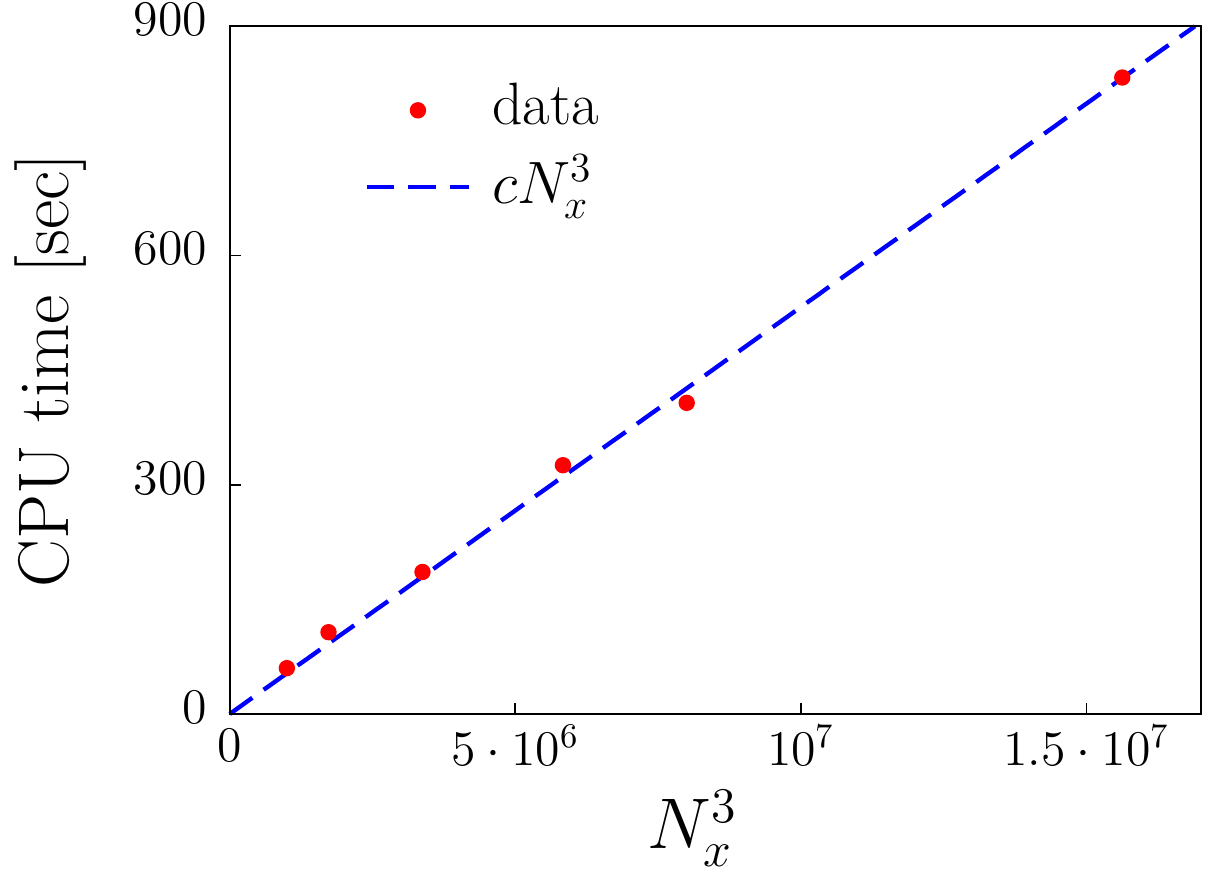}\hspace{5mm}
\includegraphics[width=0.45\textwidth]{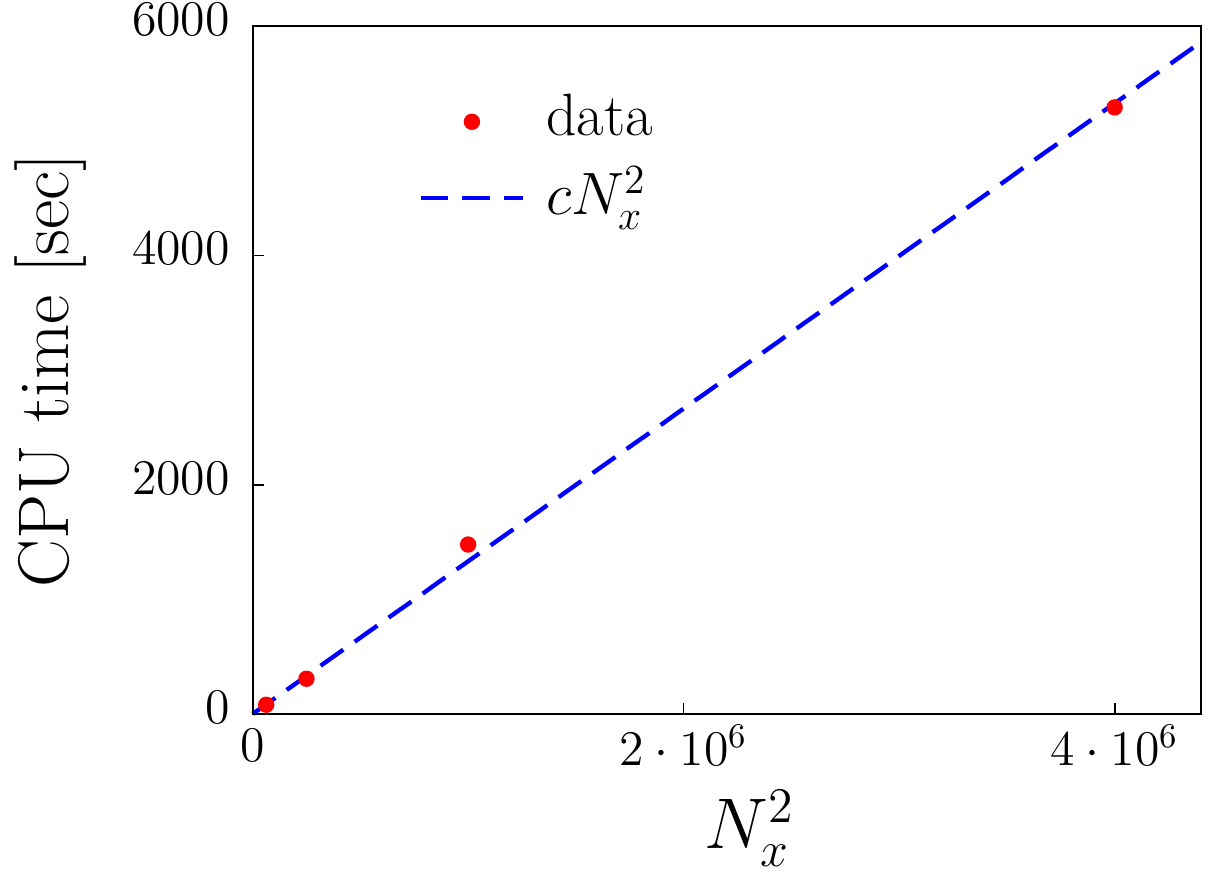}}
\caption{Scaling with the lattice size $N_x$ of the CPU time required
  for: (a) elimination of the equations in the ``slow'' algorithm, (b)
  computation of the propagator $\Xi$ in the ``fast'' one. Eight
  processors are used. Regard drastically different lattice ranges in
  Figs.~(a), (b): $N_t = 3200$, $N_x = 100\div 250$ and $N_t = 11000$,
  $N_x = 250\div 2000$, respectively.\label{fig:CPUtime}}   
\end{figure}
\begin{figure}[h]
\centerline{\includegraphics[width=0.45\textwidth]{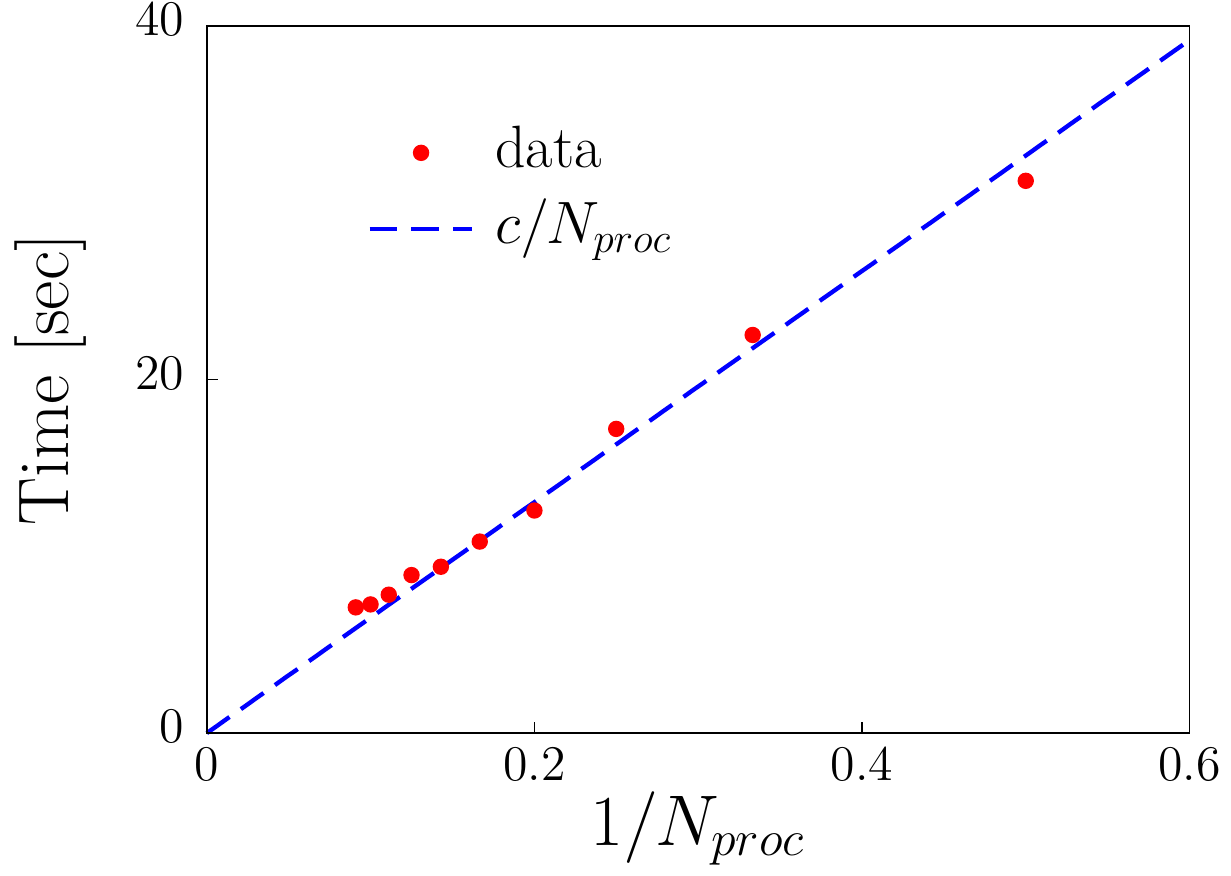}}
\caption{The time for computing the propagator $\Xi$ at
  $N_{proc} =   2\div 11$ processors; $N_t \times N_x = 11000 \times
  250$. The data points scale as $c/N_{proc}$ (dashed
  line). \label{fig:CPU_proc}}
\end{figure}

In the ``fast'' method we arrive at the solution in  $cN_t N_x^2$
operations, a factor of $N_x$ faster than in the ``slow''
algorithm. The $N_x^3$  and $N_x^2$ scalings of the CPU times in our
``slow'' and ``fast'' codes are illustrated in
Figs.~\ref{fig:CPUtime}a,b.

Let us point at the reason behind the poor stability properties of the
``fast'' algorithm at low energies. We argued in Sec.~\ref{sec:3}
that the distinctive property of the semiclassical solutions
at $E<E_{cb}$ is long periods $T$ of their Euclidean
evolutions. Linear  perturbations $\delta
\phi_j$ grow exponentially in Euclidean spacetime magnifying
the initial round--off errors. They cannot be correctly evolved within the
``fast'' Cauchy approach as opposed to the homogeneous ``slow''
algorithm. As a consequence, we exploit the ``fast'' procedure only for
almost--Minkowskian solutions  at $E>E_{cb}$. 

Both our algorithms can be performed in parallel. To
this end one divides equations~(\ref{eq:37}) into the subsets 
with index ranges $j   = 0 \dots N_k$, $N_k \dots 2N_k$, etc., and
performs computations in every subset at a separate processor. At the 
second stage of the algorithm one combines the data from all
processors. For example, in the parallel version of the ``fast''
algorithm the propagators $\Xi^{(k)}_j$ in all subsets are
computed, and the original propagator $\Xi$ is restored as their
product. In Fig.~\ref{fig:CPU_proc} we demonstrate that the elapsed
real time for obtaining $\Xi$ scales as $c/N_{proc}$ with the number
of processors.

\section{Thin--wall approximation}
\label{App:C}
The semiclassical solutions describing false vacuum decay can be
found analytically if the sizes of their true vacuum bubbles are
parametrically large compared to the widths of the bubble walls. We
will see that this thin--wall regime~\cite{Coleman, 
  Voloshin:1986zq, voloshin_induced, Rubakov:1992gi} occurs at 
$\delta \rho \to 0$ and~${E<E_{cb}}$.

\begin{figure}[t]
\centerline{\includegraphics[height=0.35\textwidth]{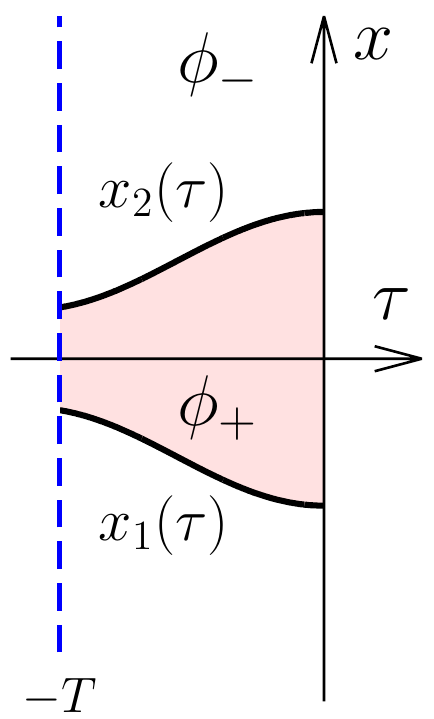}\hspace{1cm}
\includegraphics[height=0.35\textwidth]{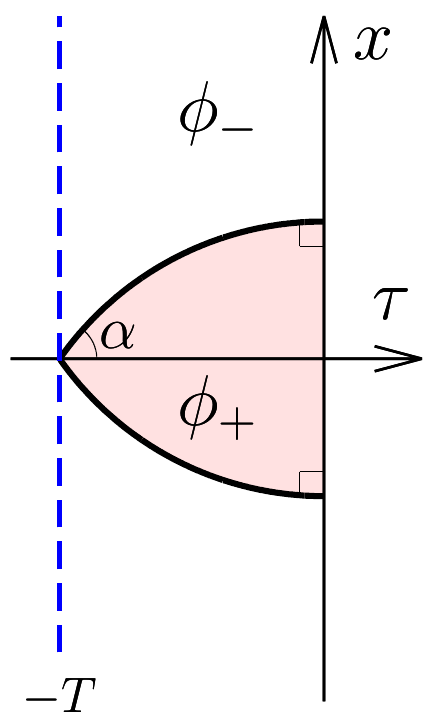} \hspace{1cm}
\includegraphics[height=0.35\textwidth]{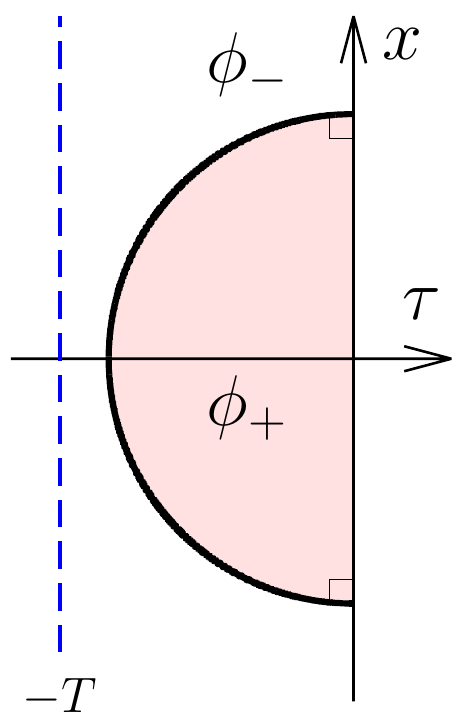}\hspace{1cm}}
\hspace{2.5cm}(a) \hspace{3.8cm} (b)\hspace{4.3cm} (c)
\caption{Thin--wall configurations in Euclidean spacetime
  $(\tau,\, x)$: general configuration (a) and the stationary
  solutions with (b) $T<R_b$ and  (c) $T>R_b$. Pink/gray and white regions are
  filled with the true and false vacua,
  respectively.\label{thin-bubbles}}   
\end{figure}
Consider some Euclidean field configuration with a large true
vacuum bubble, see Fig.~\ref{thin-bubbles}a. At a crude level it can be  
characterized with the positions of the bubble walls $x_1(\tau)$ and 
$x_2(\tau)$, where $\tau \equiv it$ is a Euclidean time. The
corresponding Euclidean action  receives large contributions from the
constant energy density 
$(-\delta \rho/g^2)$ inside the bubble and tension of its walls,
cf.\ Eq.~(\ref{eq:1}), 
\begin{equation}
\label{thin-action}
S_{E} \approx M_{S}L - \delta\rho A/g^2 = \int_{-T}^0 d\tau \left[  M_S
\sqrt{1 + \dot{x}_1^2} + M_S \sqrt{1 + \dot{x}_2^2} - \delta \rho \,(x_2
- x_1)/g^2 \right] \;.
\end{equation}
Here $A$ is the spacetime area occupied by the true vacuum, $L$ is the
length of its boundaries; in the second equality we expressed $A$ and $L$
in terms $x_1(\tau)$, $x_2(\tau)$ denoting the $\tau$--derivatives
with dots. Anticipating the limit $\delta \rho\to  
0$, we identified the wall tension with the soliton mass 
$M_S$. In what follows we treat the action (\ref{thin-action}) as a
functional of $x_1(\tau)$ and~$x_2(\tau)$. 

Now, let us construct the periodic instantons, i.e.\
solutions to the Euclidean field equations with the boundary conditions
(\ref{period}). One notes that these are the extrema of the Euclidean
action in the interval $\tau \in [-T,\, 0]$. Indeed, the Neumann
conditions (\ref{period}) mean that the action is 
stationary with respect to the boundary values of the fields at $\tau  = -T$
and~$0$.  Working in the thin--wall approximation, we extremize the
action~(\ref{thin-action}) by varying $x_1(\tau)$ and
$x_2(\tau)$. One can draw some intuition here by noting that the latter
action is similar to the static energy of a soap bubble in two space
dimensions. Its extremum is achieved if the bubble walls $x_1$ and
$x_2$ form circular arcs with fixed curvature radius 
$$
R_b =  \frac{g^2 M_{S}}{\delta\rho}\;.
$$
Extremization with respect to the boundary
values of $x_1$ and $x_2$ gives conditions $\dot{x}_1 = \dot{x}_2 = 0$
at $\tau = -T$ and $0$. The only exception appears if the arcs $x_1$
and $x_2$ meet at
one point as in Fig.~\ref{thin-bubbles}b; then the boundary
condition changes to $\dot{x}_1 = -\dot{x}_2$. 

The stationary thin--wall configurations at $T < R_{b}$
and $T > R_b$ are shown in Figs.~\ref{thin-bubbles}b
and~\ref{thin-bubbles}c, respectively. Substituting them into the
action~(\ref{thin-action}), one obtains,
\begin{equation} 
\label{eq:44}
S_{E} \approx \frac{g^2M_{S}^2}{2\delta\rho} \left\{ \begin{array}{ll} 
2\alpha + \sin{2\alpha}  & \mbox{at} \;\;\;\; T \equiv R_b\sin\alpha < R_b\;,\\
\pi &\mbox{at}\;\;\;\; T > R_b\;,
\end{array}
\right.
\end{equation}
where $\alpha$ is the half--angle between the arcs in
Fig.~\ref{thin-bubbles}b. One obtains Eq.~\eqref{eq:45} for the
Minkowski action $S \equiv iS_E$ in terms of the half--period $T$.  

Let us find out when the thin--wall approximation is trustworthy. One
naively expects this regime to be solid at small $\delta \rho$ because all
bubble sizes are proportional to $R_b \propto 1/\delta\rho$. However,
the distance between the bubble walls in Fig.~\ref{thin-bubbles}b
vanishes at $\alpha\to 0$ as $\alpha^2 R_b$. Comparing it with
the typical wall width $m_+^{-1}$, one obtains the correct condition
\begin{equation}
\label{eq:47}
\alpha \gg (m_+ R_b)^{-1/2}\;,
\end{equation}
for the thin--wall approximation.  This inequality 
breaks down when the solution approaches the critical 
bubble. Indeed, consider the thin--wall energy 
\begin{equation}
\label{eq:48}
E = 2M_{S}\cos{\alpha}\;,
\end{equation}
which is easily deduced from Eq.~(\ref{thin-action}) and Fig.~\ref{thin-bubbles}b. It
tends to $2M_S$ as $\alpha,\,  T \to  0$. Thus, at small
$\delta \rho$ the thin--wall approximation is trustworthy below the
barrier height $E_{cb}$ but breaks at~$E\gtrsim E_{cb}$. 

Next, we consider transitions from the semi--inclusive initial states with
fixed energy $E$ and multiplicity $N$. Note that the periodic
instantons describe particular processes of this kind with ${N =
  N_{PI}(E)}$, see Fig.~\ref{EN}. Their in--states are  obtained by
continuing the solutions from  
${\tau \equiv it = -T}$ to the initial part AB of the time contour in
Fig.~\ref{fig3}a. One notices~\cite{Rubakov:1992gi}, however, that the
$\tau = -T$ sections of the thin--wall configurations in
Figs.~\ref{thin-bubbles}b,c do not depend on $\delta
\rho$. This suggests that the initial states of the periodic instantons
and their matrix elements with the states of smaller multiplicities
are also independent of $\delta \rho$. Thus, decreasing the initial
multiplicity from $N=N_{PI}$ to $N=2$ one at best gets $O(\delta
\rho^0)$ correction to the exponent $F_N(E)$. We
conclude\footnote{Direct 
  calculation~\cite{voloshin_induced} of the thin--wall exponent at 
  $N=2$ confirms this conclusion.} that the  
leading $1/\delta \rho$ part of $F_N(E)$ does not depend on
$N$.  Expressing the action~(\ref{eq:44}) in terms of energy~(\ref{eq:48})
and substituting it into Eq.~(\ref{s_exp}), one obtains the thin--wall
result~\eqref{eq:61}, \eqref{eq_4.2} for the suppression exponent
at $E<E_{cb}$. 

Finally, let us guess the behavior of  $F_N(E)$ near the
point $E\approx 2M_S$ where the asymptotic expansion in $\delta \rho$
breaks down. Inequality (\ref{eq:47}) shows that the thin--wall
approximation is valid at
\begin{equation}
\label{thin_breaking_2}
|E-2M_S| \gg \frac{\delta \rho}{g^4 M_S}\;,
\end{equation}
where we used Eq.~(\ref{eq:48}) and $m_+ \sim g^2 M_S$. One
expects  that the terms of the thin--wall expansion~\eqref{eq:61}
become comparable at the boundary of this interval. In particular,
\begin{equation}
\notag
F_{N,\, 0}\sim \frac{F_{N,\, -1}}{\delta \rho}\approx 
\frac{4 g^4 M_S^{1/2}}{3\delta\rho }\left|2M_s-E\right|^{3/2} \;,
\end{equation}
where the explicit form~\eqref{eq_4.2} was used. We express $\delta
\rho$ from Eq.~(\ref{thin_breaking_2}), and obtain 
$F_{N,0} \sim \mathrm{const}\cdot |E - 2M_S|^{1/2}$. Assuming some
regular contribution 
besides this singular term, we arrive at Eq.~(\ref{eq:49}).  

\bibliographystyle{asmplain}

\end{document}